\documentclass[pdflatex,sn-nature]{sn-jnl}

\usepackage{graphicx}
\usepackage{float}
\usepackage{capt-of}
\usepackage{multirow}
\usepackage{amsmath,amssymb,amsfonts}
\usepackage[mathlines]{lineno}

\usepackage{bm}
\usepackage{amsthm}
\usepackage{mathrsfs}
\usepackage{xcolor}

\usepackage[normalem]{ulem}

\usepackage{textcomp}
\usepackage{manyfoot}
\usepackage{booktabs}
\usepackage{tabularray}
\usepackage{algorithm}
\usepackage{algorithmicx}
\usepackage{algpseudocode}
\hypersetup{hypertexnames=false}

\newcommand{\A}{\mathrm{A}}
\newcommand{\B}{\mathrm{B}}
\newcommand{\C}{\mathrm{C}}
\newcommand{\D}{\mathrm{D}}
\newcommand{\CC}{\mathrm{CC}}
\newcommand{\CD}{\mathrm{CD}}
\newcommand{\DC}{\mathrm{DC}}
\newcommand{\DD}{\mathrm{DD}}
\newcommand{\G}{\mathrm{G}}

\begin{document}

\title{Learning to cooperate in a changing world:\\How caring about the future promotes cooperation across scales}

\author[1,2]{\fnm{Yuxin} \sur{Geng}}
\author*[1,4]{\fnm{Xingru} \sur{Chen}}
\email{xingrucz@gmail.com}
\author[1,2,3,4,5]{\fnm{Xin} \sur{Wang}}
\author[8]{\fnm{Hongwei} \sur{Zheng}}
\author*[1,2,3,4,5]{\fnm{Longzhao} \sur{Liu}}
\email{longzhao@buaa.edu.cn}
\author*[1,2,3,4,5,6,7]{\fnm{Shaoting} \sur{Tang}}
\email{tangshaoting@buaa.edu.cn}
\author[9,10]{\fnm{Feng} \sur{Fu}}

\affil[1]{\orgdiv{School of Artificial Intelligence}, \orgname{Beihang University}, \orgaddress{\city{Beijing}, \postcode{100191}, \country{China}}}
\affil[2]{\orgname{Zhongguancun Laboratory}, \orgaddress{\city{Beijing}, \postcode{100094}, \country{China}}}
\affil[3]{\orgdiv{Hangzhou International Innovation Institute}, \orgname{Beihang University}, \orgaddress{\city{Hangzhou}, \postcode{311115}, \country{China}}}
\affil[4]{\orgdiv{Key Laboratory of Mathematics, Informatics and Behavioral Semantics}, \orgname{Beihang University}, \orgaddress{\city{Beijing}, \postcode{100191}, \country{China}}}
\affil[5]{\orgdiv{Beijing Advanced Innovation Center for Future Blockchain and Privacy Computing}, \orgname{Beihang University}, \orgaddress{\city{Beijing}, \postcode{100191}, \country{China}}}
\affil[6]{\orgdiv{Institute of Trustworthy Artificial Intelligence}, \orgname{Zhejiang Normal University}, \orgaddress{\city{Hangzhou}, \postcode{310012}, \country{China}}}
\affil[7]{\orgdiv{Shandong Key Lab of Complex Medical Intelligence and Aging}, \orgname{Shandong Medical and Pharmaceutical University}, \orgaddress{\city{Yantai}, \postcode{264003}, \country{China}}}
\affil[8]{\orgname{Beijing Academy of Blockchain and Edge Computing}, \orgaddress{\city{Beijing}, \postcode{100085}, \country{China}}}
\affil[9]{Department of Mathematics, Dartmouth College, Hanover, NH 03755, USA}
\affil[10]{Department of Biomedical Data Science, Geisel School of Medicine at Dartmouth, Lebanon, NH 03756, USA}

\abstract{
In social dilemmas, individuals need to forgo short-term temptations to achieve synergistic collective outcomes through cooperation. Previous work has examined mechanisms through which cooperation can evolve, including direct reciprocity, indirect reciprocity, environmental stochasticity, network reciprocity, and demographic stochasticity. These mechanisms have largely been studied under natural selection or social learning, where strategies with higher payoffs are more likely to spread. It is equally important to study cooperation under self-learning, where individuals adapt through their own experiences. Here, we focus on multi-agent reinforcement learning and derive analytical conditions under which these mechanisms stabilize cooperation under learning dynamics across scales. We find that reinforcement learning can steer self-interested individuals toward cooperation when they value the future over short-term temptation. Our work provides a unified approach to identifying the determinants of learning to cooperate in a changing world, thereby paving the way for the advancement of cooperative artificial intelligence.
}


\maketitle

Cooperation is found across diverse levels of biological and social organization \cite{nowak2006five}. In human societies, division of labor, the management of common-pool resources, and the provision of public goods all rely on individuals forgoing short-term benefits \cite{ostrom1990governing,henrich2016secret,olson1971logic}. As decision-making is increasingly delegated to artificial intelligence (AI), autonomous agents are becoming increasingly embedded in social and economic systems, and they face the same challenge of cooperation \cite{calvano2020artificial,banchio2022artificial,barfuss2025collective}. Developing cooperative AI can strengthen collaboration among machines and between humans and machines, expanding humanity's capacity to address shared challenges through collective action \cite{dafoe2021cooperative}.

Research on the evolution of cooperation has examined several mechanisms, including direct reciprocity (repeated interactions) \cite{trivers1971evolution,glynatsi2024conditional}, indirect reciprocity (reputation-based interactions) \cite{nowak1998evolution,santos2018social}, environmental stochasticity (stochastic games) \cite{hilbe2018evolution,weitz2016oscillating}, network reciprocity (structured interactions beyond well-mixed populations) \cite{ohtsuki2006simple,allen2017evolutionary,sheng2024strategy}, and demographic stochasticity (population variations) \cite{traulsen2006stochastic}. The underlying interaction patterns emerge naturally in biological populations and human societies \cite{milinski1987tit,bshary2006image,milinski2002reputation,fowler2010cooperative,doebeli1997population}, but are absent in artificial multi-agent systems (MAS). Most crucially, these mechanisms have largely been studied under natural selection and social learning \cite{nowak2006five,hofbauer1998evolutionary}. The success of a mechanism under reproduction or imitation does not naturally establish whether individuals can learn to sustain cooperation from their own experience.

\begin{figure}[H]
    \centering
    \includegraphics[width=0.99\textwidth,height=\textheight,keepaspectratio]{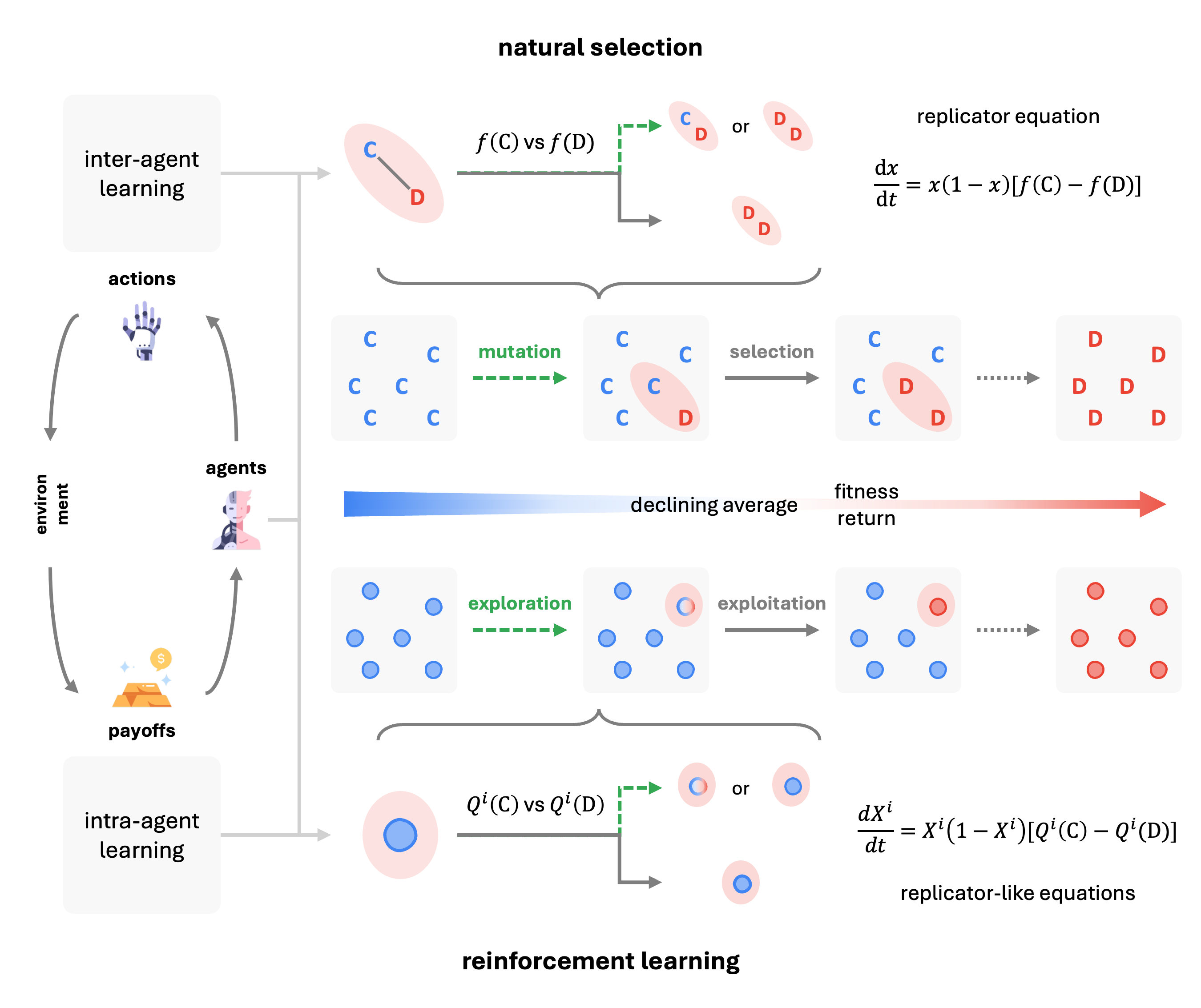}
    \caption{\textbf{Natural selection (social learning) and reinforcement learning as unified feedback systems.} In natural selection, fitness quantifies how well a strategy is adapted to its environment. Through selection and mutation, strategies with higher fitness are more likely to proliferate in the next generation. In reinforcement learning, a Q-value defines the expected cumulative reward (return) obtained by selecting an action. Through exploitation and exploration, actions with higher Q-values are more likely to be selected in the next update step. Both natural selection and reinforcement learning function as coupled feedback loops between agents and the environment, with their dynamics captured by replicator-like equations at different scales. The former models the population-level evolution of strategy frequencies, whereas the latter describes the individual-level adaptation of action probabilities. Without additional mechanisms, both systems converge from cooperation ($\C$) to defection $(\D)$. Cartoon icons in Fig.~\ref{fig:intro-evolution-rl} and Fig.~\ref{fig:intro-five-mechanisms} were adapted from icons created by max.icons and Muhammad Ali and downloaded from \href{https://www.flaticon.com}{Flaticon}.}
    \label{fig:intro-evolution-rl}
\end{figure}

Reinforcement learning (RL) is a general self-learning framework for sequential decision-making problems in complex environments \cite{sutton2018reinforcement}. RL serves as a mechanistic interpretation of how animals \cite{schultz1997neural} and humans \cite{o2004dissociable,pessiglione2006dopamine} learn to make decisions. It is also the training paradigm behind games \cite{mnih2015human}, robotics \cite{kober2013reinforcement}, and large language models (LLMs) \cite{ouyang2022training,guo2025deepseek}. RL describes adaptation of behavior as a feedback loop between agents and their environment. RL agents select actions based on their current policy, receive reward as feedback from the environment, and update their policy accordingly. An action with a higher estimated value will be selected with higher probability in the next update step. Natural selection and social learning share similar dynamics but operate on the scale of population and strategy rather than individual agents. A strategy with higher fitness will be more likely to spread and have higher frequency in the next generation \cite{nowak2006five,hofbauer1998evolutionary}. Continuous-time dynamics of both processes can take the form of replicator equations \cite{nowak2006evolutionary,borgers1997learning,bloembergen2015evolutionary}. In the absence of mechanisms, these adaptations can lead to collectively suboptimal outcomes in social dilemmas, including the tragedy of the commons \cite{hardin1968tragedy} (Fig.~\ref{fig:intro-evolution-rl}). Promoting cooperation among learning agents therefore calls for a theory that quantifies the effects of different forms of social and environmental feedback and connects individual adaptation with collective outcomes \cite{garcia2025picking,barfuss2025collective}.

Understanding the emergence of cooperation in MARL systems requires analyzing the underlying learning dynamics, which have primarily been studied through numerical or agent-based simulations \cite{barfuss2019deterministic,anastassacos2021cooperation,smit2024learning,barfuss2023intrinsic,hughes2018inequity,geng2026emergent}. Building on these foundations, we develop an analytical framework to determine when self-interested agents learn to cooperate. Specifically, we formulate MARL as a coupled dynamical system of agents and their environment, and prove that a strategy profile is asymptotically stable if every prescribed action strictly maximizes the corresponding agent's expected discounted return. Symbolically solving the Bellman equation induced from these returns gives the explicit conditions under which cooperation becomes the ``natural'' learning outcome.

We apply this framework to examine the five mechanisms across interaction scales, from dyadic encounters to well-mixed populations and higher-order graph-structured populations. The analysis also spans different levels of decision-making, from learning over primitive actions to learning over behavioral strategies. For each mechanism, we derive an analytical condition under which cooperation emerges, and validate our prediction through agent-based simulations. These conditions quantify how system hyperparameters jointly shape the emergence and stability of cooperation and the extent to which one can compensate for others. In particular, cooperation depends on the alignment between the temporal structure of reciprocal and environmental feedbacks and the horizon over which individuals evaluate their actions. When agents care enough about the future, RL can steer self-interested agents toward cooperation through these mechanisms. Our analysis brings diverse mechanisms into a unified mathematical criterion and provides a transparent interpretation of how self-interested learners can sustain collective cooperation. These findings establish a theoretical foundation for the design of cooperative and resilient AI.

\section{Results}\label{sec:results}

\subsection{Model overview}

\subsubsection{Game environments}

The game environment is modeled as a partially observable stochastic game \cite{hansen2004dynamic}. Two representative social dilemmas are the donation game for dyadic interactions and the public goods game for multiplayer interactions. In the donation game \cite{nowak2006five}, cooperation induces a cost $c$ and provides a benefit $b$ to the co-player, where $b>c>0$. In the public goods game \cite{sheng2026cooperation}, this cost $c$ is multiplied by the synergy factor $r$ and distributed equally among all $k$ participants, where $1<r<k$. In both games, defection induces no cost and provides no benefit.

\subsubsection{Reinforcement learning algorithms}

Agents optimize their behavior through RL. We consider two representative RL algorithms from the two main families of RL. For value-based methods, we consider $\varepsilon$-greedy Q-learning \cite{watkins1992q,sutton2018reinforcement}. Each agent $i$ maintains a Q-value table $Q^i(o, a)$ for each observation-action pair and selects the action with the highest Q-value with probability $1-\varepsilon$ and otherwise explores the action space randomly. The interaction samples are collected as a batch $\mathcal D^i=\{(o, a, r, o')\}$ of size $B$, where $r$ and $o'$ are the reward and next observation after taking action $a$ under observation $o$. The Q-tables are then updated using the batch-averaged temporal-difference (TD) error as
\begin{equation}
    Q^i(o,a)\leftarrow Q^i(o,a)
    +\alpha \,\mathbb I_{\mathcal D^i}(o,a) \, \mathbb{E}_{\mathcal D^i}\bigl[r+\gamma\max_{a'}Q^i(o',a')-Q^i(o,a)\bigr],
    \label{eq:Q-learning-update}
\end{equation}
where $\alpha$ is the learning rate and $\gamma$ is the discount factor, quantifying how much agents value future rewards. The indicator function $\mathbb{I}_{\mathcal D^i}(o, a)$ equals one if the observation-action pair occurs in the batch and zero otherwise.

For policy-based methods, we consider actor--critic learning \cite{konda1999actor}. Compared with Q-learning, which maintains an action-value estimate and derives its policy from it, actor--critic learning maintains a policy $X^i(o,a)$ (the actor) to make decisions and a value function $V^i(o)$ (the critic) to guide policy updates.

\subsubsection{Analytical models of MARL}

We treat MARL as a dynamical system in which the environmental state and individual behavior are coupled \cite{bloembergen2015evolutionary,barfuss2019deterministic}. For a candidate profile $\bm g=(g^i)_{i\in\mathcal N}$ with $g^i:\mathcal O^i\to\mathcal A$, the following Bellman system describes $i$'s expected discounted return from taking action $a$ at observation $o$ and following $g^i$ thereafter:
\begin{equation}
    Q_{\bm g}^i(o,a)
    =
    \tilde{R}_{\bm g}^i(o,a)
    +
    \gamma\sum_{o'}\tilde{T}_{\bm g}^i(o,a,o')Q_{\bm g}^i(o',g^i(o')).
    \label{eq:bellman-optimality-equation}
\end{equation}
Here, $\tilde{R}_{\bm g}^i(o,a)$ and $\tilde{T}_{\bm g}^i(o,a,o')$ are the expected immediate reward and the probability of observing $o'$ next under the stationary state distribution. We find that if $g^i(o)$ is best-response consistent, that is, $g^i(o)$ uniquely maximizes $Q_{\bm g}^i(o,\cdot)$ for every agent and observation, then $\bm Q_{\bm g}=(Q_{\bm g}^i)_{i\in\mathcal N}$ with greedy profile $\bm g$ is an asymptotically stable equilibrium of the Q-value dynamics under $\varepsilon$-greedy Q-learning. The same best-response criterion establishes the stability of $\bm g$ under actor--critic learning dynamics. Testing whether a mechanism promotes cooperation is therefore reduced to checking whether the cooperative profile $\bm g$ is strictly best-response consistent with its induced Bellman system. In the following, we mainly consider $\varepsilon$-greedy Q-learning, and by ``$\bm g$ is stable'' we mean that $\bm Q_{\bm g}$ is a stable equilibrium of the Q-value dynamics.

\subsection{Learning dynamics in one-shot games}

\begin{figure}[H]
    \centering
    \includegraphics[width=0.75\textwidth,height=\textheight,keepaspectratio]{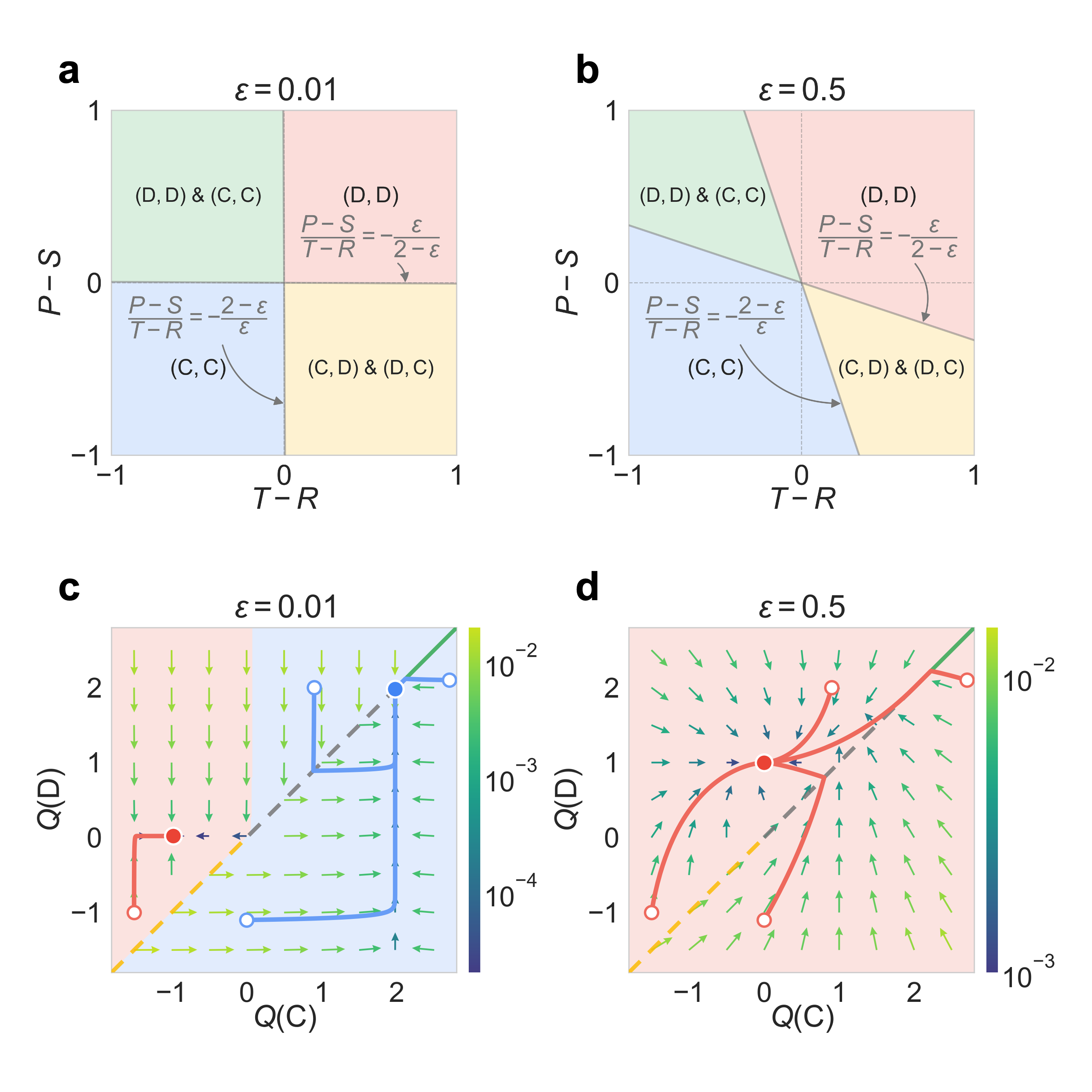}
    \caption{\textbf{Best-response equilibria and pseudo-equilibria in one-shot games.} \textbf{a}, \textbf{b}, Best-response profiles of two symmetric $\varepsilon$-greedy Q-learners in $2\times2$ one-shot games. Here, $R$, $S$, $T$ and $P$ denote the payoffs of mutual cooperation, unilateral cooperation, unilateral defection, and mutual defection. The lower-left, upper-left, lower-right, and upper-right quadrants of the $(T-R, P-S)$ plane correspond to the Harmony, Stag Hunt, Snowdrift, and Prisoner's Dilemma games. The two lines partition the plane into four colored regions with distinct sets of best-response profiles $(g^i,g^j)$ of the players' greedy actions. In the rare-exploration limit $\varepsilon\to0$, the two lines coincide with the coordinate axes, and the best-response profiles match the pure-strategy Nash equilibria. As $\varepsilon$ increases from $0.01$ in \textbf{a} to $0.5$ in \textbf{b}, the boundaries rotate toward the anti-diagonal. \textbf{c}, \textbf{d}, Q-value dynamics of two symmetric $\varepsilon$-greedy Q-learners in a one-shot donation game with $(R,S,T,P)=(b-c,-c,b,0)$, as an instance of the Prisoner's Dilemma game. Arrows encode the direction and magnitude of the Q-value drift. Curves show representative trajectories, with open and solid dots marking their initial values and limiting equilibria. The background shading indicates the corresponding basin of attraction of each equilibrium. The diagonal $Q(\C)=Q(\D)$ separates the cooperation-greedy and defection-greedy regions, and the orange dashed, gray dashed, and green solid portions denote the escaping, crossing, and attracting sliding segments of the Filippov differential inclusion induced by the discontinuous switch between the dynamics in the two regions. In panel \textbf{c}, under $\varepsilon=0.01<1-\sqrt{c/b}$, the boundary pseudo-equilibrium coexists with the mutual-defection equilibrium and attracts trajectories from all sufficiently optimistic Q-value initializations. In panel \textbf{d}, as $\varepsilon=0.5$ exceeds the threshold $1-\sqrt{c/b}$, mutual defection is the unique equilibrium. Parameters: \textbf{c}, \textbf{d}, $\gamma=0.5$, $\alpha=0.01$, $B=1$, and the payoff matrix is that of a donation game with $b=2$ and $c=1$.}
    \label{fig:intro-normal-form}
\end{figure}

In Fig.~\ref{fig:intro-normal-form}a,b, we characterize the best-response equilibria in symmetric $2 \times 2$ one-shot games. For the donation game, mutual defection is always the unique best-response equilibrium at any exploration rate. However, in Fig.~\ref{fig:intro-normal-form}c,d, an additional attracting pseudo-equilibrium exists on the Q-value boundary $Q(\C)=Q(\D)$ when
\begin{equation}
    \varepsilon<1-\sqrt{c/b}.
    \label{eq:main-pseudo-equilibrium}
\end{equation}
At this equilibrium, agents select cooperation more frequently than defection, and the cooperation rate tends to one as $\varepsilon\to0$. Trajectories starting from all sufficiently optimistic initial states
converge to this equilibrium. However, this equilibrium arises from insufficient exploration and disappears when $\varepsilon$ exceeds the threshold. In the following, we focus on mechanisms through which cooperation can emerge endogenously.

\subsection{Direct reciprocity}

When the same individuals interact repeatedly, each can condition its behavior on their interaction history. We consider a repeated donation game between two memory-one agents. Each one observes the previous joint action and decides whether to cooperate. 

We identify two cooperative equilibria whose greedy policies correspond to the Grim Trigger (GRIM) or the Win-Stay, Lose-Shift (WSLS) strategy. GRIM cooperates after mutual cooperation but defects permanently after the co-player's defection. WSLS cooperates after the matching outcomes $\CC$ and $\DD$, and defects after the mismatching outcomes $\CD$ and $\DC$ \cite{nowak1993strategy}. The stability conditions of these two equilibria are given by
\begin{equation}
    \gamma > c/b \quad \text{ and } \quad \gamma > c/(b-c). \label{eq:dr-rule}
\end{equation}
In the Supplementary Information, we show that GRIM becomes unstable for large $\gamma$ and small $\varepsilon$, whereas WSLS remains stable and is robustly learned in the simulation, as shown in Fig.~\ref{fig:intro-five-mechanisms}a. 

In contrast, the strategy of Tit-for-Tat (TFT) \cite{axelrod1981evolution}, which cooperates after the co-player cooperates and defects after the co-player defects, does not correspond to an equilibrium. TFT induces its co-player to deviate from TFT by adopting Always Cooperate (ALLC) when $\gamma>c/b$ and Always Defect (ALLD) otherwise.

\subsection{Indirect reciprocity}

Reciprocity does not require the same individuals to meet repeatedly. An action observed by others can change their evaluation of the actor, which in turn shapes the responses of future co-players \cite{alexander2017biology}. We consider a population in which each individual carries one of the two reputations, good ($\G$) or bad ($\B$). Agents can observe their co-player's reputation with probability $q$ (with an unobserved co-player treated as good) and make decisions based on both players' reputations in a donation game. After each interaction, the reputation is updated through a given social norm, which maps the donor's action and the two players' reputations to the donor's new reputation. 

Under the social norm of image scoring, cooperation and defection are assigned $\G$ and $\B$ \cite{nowak1998evolution}. In Fig.~\ref{fig:intro-five-mechanisms}b, we run the simulation in which agents always defect against bad recipients and learn to decide whether to cooperate with a good one. The agents' greedy policies change from ALLD to the Discriminator (DISC) strategy. DISC cooperates with recipients it assesses as $\G$ and defects otherwise. Within this strategy space, we find that DISC is stable if and only if
\begin{equation}
    \gamma q>c/b. \label{eq:ir-rule-1}
\end{equation}

As image scoring assigns reputations based solely on donors' actions, it cannot distinguish punitive defection against a bad recipient from exploitation of a good one. Richer contextual information can be encoded in social norms by conditioning reputation updates on the donor's and recipient's reputations as well as the donor's action. Eight such norm--strategy pairs, known as the ``leading eight,'' encode patterns of judgment and behavior that are widely recognized in human societies \cite{ohtsuki2004should,ohtsuki2006leading}. For example, cooperation with a good recipient is good, defection against a good recipient is bad, and defection by a good donor against a bad recipient is viewed as a justified sanction. Under $q\to 1$, we find that all eight cooperative strategies can be stably sustained under their associated social norms if $\gamma > c/b$. In Supplementary Fig.~S3, we conduct agent-based simulations to show that agents initialized at ALLD can converge to each of the eight strategies.

\subsection{Environmental stochasticity}

Agents' actions affect not only their co-players but also the environment, which in turn shapes their subsequent behavior. Such environmental feedback is captured by a two-state donation game \cite{hilbe2018evolution}. In both states, cooperation incurs the same cost $c$ but yields different benefits, which are $b_{\A}$ in the productive state $\A$ and $b_{\B}$ in the degraded state $\B$, with $b_{\A}>b_{\B}$. Mutual cooperation maintains state $\A$ or restores it from $\B$, whereas any defection leads to state $\B$ in the next time step. Each agent observes the current environmental state and decides whether to cooperate accordingly.

We identify two cooperative equilibria with greedy policies $(\C,\D)$ and $(\C,\C)$, where the two components specify greedy actions in states $\A$ and $\B$. Their stability conditions are given by
\begin{equation}
\gamma > c/b_{\A} \quad \text{and} \quad
\gamma > c/(b_{\A}-b_{\B}).
\label{eq:es-rules}
\end{equation}
Similar to the GRIM strategy under direct reciprocity, the $(\C, \D)$ strategy is fragile to exploratory noise. As shown in Fig.~\ref{fig:intro-five-mechanisms}c, starting from defection, agents consistently learn to adopt the more robust $(\C, \C)$ strategy.

\subsection{Network reciprocity}

Beyond pairwise encounters and well-mixed populations, real-world interactions are often local and take place in groups \cite{mcavoy2020social}. A natural modeling approach is to describe the population as a hypergraph, in which each node represents an agent and each hyperedge contains the participants in a game \cite{sheng2024strategy}.

In normal-form public goods games, defection is the unique best-response equilibrium. We further consider a two-state public goods game with synergy factors $r_\A$ and $r_\B$ in the prosperous state $\A$ and the degraded state $\B$. In such group interactions, each defector increases the risk of degradation. Specifically, the system transitions from $\A$ to $\B$ with probability equal to the fraction of defectors. On the other hand, recovery from $\B$ to $\A$ requires all group members to cooperate. In Fig.~\ref{fig:intro-five-mechanisms}d, networks with lower hyperedge order $k$ converge to the cooperative $(\C, \C)$ strategy. We find that $(\C, \C)$ is stable if and only if
\begin{equation}
    k<r_{\A}+\gamma (r_{\A}-r_{\B}),
    \label{eq:main-network-threshold}
\end{equation}
where $\gamma(r_{\A}-r_{\B})\geq1$. In addition, the network degree $d$ changes the number of interaction samples in a batch collected between updates. In the Supplementary Fig.~S4, agents converge to cooperation sooner in sparser networks with lower $d$. This is consistent with smaller batches producing stronger sampling fluctuations and thereby accelerating escape from defection.

\subsection{Demographic stochasticity}

RL agents perform updates through finite samples of interactions. Variability in the sampled actions and state transitions introduces intrinsic fluctuations into their learning dynamics \cite{galla2009intrinsic}. In Fig.~\ref{fig:intro-five-mechanisms}a,c, although ALLD is a best-response equilibrium, agents can still leave its basin of attraction and converge to cooperative strategies. Under a non-vanishing learning rate $\alpha$, the learning dynamics are described by a stochastic differential equation (SDE) $\mathrm d\bm Q_t=\bm\mu\,\mathrm dt+\sqrt{\bm\Sigma}\,\mathrm d\bm W_t$, where $\bm\mu$ is the expected update, $\bm\Sigma$ is the covariance of the finite-batch update, and $\bm W_t$ is a standard Brownian motion. We find that the stochasticity of the system, quantified by the diffusion term $\sqrt{\bm\Sigma}$, scales with the batch size $B$ as
\begin{equation}
    \sqrt{\bm\Sigma}=O\!\Big(B^{-1/2}\Big).
    \label{eq:main-batch-noise-scaling}
\end{equation}
Decreasing $B$ strengthens sampling fluctuations and facilitates transitions between equilibria. In Fig.~\ref{fig:intro-five-mechanisms}e, we examine self-play Q-learning in a one-shot Stag Hunt game, where mutual cooperation and mutual defection are both best-response equilibria. In simulations initialized at defection, smaller batches lead to faster transitions into cooperation \cite{barfuss2023intrinsic}.

The long-run fraction of time agents spend in cooperation depends on the relative transition rates between cooperation and defection. Under self-play, the dynamics reduce to a piecewise Ornstein-Uhlenbeck process for the Q-value difference $Q(\C)-Q(\D)$. The resulting transition rates between the two equilibria decay exponentially with $B$. Under rare exploration $\varepsilon\ll1$, if cooperation is risk-dominant, $R+S>T+P$, transitions from defection to cooperation occur more frequently than the reverse. Consequently, the system spends a larger fraction of time in cooperation.

\subsection{Direct reciprocity under meta-policy learning}

Learning is not confined to primitive actions. The question remains whether mechanisms that promote cooperation remain effective in higher-level learning processes. In Fig.~\ref{fig:intro-reactive-meta-game}, we examine direct reciprocity when agents learn to choose among behavioral strategies rather than primitive actions. Each meta-action is a reactive strategy $(p_1,p_2)$, where $p_1$ and $p_2$ are the probabilities of cooperation after the co-player cooperated or defected in the previous round. The meta-policy assigns a probability to each such strategy and is updated through actor--critic learning, and the reward for each strategy is its long-run expected payoff against the current population. In a population initialized with a uniformly random meta-policy, agents transition through ALLD and TFT to Generous Tit-for-Tat (GTFT) \cite{nowak1992tit}. The meta-policy finally enters a rock-paper-scissors-like cycle among GTFT, ALLC, and Anti-TFT, during which cooperation remains close to one, and the average payoff remains near the level of mutual cooperation.

\section{Discussion}\label{sec:discussion}

In social dilemmas, self-interested learners inherently lack the incentive to cooperate, driving systems toward collectively inefficient outcomes \cite{leibo2017multi}. Even when cooperative behavior appears to emerge, such as at the pseudo-equilibrium in Fig.~\ref{fig:intro-normal-form}c, it is merely sustained by insufficient exploration and biased value estimates, while defection remains the superior action. The pivotal question is under what conditions cooperation itself becomes the more profitable action that aligns with individual self-interest. To uncover the mechanisms governing this alignment, we develop an analytical framework that characterizes learning dynamics across different scales of interaction and levels of learning.

The criterion we obtained checks whether each prescribed action delivers higher payoff than the alternatives, which makes explicit the individual interest constraints at the level of states and actions that stabilize collective cooperation. In comparison, the evolutionarily stable strategy (ESS) \cite{smith1973logic} is a population- and strategy-level equilibrium concept. A population at an ESS resists invasion by rare mutants, as determined by comparisons of the overall payoffs of resident and mutant strategies. In this sense, meta-policy learning operates at the individual- and strategy-level, which naturally bridges self-learning and social learning. In meta-policy learning, individuals adjust their distribution over behavioral policies according to the policies' long-run payoffs. This process closely parallels changes in strategy frequencies under evolutionary dynamics.

Our framework validates the five mechanisms, each of which promotes cooperation through a distinct feedback signal. Under direct reciprocity, agents cooperate in expectation of future cooperation from the same co-player. Both GRIM and WSLS respond to unilateral defection with retaliatory defection, whereas WSLS restores cooperation after mutual defection and is more robust to exploratory noise. In large anonymous populations, cooperation can be sustained through indirect reciprocity where information about individuals' past behavior is transmitted through reputation. Moreover, cooperation can be sustained even without information about one's co-players. Under environmental stochasticity, the environmental state itself mediates the feedback between individuals. The behavioral pattern of the $(\C,\D)$ strategy closely parallels that of GRIM in direct reciprocity. This correspondence can be established formally through an isomorphism between the two stochastic games. In contrast, unconditional cooperation can be sustained as $(\C,\C)$ under environmental stochasticity, but not as ALLC under direct reciprocity. Intuitively, in the productive state, an agent cooperates to avoid environmental degradation, while in the degraded state it cooperates to restore the more productive one. Direct reciprocity lacks this difference in productivity, and hence fails to stabilize unconditional cooperation. As the number of participants increases, each agent's influence on the feedback process is diluted, so cooperation is more sustainable in small groups. In general, these mechanisms all align individual interests with collective welfare by raising the long-term value of cooperation. Agents therefore need to care enough about the future to receive the corresponding signals. Accordingly, a large discount factor $\gamma$ reinforces the effects of these mechanisms by assigning greater weight to the future consequences.

Complementing mechanisms that render cooperation stable, demographic stochasticity biases the dynamics toward the cooperative attractor through fluctuations generated by finite-batch sampling, when both cooperation and defection coexist as stable equilibria. Such stochasticity explains how agents initialized in defection can escape the trap and learn to cooperate across our experiments (Fig.~\ref{fig:intro-five-mechanisms}a-d). When varying the batch size directly (Fig.~\ref{fig:intro-five-mechanisms}e), smaller batches drive rapid transitions to cooperation, whereas larger batches leave the system locked in defection. These fluctuations in the learning dynamics are analogous to demographic noise in finite-population evolutionary dynamics. In particular, the amplitude of learning noise scales with batch size as $O(B^{-1/2})$, which mirrors the $O(N^{-1/2})$ dependence of demographic noise on population size $N$ \cite{traulsen2006stochastic}. Under natural selection and social learning, network reciprocity promotes cooperation through the assortment of cooperators in sparse interaction networks \cite{ohtsuki2006simple}. Under reinforcement learning, sparsity likewise supports network reciprocity, but by amplifying sampling fluctuations that facilitate transitions from defection to cooperation.

Future work can apply this framework to a broader class of mechanisms, such as voluntary participation \cite{hauert2002volunteering} and punishment \cite{dreber2008winners}. One can also explore the effects of longer memory \cite{glynatsi2024conditional,glynatsi2024evolution} and asymmetric interactions \cite{mcavoy2015asymmetric}. Grounded in a dynamical systems perspective, our analytical approach extends beyond MARL to other self-learning algorithms, including Follow-the-Regularized-Leader \cite{shwartz2012online} and Hedge \cite{freund1997decision}. By translating the interplay between cooperation mechanisms and individual learning characteristics into explicit criteria for cooperation, our findings provide a theoretical foundation for the design of more efficient and socially intelligent MAS, and pave the way for the development of cooperative AI.

\section{Methods}\label{sec:methods}

\subsection{Environments}

We consider a finite partially observable stochastic game with agents $\mathcal N=\{1,\ldots,N\}$ and state space $\mathcal S$. In state $s\in\mathcal S$, agent $i$ receives an observation $o^i\in\mathcal O^i$ with probability $p^i(o^i\mid s)$ and selects an action $a^i$ from the finite action set $\mathcal A$ with probability $X^i(o^i,a^i)$.  We write $\bm o=(o^j)_{j\in\mathcal N}$ and $\bm a=(a^j)_{j\in\mathcal N}$ for the joint observation and action, with superscript $-i$ indicating omission of agent $i$. The reward function $R^i(s,\bm a)$ specifies agent $i$'s immediate reward, and $T(s,\bm a,s')$ gives the probability of moving from state $s$ to state $s'$ under joint action $\bm a$.

\subsection{Deterministic Learning Dynamics}

Under the joint policy $\bm X=(X^j)_{j\in\mathcal N}$, averaging over observations and actions gives the state transition probability $T(s,s')$. For agent $i$, the effective transition probability $T^i(o,a,o')$ describes the next observation $o'$ conditional on the current observation $o$ and action $a$, and $R^i(o,a)$ is the corresponding expected immediate reward:
\begin{align}
T(s,s')
&=\sum_{\bm o,\bm a}\Big[\prod_{j}p^j(o^j|s)X^j(o^j,a^j)\Big]T(s,\bm a,s'), \label{eq:methods-state-kernel}\\
T^i(o,a,o')
&=\sum_{s,s'}\sum_{\bm o^{-i},\bm a^{-i}}p^i(s|o)\Big[\prod_{j\ne i}p^j(o^j|s)X^j(o^j,a^j)\Big]T\bigl(s,(a,\bm a^{-i}),s'\bigr)p^i(o'|s'), \label{eq:methods-effective-transition}\\
R^i(o,a)
&=\sum_{s}\sum_{\bm o^{-i},\bm a^{-i}}p^i(s|o)\Big[\prod_{j\ne i}p^j(o^j|s)X^j(o^j,a^j)\Big]R^i\bigl(s,(a,\bm a^{-i})\bigr), \label{eq:methods-effective-reward}
\end{align}
Here, $p(s)$ is the current state distribution, $p^i(o)=\sum_{\bar s}p(\bar s)p^i(o\mid\bar s)$ is the probability that agent $i$ observes $o$, and $p^i(s\mid o)=p(s)p^i(o\mid s)/p^i(o)$ is the conditional state distribution for observations with $p^i(o)>0$. The state distribution evolves according to the following master equation
\begin{equation}
\frac{\mathrm d}{\mathrm dt}p(s)
=
\sum_{s'\in\mathcal S}
\Big[p(s')T(s',s)-p(s)T(s,s')\Big],
\qquad s\in\mathcal S.
\label{eq:methods-state-dynamics}
\end{equation}
For $\varepsilon$-greedy Q-learning, agent $i$ maintains action-value estimates $Q^i(o,a)$ and updates them with learning rate $\alpha$, using discount factor $\gamma\in[0,1)$. Under $\alpha\ll 1$, the state distribution $p(s)$ converges to its stationary value $\tilde p$. We use a tilde to denote evaluation of a quantity under the stationary state distribution. The probability that a batch of size $B$ visits $(o,a)$ is $\tilde{\nu}_B^i(o,a):=1-[1-\tilde p^i(o)X^i(o,a)]^B$. With time measured in batch updates, the dynamics of Q-learning are given by
\begin{equation}
\frac{\mathrm d}{\mathrm dt}Q^i(o,a)
=
\alpha\tilde{\nu}_B^i(o,a)
\Big[
\tilde{R}^i(o,a)
+\gamma\sum_{o'}\tilde{T}^i(o,a,o')\max_{a'}Q^i(o',a')
-Q^i(o,a)
\Big].
\label{eq:q-ode-main}
\end{equation}

For actor--critic learning, under $\alpha_X\ll\alpha_Q\ll1$, where $\alpha_X$ and $\alpha_Q$ are the actor and critic learning rates, the critic adapts faster and equilibrates under the current joint policy. The policy dynamics are described by the replicator-like equation
\begin{equation}
\frac{\mathrm d}{\mathrm dt}X^i(o,a)
=
\frac{\alpha_X}{\tau}\,X^i(o,a)
\Big[\bar Q^i(o,a)-\sum_{a'}X^i(o,a')\bar Q^i(o,a')\Big].
\label{actor-ode}
\end{equation}
Here, $\tau$ is the temperature of the policy, and $\bar Q^i(o,a)$ is the expected discounted return from taking action $a$ at
observation $o$ and following the current policy thereafter, satisfying
\begin{equation}
    \bar Q^i(o,a)=\tilde{R}^i(o,a)+\gamma \sum_{o',a'}\tilde{T}^i(o,a,o')X^i(o', a')\bar Q^i(o',a').
\end{equation}

\subsection{Stable Equilibria of MARL Systems}

For $\varepsilon$-greedy Q-learning, equation~\eqref{eq:q-ode-main} defines a piecewise dynamical system over the Q-value space. Each cell corresponds to one greedy action profile $\bm g=(g^i)_{i\in\mathcal N}$ with $g^i(o)$ being the unique maximizer of $Q^i(o,\cdot)$. If some profile $\bm g$ is consistent with equation~\eqref{eq:bellman-optimality-equation}, that is, $Q_{\bm g}^i(o,g^i(o))>\max_{a\ne g^i(o)}Q_{\bm g}^i(o,a)$ for all $i$ and $o$ under equation~\eqref{eq:bellman-optimality-equation}, then in a neighborhood of this solution, the Jacobian of equation~\eqref{eq:q-ode-main} is block-diagonal, with each agent $i$'s block given by $-\alpha\tilde{\bm\nu}_B^i(\bm I-\gamma\tilde{\bm M}^i)$, where $\tilde{\bm\nu}_B^i$ is the diagonal matrix with entries
$\tilde\nu_B^i(o,a)$, $\bm I$ is the identity matrix, and $\mathbb I\{\cdot\}$ equals one when its condition holds and zero otherwise. Matrix rows and columns are indexed by
observation-action pairs with entries $\tilde{M}^i_{(o, a),(o', a')}=\tilde T^i(o, a, o')\mathbb I\{a'=g^i(o')\}$. As $\bm I - \gamma\tilde{\bm M}^i$ is a nonsingular $M$-matrix and $\alpha \tilde{\bm\nu}_B^i$ is a diagonal matrix with entries in $(0,1)$, the Jacobian is Hurwitz and the equilibrium is exponentially asymptotically stable. 

In Fig.~\ref{fig:intro-normal-form}c,d we consider the one-shot donation game, batch size $B=1$ and symmetric initial Q-values. The
deterministic dynamics remain in the symmetric subspace. We hence write $q_{\C}=Q^1(\C)=Q^2(\C)$ and $q_{\D}=Q^1(\D)=Q^2(\D)$. We treat the dynamics on the boundaries of greedy regions as Filippov differential inclusions. For the one-shot donation game with symmetric agents, define the switching function $h=q_{\C}-q_{\D}$ and the switching boundary $\Gamma=\{h=0\}$, and let $\bm F_{\C}$ and $\bm F_{\D}$ denote the Q-learning fields on its two sides. The one-sided Lie derivatives $L_{\C}=\nabla h\cdot\bm F_{\C}$ and $L_{\D}=\nabla h\cdot\bm F_{\D}$ classify each boundary segment, with $L_{\C}<0<L_{\D}$ identifying attracting sliding, meaning that the vector fields on both sides point toward the boundary and slide along $\Gamma$. On such a segment, the tangent Filippov field is $\bm F_\Gamma=w\bm F_{\C}+(1-w)\bm F_{\D}$, where $w=L_{\D}/(L_{\D}-L_{\C})$. Solving $\bm F_\Gamma=0$ gives the expression for the pseudo-equilibrium on the attracting segment, which exists when $b(1-\varepsilon)^2>c$. At this equilibrium, the fraction of boundary local time for which cooperation is greedy is
\begin{equation}
    w_*
    =
    \frac{1}{2}
    +
    \frac{1}{2(1-\varepsilon)}
    \sqrt{\frac{b(1-\varepsilon)^2-c}{b-c}},
\end{equation}
which satisfies $w_*>1/2$ and $\lim_{\varepsilon\to0}w_*=1$.

For actor--critic learning, since $X^i(o,\cdot)$ is normalized for every $(i,o)$ pair, we express equation~\eqref{actor-ode} in the reduced coordinates that omit $X^i(o,g^i(o))$. The Jacobian of this reduced system is diagonal, with entries $\lambda_{\bm g}^i(o, a):=\frac{\alpha_X}{\tau}[\bar Q_{\bm g}^i(o, a)-\bar Q_{\bm g}^i(o,g^i(o))]$. Here, $\bar{\bm Q}_{\bm g}$ satisfies the same linear Bellman system as that in equation~\eqref{eq:bellman-optimality-equation} under profile $\bm g$. Therefore, if $\lambda_{\bm g}^i(o,a) < 0$ for all $i$, $o$ and $a\ne g^i(o)$, then $\bm g$ is an asymptotically stable equilibrium under equation~\eqref{actor-ode}. This stability condition is the same as that for $\varepsilon$-greedy Q-learning.

\subsection{Stochastic Learning Dynamics}

Under a non-vanishing learning rate $\alpha$, the Q-learning dynamics are modeled as a stochastic differential equation (SDE) of the following form
\begin{equation}
    \mathrm d\bm Q_t=\bm\mu\,\mathrm dt+\sqrt{\bm\Sigma}\,\mathrm d\bm W_t.
    \label{eq:methods-q-sde}
\end{equation}
Here, $\bm\mu$ is the drift given by the right-hand side of
equation~\eqref{eq:q-ode-main}, $\bm Q_t$ is the joint Q-value vector of all agents at time $t$, measured in batch updates, and $\bm W_t$ is a vector of independent standard Brownian motions of the same dimension. The matrix $\bm\Sigma$ is the covariance of the one-batch increment of Q-values. For a sampled transition $(o,a,r,o')$, where $r$ is the immediate reward, the TD error used to update the Q-value is
$r+\gamma\max_{a'}Q^i(o',a')-Q^i(o,a)$.
Conditional on the current joint Q-values and a visit to $(o,a)$, let $\delta^i(o,a)$ and $v_{oa}^i$ denote its mean and variance. The probability that one sample visits entry $(o, a)$ is $\pi_{oa}^i:=\tilde p^i(o)X^i(o, a)$. The number of visits in a batch follows a binomial distribution $\operatorname{Binomial}(B,\pi_{oa}^i)$. By the law of total variance, the diagonal entries of $\bm\Sigma$ are
\begin{equation}
\begin{aligned}
    \operatorname{Var}\bigl[\Delta Q^i(o,a)\bigr]
    ={}&
    \alpha^2v_{oa}^i
    \sum_{n=1}^B\frac1n\binom Bn
    (\pi_{oa}^i)^n(1-\pi_{oa}^i)^{B-n}
    \\
    &+
    \alpha^2\delta^i(o,a)^2
    \bigl[1-(1-\pi_{oa}^i)^B\bigr]
    (1-\pi_{oa}^i)^B
    \\
    ={}&
    \frac{\alpha^2v_{oa}^i}{B\pi_{oa}^i}
    +O(B^{-2}).
\end{aligned}
    \label{eq:methods-total-variance}
\end{equation}
A similar computation shows that the off-diagonal entries are all of order $O(B^{-2})$ in $B$. As a result, the diffusion term in equation~\eqref{eq:methods-q-sde} is of order $1/\sqrt B$.

Under self-play Q-learning in a one-shot Stag Hunt game, where the payoff entries satisfy $R>T>P>S$, equation~\eqref{eq:methods-q-sde} can be further reduced to a one-dimensional piecewise Ornstein--Uhlenbeck process for $Q(\C)-Q(\D)$. Solving its corresponding Kolmogorov backward equation gives the transition rates $\lambda_{\C\to\D}$ from the cooperation-greedy region to the defection-greedy region and $\lambda_{\D\to\C}$ in the reverse direction. Under $\varepsilon\ll1$ and $B\varepsilon \gg 1$, the two rates satisfy
\begin{equation}
    \lambda_{\C\to\D}
    \asymp
    \exp\!\left[
        -\frac{B (R-T)^2}{\alpha(T-P)^2}
    \right],
    \quad
    \lambda_{\D\to\C}
    \asymp
    \exp\!\left[
        -\frac{B (S-P)^2}{\alpha(R-S)^2}
    \right].
    \label{eq:ds-self-play-transition-rates}
\end{equation}
Here, $\asymp$ indicates the leading exponential dependence. If mutual cooperation is risk-dominant, that is, $R+S>T+P$, then $\lambda_{\D\to\C} > \lambda_{\C\to\D}$, and the system spends more time in the cooperative region.

\begin{figure}[H]
    \centering
    \includegraphics[width=0.99\textwidth,height=\textheight,keepaspectratio]{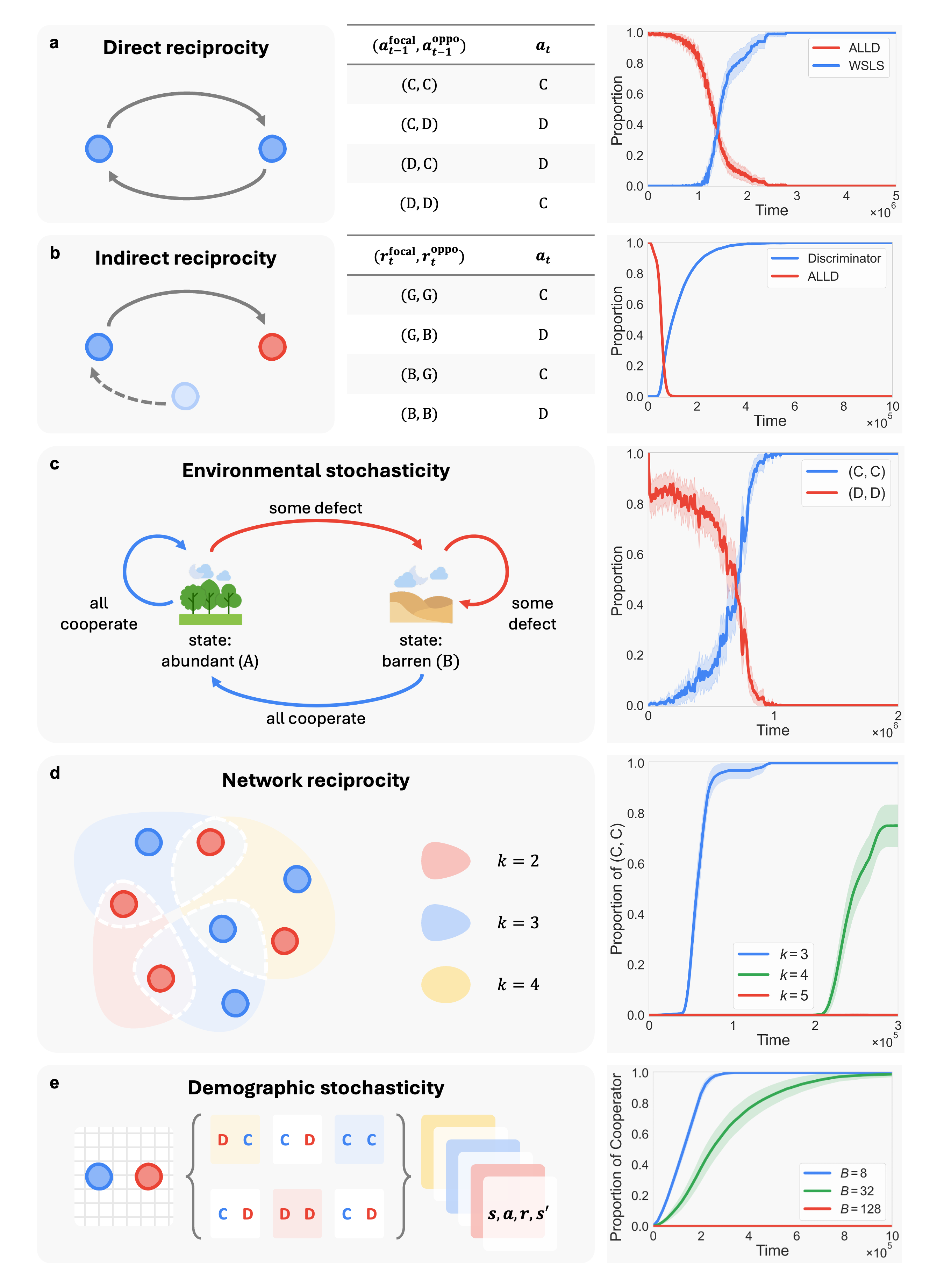}
\end{figure}
\clearpage
\begin{figure}[H]
    \caption{\textbf{Five mechanisms that promote cooperation in multi-agent reinforcement learning systems.} Each panel illustrates a mechanism and the corresponding agent-based simulations under $\varepsilon$-greedy Q-learning. Curves show means over $100$ independent runs, and shaded bands denote $95\%$ confidence intervals.
    \textbf{a}, \textbf{Direct reciprocity}. Repeated interactions can promote cooperation. In the repeated donation game, agents transition from the Always Defect (ALLD) strategy to the Win-Stay, Lose-Shift (WSLS) strategy. The table illustrates WSLS, which cooperates after mutual cooperation and mutual defection in the previous interaction and defects otherwise.
    \textbf{b}, \textbf{Indirect reciprocity}. Reputations can promote cooperation by providing information about a co-player's past behavior. We consider a population of agents with the social norm of image scoring, where cooperation and defection are assessed as good and bad. In the simulation, agents transition from ALLD to the Discriminator (DISC) strategy. The table illustrates DISC, which cooperates with good recipients and defects against bad ones.
    \textbf{c}, \textbf{Environmental stochasticity}. Cooperation can arise from agents' incentive to sustain a prosperous environment. In a two-state donation game, agents are incentivized to adopt the $(\C,\C)$ strategy, where the two entries denote the greedy actions in states $\A$ and $\B$.
    \textbf{d}, \textbf{Network reciprocity}. Local interactions among small groups of agents can promote cooperation. We represent the population as a $k$-uniform hypergraph, where each hyperedge represents a two-state public goods game among its $k$ members. In the simulations, cooperation emerges at $k=3$ but becomes less prevalent as the hyperedge order increases.
    \textbf{e}, \textbf{Demographic stochasticity}. Stochasticity induced by finite-batch sampling can drive transitions to cooperation. We consider a Stag Hunt game, in which cooperation and defection are both best-response-consistent equilibria. In simulations initialized with defection, transitions to cooperation occur more frequently as the batch size $B$ decreases.
    Parameters: All simulations use a learning rate of $\alpha=0.1$ and a discount factor of $\gamma=0.999$. Q-values are initialized so that defection is the greedy action at every observation. In \textbf{a}--\textbf{d}, the exploration rate $\varepsilon$ decays from $1$ to $0.01$, and the batch size is $B=64$. In \textbf{e}, the exploration rate remains fixed at $\varepsilon=0.1$. Panel-specific parameters: \textbf{a}, donation game with $b=5$ and $c=1$; \textbf{b}, population size $N=100$, reputation-observation accuracy $q=0.9$, and donation game with $b=5$ and $c=1$; \textbf{c}, two-state donation game with benefits $(b_{\A},b_{\B})=(5,2)$ in states $\A$ and $\B$, and cost $c=1$; \textbf{d}, population size $N=105$, and two-state public goods game with synergy factors $(r_{\A},r_{\B})=(2.8,1.2)$ in states $\A$ and $\B$, and cost $c=1$; \textbf{e}, payoff entries $(R,S,T,P)=(4,0,2,1)$.}
    \label{fig:intro-five-mechanisms}
\end{figure}
\clearpage

\begin{figure}[H]
    \centering
    \includegraphics[width=0.99\textwidth,height=\textheight,keepaspectratio]{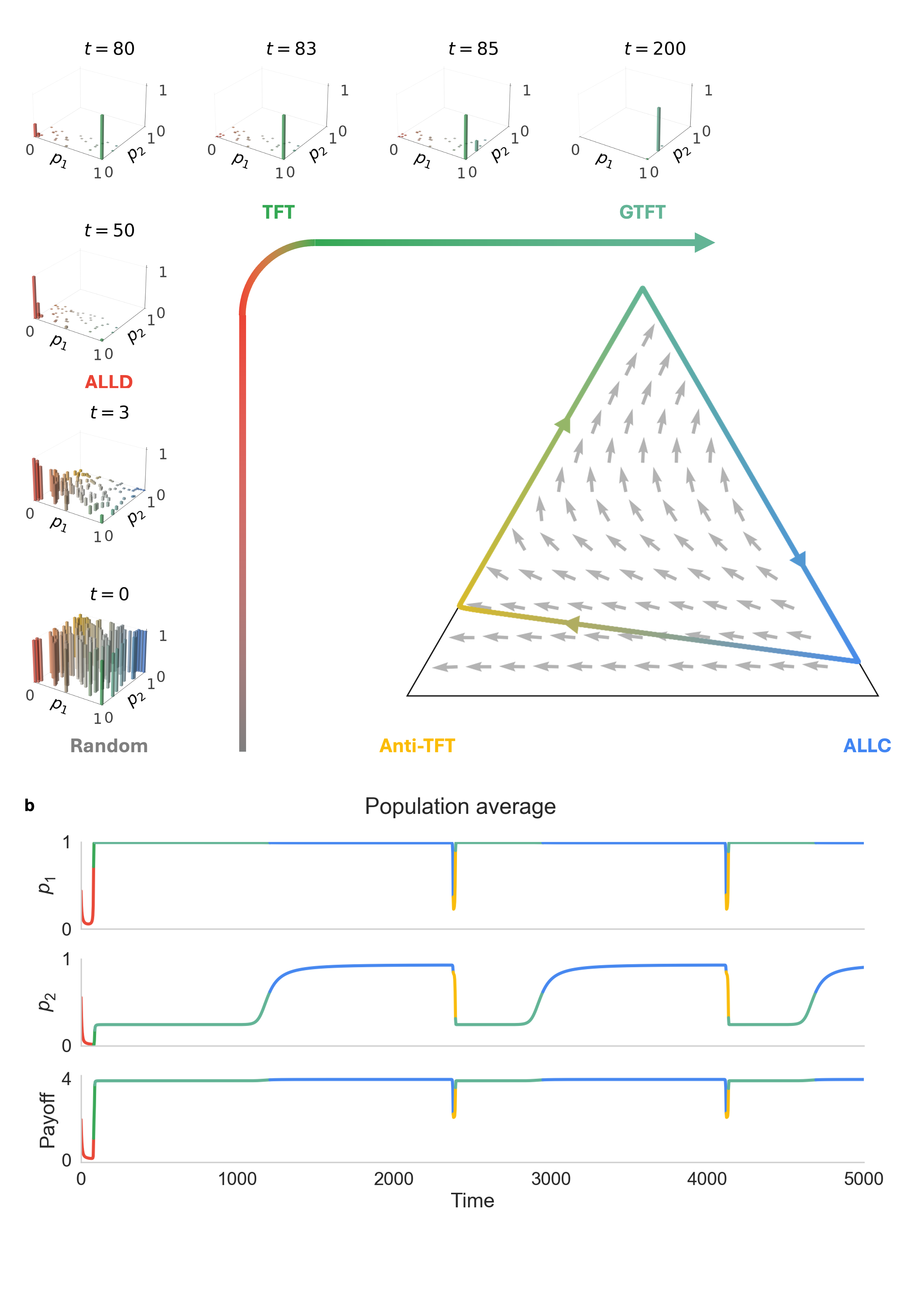}
\end{figure}
\clearpage
\begin{figure}[H]
    \caption{\textbf{Meta-policy learning dynamics under direct reciprocity.} Each meta-action is a reactive strategy $(p_1,p_2)$, where $p_1$ and $p_2$ denote the probabilities of cooperation when the co-player cooperated or defected in the previous round. The meta-action space contains four near-corner points, $(0.01,0.01)$, $(0.01,0.99)$, $(0.99,0.01)$, and $(0.99,0.99)$, located near Always Defect (ALLD), Anti-Tit-for-Tat (Anti-TFT), Tit-for-Tat (TFT), and Always Cooperate (ALLC), together with $96$ randomly sampled interior strategies. Each agent in the population uses actor--critic learning to optimize its meta-policy, which specifies the probability of selecting each meta-action in this space.
    \textbf{a}, The inset bar plots show agents' average meta-policy distribution at different time steps, with the labels below the insets indicating the most prevalent strategy. Starting from the uniformly random meta-policy, the distribution first concentrates on ALLD and then passes through TFT and generous Tit-for-Tat (GTFT; $(0.99, 0.24)$). The meta-strategy subsequently enters a perpetual cycle among GTFT, ALLC, and Anti-TFT. \textbf{b}, Evolution of the population averages of $p_1$, $p_2$, and payoff. The color of each curve segment indicates the most prevalent meta-action. Parameters: payoffs are calculated from the long-run average rewards in a repeated donation game with $b=5$ and $c=1$.}
    \label{fig:intro-reactive-meta-game}
\end{figure}
\clearpage

\begingroup
\let\table\tableorg
\let\endtable\endtableorg
\begin{table}[htbp]
    \centering
    \large
    \resizebox{\textwidth}{!}{%
    \begin{tblr}{
        width=21.2cm,
        colspec={Q[c,m,wd=2.4cm] Q[c,m,wd=1.8cm]
            Q[c,m,wd=5.5cm] Q[c,m,wd=6.4cm] Q[c,m,wd=3.5cm]},
        rows={valign=m,rowsep=5pt},
        column{1}={leftsep=2pt,rightsep=2pt},
        column{2-5}={font=\fontsize{14.4}{16.8}\selectfont},
        row{1}={font=\fontsize{13}{15}\selectfont,rowsep=7pt},
        row{2}={font=\large,rowsep=3pt},
        hline{1,3,7,9,11,13}={0.6pt},
        vline{2,3,5}={0.6pt},
    }
        \SetCell[r=2]{c,m} {Mechanism\\\&\\Profile}
        & \SetCell[r=2]{c,m} State
        & \SetCell[c=2]{c,m} Q-value
        &
        & \SetCell[r=2]{c,m} {Condition} \\
        & & \SetCell{bg=black!6} $\C$ & \SetCell{bg=black!6} $\D$ & \\
        \SetCell[r=4]{c,m} {DR\\WSLS}
        & $(\C,\C)$
        & \SetCell[r=2]{c,m} $\displaystyle \frac{b-c}{1-\gamma}$
        & \SetCell[r=2]{c,m} $\displaystyle b+\gamma^2 Q(\CC,\C)$
        & \SetCell[r=4]{c,m} $\displaystyle \gamma>\frac{c}{b-c}$ \\
        & $(\D,\D)$
        &
        &
        & \\
        & \SetCell{bg=black!6} $(\C,\D)$
        & \SetCell[r=2]{c,m,bg=black!6} $\displaystyle -c+\gamma^2 Q(\CC,\C)$
        & \SetCell[r=2]{c,m,bg=black!6} $\displaystyle \gamma Q(\CC,\C)$
        & \\
        & \SetCell{bg=black!6} $(\D,\C)$
        &
        &
        & \\
        \SetCell[r=2]{c,m} {IR\\DISC}
        & $(\G,\G)$
        & $\displaystyle \frac{b-c}{1-\gamma}$
        & $\displaystyle \frac{b-\gamma c}{1-\gamma}-\gamma bq$
        & \SetCell[r=2]{c,m} $\displaystyle \gamma q>\frac{c}{b}$ \\
        & \SetCell{bg=black!6} $(\B,\G)$
        & \SetCell{bg=black!6} $\displaystyle \frac{b-c}{1-\gamma}-bq$
        & \SetCell{bg=black!6} $\displaystyle \frac{b-\gamma c}{1-\gamma}-(1+\gamma)bq$
        & \\
        \SetCell[r=2]{c,m} {ES\\$(\C, \C)$}
        & $\A$
        & $\displaystyle \frac{b_{\A}-c}{1-\gamma}$
        & $\displaystyle b_{\A}+\gamma Q(\B,\C)$
        & \SetCell[r=2]{c,m} $\displaystyle \gamma>\frac{c}{b_{\A}-b_{\B}}$ \\
        & \SetCell{bg=black!6} $\B$
        & \SetCell{bg=black!6} $\displaystyle \frac{b_{\B}-c+\gamma(b_{\A}-b_{\B})}{1-\gamma}$
        & \SetCell{bg=black!6} $\displaystyle b_{\B}+\gamma Q(\B,\C)$
        & \\
        \SetCell[r=2]{c,m} {NR\\$(\C, \C)$}
        & $\A$
        & $\displaystyle \frac{c(r_{\A}-1)}{1-\gamma}$
        & $\displaystyle
            \begin{gathered}
                \frac{(k-1)cr_{\A}}{k} \\
                + \\
                \frac{\gamma\bigl[(k-1)Q(\A,\C)+Q(\B,\C)\bigr]}{k}
            \end{gathered}$
        & \SetCell[r=2]{c,m} $\displaystyle
            k <
            \begin{gathered}
                r_{\A} \\
                + \\
                \gamma (r_{\A}-r_{\B})
            \end{gathered}$ \\
        & \SetCell{bg=black!6} $\B$
        & \SetCell{bg=black!6} $\displaystyle \frac{c\bigl(r_{\B}-1+\gamma(r_{\A}-r_{\B})\bigr)}{1-\gamma}$
        & \SetCell{bg=black!6} $\displaystyle \frac{(k-1)cr_{\B}}{k}+\gamma Q(\B,\C)$
        & \\
    \end{tblr}%
    }
    \caption{\textbf{Conditions for the emergence of cooperation across mechanisms.} The four row blocks report the Q-values and stability condition associated with each policy profile: the Win-Stay, Lose-Shift (WSLS) strategy under direct reciprocity (DR), the Discriminator (DISC) under indirect reciprocity (IR), and the strategy $(\C,\C)$ under environmental stochasticity (ES) and network reciprocity (NR). The conditions specify the hyperparameter regions in which the prescribed actions strictly maximize these values and the corresponding cooperative policy profile can thus be sustained.}
    \label{tab:intro-cooperation-rules}
\end{table}
\endgroup

\backmatter

\section*{Acknowledgements}

This work was supported by the National Cyber Security-National Science and Technology Major Project (2025ZD1503700), the National Natural Science Foundation of China (12425114, 12526528, 12201026, 12501702, 62441617), the Taishan Scholars Program of Shandong Province (tstp20230635), the Key R\&D Program of Shandong Province (2026CXPT283), the Beijing Natural Science Foundation (Z230001), the Fundamental Research Funds for the Central Universities, the Opening Project of the State Key Laboratory of General Artificial Intelligence (Project NSKLAGI2025OP16), the special funding for the “Case-by-Case-Introduction of Top Talent (Teams)”, and the Beijing Advanced Innovation Center for Future Blockchain and Privacy Computing.


\bibliography{sn-bibliography}

@article{nowak2006five,
  title={Five rules for the evolution of cooperation},
  author={Nowak, Martin A},
  journal={Science},
  volume={314},
  number={5805},
  pages={1560--1563},
  year={2006},
  publisher={American Association for the Advancement of Science}
}

@book{ostrom1990governing,
  title={Governing the commons: The evolution of institutions for collective action},
  author={Ostrom, Elinor},
  year={1990},
  publisher={Cambridge University Press}
}

@book{henrich2016secret,
  title={The secret of our success: How culture is driving human evolution, domesticating our species, and making us smarter},
  author={Henrich, Joseph},
  year={2016},
  publisher={Princeton University Press}
}

@book{olson1971logic,
  title={The logic of collective action: Public goods and the theory of groups, with a new preface and appendix},
  author={Olson, Mancur},
  volume={124},
  series={Harvard Economic Studies},
  year={1971},
  publisher={Harvard University Press}
}

@article{dafoe2021cooperative,
  title={Cooperative {AI}: machines must learn to find common ground},
  author={Dafoe, Allan and Bachrach, Yoram and Hadfield, Gillian and Horvitz, Eric and Larson, Kate and Graepel, Thore},
  journal={Nature},
  volume={593},
  number={7857},
  pages={33--36},
  year={2021},
  publisher={Nature Publishing Group UK London}
}

@article{calvano2020artificial,
  title={Artificial intelligence, algorithmic pricing, and collusion},
  author={Calvano, Emilio and Calzolari, Giacomo and Denicol{\`o}, Vincenzo and Pastorello, Sergio},
  journal={American Economic Review},
  volume={110},
  number={10},
  pages={3267--3297},
  year={2020},
  publisher={American Economic Association 2014 Broadway, Suite 305, Nashville, TN 37203}
}

@article{banchio2022artificial,
  title={Artificial intelligence and spontaneous collusion},
  author={Banchio, Martino and Mantegazza, Giacomo},
  journal={arXiv preprint arXiv:2202.05946},
  year={2022}
}

@book{sutton2018reinforcement,
  title={Reinforcement learning: An introduction},
  author={Sutton, Richard S and Barto, Andrew G},
  year={2018},
  edition={2},
  publisher={MIT Press}
}

@article{schultz1997neural,
  title={A neural substrate of prediction and reward},
  author={Schultz, Wolfram and Dayan, Peter and Montague, P Read},
  journal={Science},
  volume={275},
  number={5306},
  pages={1593--1599},
  year={1997},
  publisher={American Association for the Advancement of Science}
}

@article{o2004dissociable,
  title={Dissociable roles of ventral and dorsal striatum in instrumental conditioning},
  author={O'Doherty, John and Dayan, Peter and Schultz, Johannes and Deichmann, Ralf and Friston, Karl and Dolan, Raymond J},
  journal={Science},
  volume={304},
  number={5669},
  pages={452--454},
  year={2004},
  publisher={American Association for the Advancement of Science}
}

@article{pessiglione2006dopamine,
  title={Dopamine-dependent prediction errors underpin reward-seeking behaviour in humans},
  author={Pessiglione, Mathias and Seymour, Ben and Flandin, Guillaume and Dolan, Raymond J and Frith, Chris D},
  journal={Nature},
  volume={442},
  number={7106},
  pages={1042--1045},
  year={2006},
  publisher={Nature Publishing Group UK London}
}

@article{mnih2015human,
  title={Human-level control through deep reinforcement learning},
  author={Mnih, Volodymyr and Kavukcuoglu, Koray and Silver, David and Rusu, Andrei A and Veness, Joel and Bellemare, Marc G and Graves, Alex and Riedmiller, Martin and Fidjeland, Andreas K and Ostrovski, Georg and others},
  journal={Nature},
  volume={518},
  number={7540},
  pages={529--533},
  year={2015},
  publisher={Nature Publishing Group UK London}
}

@article{kober2013reinforcement,
  title={Reinforcement learning in robotics: A survey},
  author={Kober, Jens and Bagnell, J Andrew and Peters, Jan},
  journal={The International Journal of Robotics Research},
  volume={32},
  number={11},
  pages={1238--1274},
  year={2013},
  publisher={SAGE Publications Sage UK: London, England}
}

@inproceedings{ouyang2022training,
  title={Training language models to follow instructions with human feedback},
  author={Ouyang, Long and Wu, Jeffrey and Jiang, Xu and Almeida, Diogo and Wainwright, Carroll and Mishkin, Pamela and Zhang, Chong and Agarwal, Sandhini and Slama, Katarina and Ray, Alex and others},
  booktitle={Advances in Neural Information Processing Systems},
  volume={35},
  pages={27730--27744},
  year={2022}
}

@article{guo2025deepseek,
  title={{DeepSeek-R1} incentivizes reasoning in {LLMs} through reinforcement learning},
  author={Guo, Daya and Yang, Dejian and Zhang, Haowei and Song, Junxiao and Wang, Peiyi and Zhu, Qihao and Xu, Runxin and Zhang, Ruoyu and Ma, Shirong and Bi, Xiao and others},
  journal={Nature},
  volume={645},
  number={8081},
  pages={633--638},
  year={2025},
  publisher={Nature Publishing Group UK London}
}

@book{hofbauer1998evolutionary,
  title={Evolutionary games and population dynamics},
  author={Hofbauer, Josef and Sigmund, Karl},
  year={1998},
  publisher={Cambridge University Press}
}

@book{nowak2006evolutionary,
  title={Evolutionary dynamics: exploring the equations of life},
  author={Nowak, Martin A},
  volume={1115},
  year={2006},
  publisher={Belknap Press of Harvard University Press Cambridge, MA}
}

@article{borgers1997learning,
  title={Learning through reinforcement and replicator dynamics},
  author={B{\"o}rgers, Tilman and Sarin, Rajiv},
  journal={Journal of Economic Theory},
  volume={77},
  number={1},
  pages={1--14},
  year={1997},
  publisher={Elsevier}
}

@article{bloembergen2015evolutionary,
  title={Evolutionary dynamics of multi-agent learning: A survey},
  author={Bloembergen, Daan and Tuyls, Karl and Hennes, Daniel and Kaisers, Michael},
  journal={Journal of Artificial Intelligence Research},
  volume={53},
  pages={659--697},
  year={2015}
}

@article{hardin1968tragedy,
  title={The tragedy of the commons: the population problem has no technical solution; it requires a fundamental extension in morality.},
  author={Hardin, Garrett},
  journal={Science},
  volume={162},
  number={3859},
  pages={1243--1248},
  year={1968},
  publisher={American Association for the Advancement of Science}
}

@inproceedings{hughes2018inequity,
  title={Inequity aversion improves cooperation in intertemporal social dilemmas},
  author={Hughes, Edward and Leibo, Joel Z and Phillips, Matthew and Tuyls, Karl and Due{\~n}ez-Guzman, Edgar and Garc{\'\i}a Casta{\~n}eda, Antonio and Dunning, Iain and Zhu, Tina and McKee, Kevin and Koster, Raphael and others},
  booktitle={Advances in Neural Information Processing Systems},
  volume={31},
  pages={3326--3336},
  year={2018}
}

@inproceedings{anastassacos2021cooperation,
author = {Anastassacos, Nicolas and Garc{\'i}a, Julian and Hailes, Stephen and Musolesi, Mirco},
title = {Cooperation and reputation dynamics with reinforcement learning},
booktitle= {Proceedings of the 20th International Conference on Autonomous Agents and MultiAgent Systems},
pages = {115--123},
year = {2021}
}

@inproceedings{smit2024learning,
  title={Learning fair cooperation in mixed-motive games with indirect reciprocity},
  author={Smit, Martin and Santos, Fernando P},
  booktitle={Proceedings of the Thirty-Third International Joint Conference on Artificial Intelligence},
  pages={220--228},
  year={2024}
}

@article{barfuss2023intrinsic,
  title={Intrinsic fluctuations of reinforcement learning promote cooperation},
  author={Barfuss, Wolfram and Meylahn, Janusz M},
  journal={Scientific Reports},
  volume={13},
  number={1},
  pages={1309},
  year={2023},
  publisher={Nature Publishing Group UK London}
}

@inproceedings{geng2026emergent,
  title={Emergent fast-slow dynamics in multi-agent {Q}-learning for networked stochastic games},
  author={Geng, Yuxin and Barfuss, Wolfram and Chen, Xingru},
  booktitle={Proceedings of the AAAI Conference on Artificial Intelligence},
  volume={40},
  number={35},
  pages={29450--29458},
  year={2026}
}

@article{barfuss2019deterministic,
  title={Deterministic limit of temporal difference reinforcement learning for stochastic games},
  author={Barfuss, Wolfram and Donges, Jonathan F and Kurths, J{\"u}rgen},
  journal={Physical Review E},
  volume={99},
  number={4},
  pages={043305},
  year={2019},
  publisher={APS}
}

@article{trivers1971evolution,
  title={The evolution of reciprocal altruism},
  author={Trivers, Robert L},
  journal={The Quarterly Review of Biology},
  volume={46},
  number={1},
  pages={35--57},
  year={1971},
  publisher={Stony Brook Foundation, Inc.}
}

@article{glynatsi2024conditional,
  title={Conditional cooperation with longer memory},
  author={Glynatsi, Nikoleta E and Akin, Ethan and Nowak, Martin A and Hilbe, Christian},
  journal={Proceedings of the National Academy of Sciences},
  volume={121},
  number={50},
  pages={e2420125121},
  year={2024},
  publisher={National Academy of Sciences}
}

@article{glynatsi2024evolution,
  title={Evolution of reciprocity with limited payoff memory},
  author={Glynatsi, Nikoleta E and McAvoy, Alex and Hilbe, Christian},
  journal={Proceedings of the Royal Society B: Biological Sciences},
  volume={291},
  number={2025},
  pages={20232493},
  year={2024}
}

@article{nowak1998evolution,
  title={Evolution of indirect reciprocity by image scoring},
  author={Nowak, Martin A and Sigmund, Karl},
  journal={Nature},
  volume={393},
  number={6685},
  pages={573--577},
  year={1998},
  publisher={Nature Publishing Group UK London}
}

@article{ohtsuki2006leading,
  title={The leading eight: social norms that can maintain cooperation by indirect reciprocity},
  author={Ohtsuki, Hisashi and Iwasa, Yoh},
  journal={Journal of Theoretical Biology},
  volume={239},
  number={4},
  pages={435--444},
  year={2006},
  publisher={Elsevier}
}

@article{santos2018social,
  title={Social norm complexity and past reputations in the evolution of cooperation},
  author={Santos, Fernando P and Santos, Francisco C and Pacheco, Jorge M},
  journal={Nature},
  volume={555},
  number={7695},
  pages={242--245},
  year={2018},
  publisher={Nature Publishing Group UK London}
}

@article{hilbe2018evolution,
  title={Evolution of cooperation in stochastic games},
  author={Hilbe, Christian and {\v{S}}imsa, {\v{S}}t{\v{e}}p{\'a}n and Chatterjee, Krishnendu and Nowak, Martin A},
  journal={Nature},
  volume={559},
  number={7713},
  pages={246--249},
  year={2018},
  publisher={Nature Publishing Group UK London}
}

@article{weitz2016oscillating,
  title={An oscillating tragedy of the commons in replicator dynamics with game-environment feedback},
  author={Weitz, Joshua S and Eksin, Ceyhun and Paarporn, Keith and Brown, Sam P and Ratcliff, William C},
  journal={Proceedings of the National Academy of Sciences},
  volume={113},
  number={47},
  pages={E7518--E7525},
  year={2016},
  publisher={National Academy of Sciences}
}

@article{ohtsuki2006simple,
  title={A simple rule for the evolution of cooperation on graphs and social networks},
  author={Ohtsuki, Hisashi and Hauert, Christoph and Lieberman, Erez and Nowak, Martin A},
  journal={Nature},
  volume={441},
  number={7092},
  pages={502--505},
  year={2006},
  publisher={Nature Publishing Group UK London}
}

@article{allen2017evolutionary,
  title={Evolutionary dynamics on any population structure},
  author={Allen, Benjamin and Lippner, Gabor and Chen, Yu-Ting and Fotouhi, Babak and Momeni, Naghmeh and Yau, Shing-Tung and Nowak, Martin A},
  journal={Nature},
  volume={544},
  number={7649},
  pages={227--230},
  year={2017},
  publisher={Nature Publishing Group UK London}
}

@article{sheng2024strategy,
  title={Strategy evolution on higher-order networks},
  author={Sheng, Anzhi and Su, Qi and Wang, Long and Plotkin, Joshua B},
  journal={Nature Computational Science},
  volume={4},
  number={4},
  pages={274--284},
  year={2024},
  publisher={Nature Publishing Group US New York}
}

@article{traulsen2006stochastic,
  title={Stochastic dynamics of invasion and fixation},
  author={Traulsen, Arne and Nowak, Martin A and Pacheco, Jorge M},
  journal={Physical Review E},
  volume={74},
  number={1},
  pages={011909},
  year={2006},
  publisher={APS}
}

@article{doebeli1997population,
  title={Population dynamics, demographic stochasticity, and the evolution of cooperation},
  author={Doebeli, Michael and Blarer, Albert and Ackermann, Martin},
  journal={Proceedings of the National Academy of Sciences},
  volume={94},
  number={10},
  pages={5167--5171},
  year={1997},
  publisher={The National Academy of Sciences of the USA}
}

@article{milinski1987tit,
  title={Tit for tat in sticklebacks and the evolution of cooperation},
  author={Milinski, Manfred},
  journal={Nature},
  volume={325},
  number={6103},
  pages={433--435},
  year={1987},
  publisher={Nature Publishing Group UK London}
}

@article{bshary2006image,
  title={Image scoring and cooperation in a cleaner fish mutualism},
  author={Bshary, Redouan and Grutter, Alexandra S},
  journal={Nature},
  volume={441},
  number={7096},
  pages={975--978},
  year={2006},
  publisher={Nature Publishing Group UK London}
}

@article{milinski2002reputation,
  title={Reputation helps solve the `tragedy of the commons'},
  author={Milinski, Manfred and Semmann, Dirk and Krambeck, Hans-J{\"u}rgen},
  journal={Nature},
  volume={415},
  number={6870},
  pages={424--426},
  year={2002},
  publisher={Nature Publishing Group UK London}
}

@article{fowler2010cooperative,
  title={Cooperative behavior cascades in human social networks},
  author={Fowler, James H and Christakis, Nicholas A},
  journal={Proceedings of the National Academy of Sciences},
  volume={107},
  number={12},
  pages={5334--5338},
  year={2010},
  publisher={National Academy of Sciences}
}

@inproceedings{hansen2004dynamic,
  title={Dynamic programming for partially observable stochastic games},
  author={Hansen, Eric A and Bernstein, Daniel S and Zilberstein, Shlomo},
  booktitle={Proceedings of the 19th National Conference on Artificial Intelligence},
  pages={709--715},
  year={2004}
}

@article{sheng2026cooperation,
  title={Cooperation conflicts with equality when allocating public goods},
  author={Sheng, Anzhi and Su, Qi and McAvoy, Alex and Wang, Long and Plotkin, Joshua B},
  journal={Nature},
  volume={655},
  pages={1217--1223},
  year={2026},
  publisher={Nature Publishing Group}
}

@article{watkins1992q,
  title={{Q}-learning},
  author={Watkins, Christopher J. C. H. and Dayan, Peter},
  journal={Machine Learning},
  volume={8},
  number={3--4},
  pages={279--292},
  year={1992},
  publisher={Springer}
}

@inproceedings{konda1999actor,
  title={Actor-critic algorithms},
  author={Konda, Vijay R. and Tsitsiklis, John N.},
  booktitle={Advances in Neural Information Processing Systems},
  volume={12},
  pages={1008--1014},
  year={1999}
}

@article{nowak1993strategy,
  title={A strategy of win-stay, lose-shift that outperforms tit-for-tat in the {Prisoner's Dilemma} game},
  author={Nowak, Martin and Sigmund, Karl},
  journal={Nature},
  volume={364},
  number={6432},
  pages={56--58},
  year={1993},
  publisher={Nature Publishing Group UK London}
}

@article{axelrod1981evolution,
  title={The evolution of cooperation},
  author={Axelrod, Robert and Hamilton, William D},
  journal={Science},
  volume={211},
  number={4489},
  pages={1390--1396},
  year={1981},
  publisher={American Association for the Advancement of Science}
}

@book{alexander2017biology,
  title={The biology of moral systems},
  author={Alexander, Richard},
  year={2017},
  publisher={Routledge}
}

@article{ohtsuki2004should,
  title={How should we define goodness?—reputation dynamics in indirect reciprocity},
  author={Ohtsuki, Hisashi and Iwasa, Yoh},
  journal={Journal of Theoretical Biology},
  volume={231},
  number={1},
  pages={107--120},
  year={2004},
  publisher={Elsevier}
}

@article{mcavoy2020social,
  title={Social goods dilemmas in heterogeneous societies},
  author={McAvoy, Alex and Allen, Benjamin and Nowak, Martin A},
  journal={Nature Human Behaviour},
  volume={4},
  number={8},
  pages={819--831},
  year={2020},
  publisher={Nature Publishing Group UK London}
}

@article{galla2009intrinsic,
  title={Intrinsic noise in game dynamical learning},
  author={Galla, Tobias},
  journal={Physical Review Letters},
  volume={103},
  number={19},
  pages={198702},
  year={2009},
  publisher={APS}
}

@article{nowak1992tit,
  title={Tit for tat in heterogeneous populations},
  author={Nowak, Martin A and Sigmund, Karl},
  journal={Nature},
  volume={355},
  number={6357},
  pages={250--253},
  year={1992},
  publisher={Nature Publishing Group UK London}
}

@inproceedings{leibo2017multi,
  title={Multi-agent reinforcement learning in sequential social dilemmas},
  author={Leibo, Joel Z and Zambaldi, Vinicius and Lanctot, Marc and Marecki, Janusz and Graepel, Thore},
  booktitle={Proceedings of the 16th International Conference on Autonomous Agents and MultiAgent Systems},
  pages={464--473},
  year={2017}
}

@article{smith1973logic,
  title={The logic of animal conflict},
  author={Maynard Smith, J. and Price, George R.},
  journal={Nature},
  volume={246},
  number={5427},
  pages={15--18},
  year={1973},
  publisher={Nature Publishing Group UK London}
}

@article{hauert2002volunteering,
  title={Volunteering as {Red Queen} mechanism for cooperation in public goods games},
  author={Hauert, Christoph and De Monte, Silvia and Hofbauer, Josef and Sigmund, Karl},
  journal={Science},
  volume={296},
  number={5570},
  pages={1129--1132},
  year={2002},
  publisher={American Association for the Advancement of Science}
}

@article{dreber2008winners,
  title={Winners don't punish},
  author={Dreber, Anna and Rand, David G and Fudenberg, Drew and Nowak, Martin A},
  journal={Nature},
  volume={452},
  number={7185},
  pages={348--351},
  year={2008},
  publisher={Nature Publishing Group UK London}
}

@article{mcavoy2015asymmetric,
  title={Asymmetric evolutionary games},
  author={McAvoy, Alex and Hauert, Christoph},
  journal={PLoS Computational Biology},
  volume={11},
  number={8},
  pages={e1004349},
  year={2015},
  publisher={Public Library of Science San Francisco, CA USA}
}

@article{shwartz2012online,
author = {Shalev-Shwartz, Shai},
title = {Online learning and online convex optimization},
journal = {Foundations and Trends in Machine Learning},
volume = {4},
number = {2},
pages = {107--194},
year = {2012},
doi = {10.1561/2200000018},
publisher = {Now Publishers Inc.}
}

@article{freund1997decision,
  title={A decision-theoretic generalization of on-line learning and an application to boosting},
  author={Freund, Yoav and Schapire, Robert E},
  journal={Journal of Computer and System Sciences},
  volume={55},
  number={1},
  pages={119--139},
  year={1997},
  publisher={Elsevier}
}

@article{barfuss2025collective,
  title={Collective cooperative intelligence},
  author={Barfuss, Wolfram and Flack, Jessica and Gokhale, Chaitanya S and Hammond, Lewis and Hilbe, Christian and Hughes, Edward and Leibo, Joel Z and Lenaerts, Tom and Leonard, Naomi and Levin, Simon and others},
  journal={Proceedings of the National Academy of Sciences},
  volume={122},
  number={25},
  pages={e2319948121},
  year={2025},
  publisher={National Academy of Sciences}
}

@article{garcia2025picking,
  title={Picking strategies in games of cooperation},
  author={Garc{\'\i}a, Julian and Traulsen, Arne},
  journal={Proceedings of the National Academy of Sciences},
  volume={122},
  number={25},
  pages={e2319925121},
  year={2025},
  publisher={National Academy of Sciences}
}

\end{document}


\maketitle
\newpage
\tableofcontents
\newpage

\section{Introduction}

Reinforcement learning (RL) provides a canonical framework for sequential decision-making problems~\cite{sutton2018reinforcement}.  In neuroscience, RL is used to model how humans and animals learn to make decisions through trial and error~\cite{schultz1997neural,pessiglione2006dopamine}. In computer science and artificial intelligence, RL underlies learning systems for video games, robotics, and the alignment of large language models~\cite{mnih2015human,kober2013reinforcement,ouyang2022training}.

In systems involving multiple interacting RL agents, the agents' learning processes are coupled through their interactions. In such multi-agent reinforcement learning (MARL) systems, interactions among learning agents can produce collective phenomena including cooperation~\cite{leibo2017multi}, collusion~\cite{calvano2020artificial}, communication, and consensus. We focus on the emergence of cooperation. In this Supplementary Information, we develop a theoretical framework to investigate decentralized cooperation in MARL systems. We further propose analytical models to study mechanisms that promote the emergence of cooperation.

Section~\ref{sec:marl-framework} formalizes the game environment and the learning algorithms. Section~\ref{sec:general-theory} derives the learning dynamics and the stability criterion on which the later sections rely. Section~\ref{sec:normal-form-games} studies learning dynamics in one-shot games. Sections~\ref{sec:direct-reciprocity}--\ref{sec:demographic-stochasticity} analyze the mechanisms of direct reciprocity, indirect reciprocity, environmental stochasticity, network reciprocity, and demographic stochasticity. Section~\ref{sec:reactive-meta-policy} extends the analysis to learning over behavioral strategies rather than primitive actions. 

\clearpage

\section{Multi-Agent Reinforcement Learning Framework}
\label{sec:marl-framework}

\subsection{Game Environment}
\label{sec:game-environment}

The game environment is modeled as a general $N$-player partially observable stochastic game~\cite{hansen2004dynamic}
\begin{equation}
    \mathcal{G} = \left(\mathcal{N}, \mathcal{S}, \mathcal{O}, \{p^i\}_{i\in\mathcal{N}}, \mathcal{A}, T, \{R^i\}_{i\in\mathcal{N}}, \rho, \gamma\right).
\end{equation}
Here,
\begin{itemize}
    \item $\mathcal{N} = \{1, 2, \ldots, N\}$ is the set of players.
    \item $\mathcal{S} = \{s_1, s_2, \ldots, s_K\}$ is the finite state space.
    \item For each player $i\in\mathcal{N}$, $\mathcal{O}=\{o_1, o_2, \ldots, o_{K'}\}$ is its finite observation space and
          $p^i: \mathcal{S} \times \mathcal{O}\to [0, 1]$ is its observation kernel, where
          $p^i(o \mid s)$ denotes the probability that player $i$ observes $o$ in state $s$.
          Observations are drawn independently across players given the state.
    \item $\mathcal{A} = \{a_1, a_2, \ldots, a_M\}$ is the finite action set, and $\mathcal{A}^N$ is the joint action space. Where explicitly stated, an observation-dependent admissible action set $\mathcal{A}(o) \subseteq \mathcal{A}$ denotes the available actions under observation $o$.
    \item $T: \mathcal{S} \times \mathcal{A}^N \to \Delta(\mathcal{S})$ is the state transition kernel,
          where $T(s, \bm{a}, s')$ denotes the probability of moving to state $s'$ given the current state $s$ and joint action $\bm{a}$.
    \item $R^i: \mathcal{S} \times \mathcal{A}^N \to \mathbb{R}$ is player $i$'s reward function, where $R^i(s, \bm{a})$ denotes the reward that $i$ receives when the state is $s$ and the joint action is $\bm{a}$.
    \item $\rho \in \Delta(\mathcal{S})$ is the initial state distribution.
    \item $\gamma \in [0,1)$ is the discount factor, quantifying the degree of agents' farsightedness.
\end{itemize}
The policy $X^i(o, a)$ denotes the probability that player $i$ takes action $a\in \mathcal{A}(o)$ given the observation $o$.

\subsection{Learning Algorithms}
\label{sec:algorithms}

\subsubsection{Q-Learning}

Value-based methods learn estimates of expected discounted return. Because higher state or action values indicate better observations or actions, an agent's policy is derived directly from these estimates. Here we consider Q-learning as a representative value-based algorithm~\cite{watkins1992q}. Each Q-learning agent $i$ maintains a Q-table $\bm{Q}^i$ in which each entry $Q^i(o, a)$ is $i$'s estimate of the value of taking action $a$ under observation $o$. The policy $X^i$ is derived from $\bm{Q}^i$ through the $\varepsilon$-greedy mapping. Specifically, the agent chooses uniformly among the actions with the highest Q-value with probability $1-\varepsilon$, and chooses an action uniformly at random with probability $\varepsilon$. Let $\mathcal A^{i,*}(o)=
\arg\max_{a'\in\mathcal A(o)}Q^i(o,a')$ denote the set of greedy actions. The $\varepsilon$-greedy policy is formalized as
\begin{equation}
\label{eq:epsilon-greedy}
    X^i(o, a) = \begin{cases}
        \displaystyle (1 - \varepsilon) / |\mathcal A^{i,*}(o)| + \varepsilon / M & \text{if } a \in \mathcal A^{i,*}(o), \\
        \displaystyle \varepsilon / M & \text{otherwise}.
    \end{cases}
\end{equation}
Note that if an admissible action set $\mathcal{A}(o)$ is specified, the maximization and the uniform exploration in \eqref{eq:epsilon-greedy} range over $\mathcal{A}(o)$ only, with $M$ replaced by $|\mathcal{A}(o)|$. The same convention applies throughout this Supplementary Information and the main text. Every maximization, summation, and quantifier over actions should range over the admissible set $\mathcal{A}(o)$, and Q-tables carry entries only for admissible $(o, a)$ pairs.

Agents update their Q-tables in batches. During interaction, agent $i$ collects $B$ interaction samples into a batch $\mathcal D^i
=\bigl((o_\ell,a_\ell,r_\ell^i,o'_\ell)\bigr)_{\ell=1}^{B}$, where $r$ is the immediate reward received after taking action $a$ under observation $o$, and $o^\prime$ is the subsequent observation. The integer $B \ge 1$ is therefore referred to as the \textit{batch size}. The temporal difference (TD) error quantifies the discrepancy between the current Q-value and the one-step lookahead estimate of the value, and is computed as
\begin{equation}
    \delta^i(o,a)=
\begin{cases}
\dfrac{1}{N_{oa}^i}
\displaystyle\sum_{\ell:o_\ell=o,\ a_\ell=a}
\left[
r_\ell^i+
\gamma\max_{a'}
Q^i(o'_\ell,a')-Q^i(o,a)
\right],
&N_{oa}^i>0,\\[2mm]
0,&N_{oa}^i=0.
\end{cases}
\end{equation}
Here, $N_{oa}^i=\sum_{\ell=1}^{B}
\mathbf1\{o_\ell=o,\ a_\ell=a\}$. Each agent's Q-table is then updated by
\begin{equation}
\label{eq:batch-q-update}
    Q^i(o, a) \leftarrow Q^i(o, a) + \alpha\, \delta^i(o, a),
\end{equation}
where $\alpha \in (0, 1]$ is the learning rate.

\subsubsection{Actor-Critic Learning}

As a counterpart to value-based methods, actor-critic methods maintain a value estimate as the critic and a parameterized policy as the actor~\cite{konda1999actor}. The actor is a policy $X^i(o, a)$ with parameter $\bm{\theta}^i$ and temperature $\tau>0$:
\begin{equation}
    X^i(o, a) = \frac{\exp\!\left(\theta^i(o, a)/\tau\right)}{\sum_{a'} \exp\!\left(\theta^i(o, a')/\tau\right)}.
\end{equation}
The critic $Q^i(o, a)$ estimates the values of actions and provides learning signals to guide updates of the actor. Given a batch $\mathcal D^i=\bigl((o_\ell,a_\ell,r_\ell^i,o'_\ell,a_\ell')\bigr)_{\ell=1}^{B}$, the TD error is computed as
\begin{equation}
    \delta^i(o,a)=
\begin{cases}
\dfrac{1}{N_{oa}^i}
\displaystyle\sum_{\ell:o_\ell=o,\ a_\ell=a}
\left[
r_\ell^i+
\gamma
Q^i(o'_\ell,a_\ell')-Q^i(o,a)
\right],
&N_{oa}^i>0,\\[2mm]
0,&N_{oa}^i=0.
\end{cases}
\end{equation}
The critic and the actor are updated by 
\begin{align}
    Q^i(o, a)
    &\leftarrow Q^i(o, a)
    + \alpha_Q \delta^i(o, a), \\
    \theta^i(o,a)
    &\leftarrow \theta^i(o,a)
    + \alpha_X
    \left[
        Q^i(o,a)-V^i(o)
    \right].
\end{align}
Here, $\alpha_Q > 0$ and $\alpha_X > 0$ are the critic and actor learning rates, and $V^i(o)=\sum_{a}X^i(o,a)Q^i(o,a)$ is the value estimate for observation $o$ derived from the critic. 

\section{Dynamics of Multi-Agent Learning}
\label{sec:learning-dynamics}
\label{sec:general-theory}

To analyze the learning dynamics, we adopt a continuous-time description~\cite{barfuss2019deterministic} and derive the dynamical systems for the agents' values and policies. Specifically, we first compute the state transition probability of the state from $s$ to $s'$ under $i$'s action $a$ after averaging over the co-players' policies, and after averaging over $i$'s policy as
\begin{align}
    T^i(s, a, s')
    &=
    \sum_{\bm{o}^{-i}\in\mathcal{O}^{N-1}}
    \sum_{\bm{a}^{-i}\in\mathcal{A}^{N-1}}
    \Bigg[
        \prod_{j\ne i} p^j(o^j\mid s) X^j(o^j,a^j)
    \Bigg]
    T\!\left(s,(a,\bm{a}^{-i}),s'\right),
    \\
    T(s, s')
    &=
    \sum_{o^i\in\mathcal{O}}
    \sum_{a\in\mathcal{A}}
    p^i(o^i\mid s) X^i(o^i,a) T^i(s,a,s').
\end{align}
Let $p_t(s)$ denote the probability that the system is in state $s$ at time $t$. Its evolution is described by the master equation
\begin{equation}\label{eq:master-equation}
    \frac{\mathrm{d}}{\mathrm{d}t}p(s)
    =
    \sum_{s'\in\mathcal{S}}
    \left[
        p(s')T(s',s)-p(s)T(s,s')
    \right],
    \qquad s\in\mathcal{S},
\end{equation}
or equivalently $\dot{\bm p}=\bm p(\bm T- \bm I)$ in vector form. By analogy with the state transition, the average reward received by agent $i$ when the state is $s$ and it uses action $a$ is
\begin{equation}
    R^i(s, a)
    =
    \sum_{\bm{o}^{-i}\in\mathcal{O}^{N-1}}
    \sum_{\bm{a}^{-i}\in\mathcal{A}^{N-1}}
    \Bigg[
        \prod_{j\ne i} p^j(o^j\mid s) X^j(o^j,a^j)
    \Bigg]
    R^i\!\left(s,(a,\bm{a}^{-i})\right).
\end{equation}

The Q-table of agent $i$ is indexed by its own observation rather than by the latent state. Therefore, the TD error for the entry $(o,a)$ is obtained by conditioning the latent state on the event that agent $i$ observes $o$. Let
\begin{equation}
    p^i(o)
    =
    \sum_{s\in\mathcal{S}} p(s)p^i(o\mid s)
\end{equation}
be the marginal probability that agent $i$ observes $o$, and define
\begin{equation}
    p^i(s\mid o)
    =
    \frac{p(s)p^i(o\mid s)}{p^i(o)}
\end{equation}
whenever $p^i(o)>0$. The reward and transition kernel conditional on observations are then
\begin{align}
    R^i(o,a)
    &=
    \sum_{s\in\mathcal{S}} p^i(s\mid o) R^i(s,a),
    \\
    T^i(o,a,o')
    &=
    \sum_{s\in\mathcal{S}}
    \sum_{s'\in\mathcal{S}}
    p^i(s\mid o) T^i(s,a,s') p^i(o'\mid s').
\end{align}

\subsection{Dynamics of Multi-Agent $Q$-learning}
\label{sec:q-ode}

For Q-learning, the expected TD error is computed as
\begin{align}
    \delta^i(o,a)
    &=
    R^i(o,a)
    +
    \gamma \sum_{o'\in\mathcal{O}^i}
    T^i(o,a,o') \max_{a'\in\mathcal{A}} Q^i(o',a')
    -
    Q^i(o,a).
\end{align}
For a batch of size $B$, the Q-entry $(o,a)$ is updated if the pair appears at least once in the batch. Let
\begin{equation}
    \nu^i(o,a)
    =
    \mathbb{P}\!\left((o,a) \text{ appears at least once in } \mathcal{D}^i\right)
\end{equation}
be the corresponding activation probability under the current observation distribution and policy profile. Measuring time in units of batches, the continuous-time expected Q-learning dynamics for each $(i, o, a)$-triplet are
\begin{equation} \label{eq:q_ode}
    \frac{\mathrm{d}}{\mathrm{d}t}Q^i(o,a)
    =
    \alpha\,\nu_{B}^i(o,a)\,\delta^i(o,a) := \mu^i(o,a),
    \qquad i\in\mathcal{N},\ o\in\mathcal{O}^i,\ a\in\mathcal{A}.
\end{equation}
For small learning rates $\alpha\to 0$, Eq.~\ref{eq:master-equation} evolves on a faster time scale than the Q-values and the ergodic Markov chain of states quickly converges to its stationary distribution $\tilde{p}(s)$, which is
\begin{equation}
    \label{eq:stationary-ps}
    \tilde{p}(s)
    =
    \sum_{s'\in\mathcal{S}}
    \tilde{p}(s')T(s',s),
    \quad
    \text{or equivalently}
    \quad
    \bm \tilde{p}\left(\bm T-\bm I\right)=\bm 0.
\end{equation}
Note that in the notation we omit the dependence of $\tilde{p}$ on the joint Q-profile $\bm{Q}$. Hereafter, we use a tilde to denote quantities evaluated under the stationary state distribution $\tilde{p}(s)$ induced by $\bm{Q}$. On the slow time scale, the Q-values of each player evolve along the stable manifold defined by $\tilde{p}$, under which the activation probability and the reduced slow dynamics read
\begin{align}
    \tilde{\nu}^{i}(o,a)
    &=
    1-
    \Big\{1-\sum_{s\in\mathcal{S}}\big[\tilde{p}(s) p^i(o \mid s)\big] X^i(o,a)\Big\}^B,
    \label{eq:reduced-activation-probability}
    \\
    \frac{\mathrm{d}}{\mathrm{d}t} Q^i(o,a)
    &=
    \alpha\,\tilde{\nu}^{i}(o,a)\,
    \tilde{\delta}^{i}(o,a)= \tilde{\mu}^i(o,a).
    \label{eq:q-ode-stationary}
\end{align}


For $\varepsilon$-greedy Q-learning, the greedy maximizers partition agents' joint Q-value space into finitely many regions, within each of which the induced joint policy profile is fixed. Eq.~\eqref{eq:q-ode-stationary} defines a piecewise dynamical system over this space. With a non-zero exploration rate $\varepsilon > 0$, every action has probability at least $\varepsilon/M$ and every observation has positive stationary probability, so $\tilde{\nu}_{B}^{i}(o, a; \bm Q) > 0$ for each $(i, o, a)$--triplet. Consequently, within each region, a point is an equilibrium of Eq.~\eqref{eq:q-ode-stationary} if and only if it solves the corresponding coupled Bellman optimality equations. We further show in the following theorem that these equilibria are stable.

\begin{theorem}[Strict Bellman solutions are asymptotically stable]
\label{thm:q-ode-stability}
For $\varepsilon$-greedy Q-learning with $\varepsilon > 0$, let $\bm Q^\star$ be a solution of the coupled Bellman optimality equations
\begin{equation}
    \tilde{\delta}^i(o, a) = 0,
    \qquad \text{for all } i\in\mathcal N,\; o\in \mathcal{O}^i,\; a\in\mathcal A.
\end{equation}
If for each $i\in\mathcal N$ and $o\in \mathcal{O}^i$, the following maximizer
\begin{equation}
    a_i^\star(o)
    =
    \arg\max_{a\in\mathcal A} Q^i(o,a)
\end{equation}
is unique, then $\bm Q^\star$ is an asymptotically stable equilibrium of Eq.~\eqref{eq:q-ode-stationary}.
\end{theorem}

\begin{proof}
Since $\tilde{\delta}^i(o, a) = 0$ for all $i,o,a$, the point $\bm Q^\star$ is an equilibrium of \eqref{eq:q-ode-stationary}. Define the greedy gap, the minimal greedy gap, and the neighborhood as
\begin{align}
    \Delta_i(o)
    &:=
    Q^{i\star}(o,a_i^\star(o))
    -
    \max_{a\neq a_i^\star(o)} Q^{i\star}(o,a),
    \\
    \Delta
    &:=
    \min_{i,o}\Delta_i(o),
    \\
    \mathcal U
    &:=
    \left\{
        \bm Q:
        \|\bm Q-\bm Q^\star\|_\infty < \frac{\Delta}{2}
    \right\}.
\end{align}
The gaps $\Delta_i(o)$ and the minimal gap $\Delta$ are strictly positive by the uniqueness of the maximizer. Within $\mathcal U$, every agent's greedy action uniquely remains $a_i^\star(o)$, as

\begin{equation}
\begin{aligned}
    Q^i(o,a_i^\star(o)) - Q^i(o,a)
    &>
    \left(
        Q^{i\star}(o,a_i^\star(o)) - \frac{\Delta}{2}
    \right)
    -
    \left(
        Q^{i\star}(o,a) + \frac{\Delta}{2}
    \right)
    \\
    &=
    Q^{i\star}(o,a_i^\star(o)) - Q^{i\star}(o,a) - \Delta
    \\
    &\ge
    \Delta_i(o)-\Delta \ge 0.
\end{aligned}
\end{equation}
This uniqueness ensures that the stationary distribution $\tilde{p}(s)$, the reward tensor $\tilde{R}^i(o, a)$, and the transition tensor $\tilde{T}^i(o, a, o')$ are all constant in $\mathcal{U}$. Therefore, in $\mathcal{U}$, the activation factor $\tilde{\nu}_B^i$ is constant in $\bm{Q}$, and the Bellman residual $\tilde{\delta}^i(o, a)$ is affine in $i$'s Q-values $\bm{Q}^i$ and constant in the co-players' Q-values $\bm{Q}^{-i}$. For each agent $i$, define the matrix $\tilde{\bm{M}}^i$ indexed by $(o,a)\in\mathcal O^i\times\mathcal A$ as
\begin{equation}
    \tilde M^i_{(o,a),(o^\prime,a^\prime)}
    =
    \tilde T^i(o,a,o^\prime)\,\mathbb{I}\{a^\prime = a_i^\star(o^\prime)\}.
\end{equation}
Viewing the Q-values $\bm{Q}^i$ and the reward tensor $\tilde{\bm{R}}^i$ as vectors and the activation factor $\tilde{\bm{\nu}}_B^i$ as a positive diagonal matrix indexed by $(o,a)$, the dynamics on $\mathcal U$ take the affine, agent-wise decoupled form
\begin{equation}
    \frac{\mathrm{d}}{\mathrm{d}t} \bm{Q}^i
    =
    \alpha \tilde{\bm{\nu}}_B^i \left[ \tilde{\bm{R}}^i + (\gamma \tilde{\bm{M}}^i - \bm{I}) \bm{Q}^i \right].
\end{equation}
The Jacobian of the full system on $\mathcal U$ is therefore block diagonal across agents, with blocks
\begin{equation}
    \bm{J}^i
    =
    -\alpha \tilde{\bm{\nu}}_B^i(\bm{I}-\gamma \tilde{\bm{M}}^i).
\end{equation}

As each $\tilde{\bm{M}}^i$ is row-stochastic, $\rho(\tilde{\bm{M}}^i)=1$ and $\bm{I}-\gamma \tilde{\bm{M}}^i$ is a nonsingular $M$-matrix. Left multiplication by the positive diagonal matrix $\tilde{\bm{\nu}}_B^i$ preserves the nonsingular $M$-matrix property. Therefore, $\bm{J}^i$ is Hurwitz, as is the full block-diagonal Jacobian $\bm{J}$. Hence $\bm Q^\star$ is asymptotically stable.
\end{proof}

Theorem~\ref{thm:q-ode-stability} states the criteria for the joint Q-value profile to be stable. We next move our focus to the stability of policies. The next proposition shows that the strict solutions derived from a joint policy profile $\bm{X}$ are exactly the Bellman-optimality solutions that are self-consistent with $\bm g$.

\begin{proposition}
\label{prop:self-consistency}
Consider the joint greedy action profile $\bm{g}=(g^i)_{i\in\mathcal N}$ with $g^i: \mathcal{O}^i\to\mathcal A$ and let
$\bm X_{\bm g}$ be its induced
$\varepsilon$-greedy joint policy profile. Consider the linear Bellman system
\begin{equation}
\label{eq:bellman-system-g}
    Q_{\bm g}^i(o,a)
    =
    \tilde R^i(o,a)
    +
    \gamma \sum_{o'\in \mathcal{O}^i} \tilde T^i(o,a,o')\, Q_{\bm g}^i(o',g^i(o'))
\end{equation}
for all $i\in\mathcal N$, $o\in \mathcal{O}^i$, and $a\in\mathcal A$. Then the following are equivalent:
\begin{enumerate}[label=(\arabic*)]
    \item the Bellman optimality equations are self-consistent with \(g\). That is, there exists \(\bm Q^\star\) such that
    \[
        \tilde{\delta}^i(o, a)=0, \quad \text{and}
        \quad
        g^i(o)=\arg\max_{a\in\mathcal A} Q^{i\star}(o,a)
    \]
    with a unique maximizer for every \(i,o\);
    \item the unique solution \(\bm Q\) of \eqref{eq:bellman-system-g} satisfies
    \[
        Q_{\bm g}^i(o,g^i(o))>Q_{\bm g}^i(o,a),
        \qquad \forall i\in\mathcal N,\; o\in \mathcal{O}^i,\; a\neq g^i(o).
    \]
\end{enumerate}
\end{proposition}

\begin{proof}
For each agent \(i\), define the matrix \(\tilde{\bm{M}}^i_{\bm g}\) indexed by \((o,a)\in \mathcal{O}^i\times\mathcal A\) as
\[
    (\tilde{\bm{M}}^i_{\bm g})_{(o,a),(o^\prime,a^\prime)}
    =
    \tilde T^i(o,a,o^\prime)\,\mathbb{I}\{a^\prime=g^i(o^\prime)\}.
\]
Then \eqref{eq:bellman-system-g} can be written as
\[
    \bm{Q}_{\bm g}^i=\tilde{\bm{R}}^i+\gamma \tilde{\bm{M}}^i_{\bm g} \bm{Q}_{\bm g}^i.
\]
Each \(\tilde M^i_{\bm g}\) is row-stochastic, so \(\rho(\gamma \tilde{\bm{M}}^i_{\bm g})=\gamma<1\). Hence \(I-\gamma \tilde{\bm{M}}^i_{\bm g}\) is invertible, and \eqref{eq:bellman-system-g} has a unique solution \(\bm Q_{\bm g}\).

\((1)\Rightarrow(2)\).~~Suppose \(\bm Q^\star\) solves the Bellman optimality equations and is self-consistent with \(g\), with a unique greedy action \(g^i(o)\) at every \(i,o\). Then
\[
    \max_{a'\in\mathcal A} Q^{i\star}(o',a')
    =
    Q^{i\star}(o',g^i(o')),
    \qquad \forall i,o'.
\]
Substituting this identity into the Bellman optimality equations shows that \(\bm Q^\star\) satisfies \eqref{eq:bellman-system-g}. By uniqueness of the solution to \eqref{eq:bellman-system-g}, we must have \(\bm Q^\star=\bm Q_{\bm g}\). The strict inequalities
\[
    Q^i_{\bm g}(o,g^i(o))>Q^i_{\bm g}(o,a),
    \qquad a\neq g^i(o),
\]
then follow immediately from the uniqueness of the maximizer.

\((2)\Rightarrow(1)\).~~Suppose the unique solution \(\bm Q_{\bm g}\) of \eqref{eq:bellman-system-g} satisfies
\[
    Q^i_{\bm g}(o,g^i(o))>Q^i_{\bm g}(o,a),
    \qquad \forall i,o,\ a\neq g^i(o).
\]
Then for every \(i,o'\),
\[
    \max_{a'\in\mathcal A} Q^i_{\bm g}(o',a')
    =
    Q^i_{\bm g}(o',g^i(o')).
\]
Substituting this identity into \eqref{eq:bellman-system-g} yields
\[
    Q^i_{\bm g}(o,a)
    =
    \tilde R^i_{\bm g}(o,a)
    +
    \gamma \sum_{o'\in \mathcal{O}^i} \tilde T^i_{\bm g}(o,a,o') \max_{a'\in\mathcal A} Q^i_{\bm g}(o',a'),
\]
which is exactly the Bellman optimality equation \(\tilde{\delta}^i_{o,a}=0\). Since \(g^i(o)\) is also the unique maximizer of \(Q^i_{\bm g}(o,\cdot)\), the Bellman optimality equations are self-consistent with \(g\).
\end{proof}

Combining the proposition with Theorem~\ref{thm:q-ode-stability} gives the stability criterion for $\varepsilon$-greedy Q-learning used in the later sections. To state this criterion, we first define the Q-gaps for a profile $\bm g$ and the corresponding solution $\bm Q_{\bm g}$ of \eqref{eq:bellman-system-g} as
\begin{equation}
\label{eq:greedy-gap}
    \Delta^i_{\bm g}(o, a) = Q^i_{\bm g}(o, g^i(o)) - Q^i_{\bm g}(o, a), \qquad a \neq g^i(o).
\end{equation}

\begin{corollary}
\label{cor:stability-criterion-epsilon-Q}
Consider the joint greedy action profile $\bm{g}$ and let
$\bm X_{\bm g}$ be its induced
$\varepsilon$-greedy joint policy profile. The linear Bellman system \eqref{eq:bellman-system-g} induced by $\bm{X}_{\bm g}$ and $\bm{g}$ has a unique solution $\bm Q_{\bm g}$. We call $\bm g$ \emph{best-response consistent} at $\varepsilon$ if all Q-gaps are strictly positive,
\[
    \Delta^i_{\bm g}(o, a) > 0 \qquad \forall i\in\mathcal N,\; o\in \mathcal{O}^i,\; a\neq g^i(o).
\]
If $\bm g$ is best-response consistent, then $\bm Q_{\bm g}$ is a strict solution of the coupled Bellman optimality equations and a locally exponentially stable equilibrium of the learning dynamics \eqref{eq:q-ode-stationary}. On the other hand, if instead $\Delta^i_{\bm g}(o, a) < 0$ for some $i, o, a$, then no equilibrium of \eqref{eq:q-ode-stationary} has $\bm g$ as its unique greedy profile.
\end{corollary}

\begin{proof}
By Proposition~\ref{prop:self-consistency} the solution $\bm Q_{\bm g}$ exists and is unique. If all gaps are strictly positive, then by Proposition~\ref{prop:self-consistency} the point $\bm Q_{\bm g}$ solves the Bellman optimality equations with unique maximizers $g^i(o)$, and is asymptotically stable with respect to Eq.~\eqref{eq:q-ode-stationary}. If some gap is strictly negative, then by Proposition~\ref{prop:self-consistency} no Bellman-optimality solution is self-consistent with $\bm g$ and therefore no equilibrium of Eq.~\eqref{eq:q-ode-stationary} exists with $\bm g$ being the greedy profile.
\end{proof}

Degenerate cases with $\Delta^i_{\bm g}(o, a) = 0$, where the solution lies on the boundaries between greedy regions, are studied in Section~\ref{sec:normal-form-boundary-pseudo-equilibrium}.

Sections~\ref{sec:normal-form-games}--\ref{sec:network-reciprocity} apply Corollary~\ref{cor:stability-criterion-epsilon-Q} to derive parameter conditions for cooperative equilibria. In the rare-exploration limit, these conditions yield compact rules for the emergence of cooperation. The following proposition shows that a greedy profile satisfying all strict gap inequalities at zero exploration remains best-response consistent and asymptotically stable at small exploration rates.

\begin{proposition}
\label{prop:small-exploration-robustness}
Fix a greedy action profile $\bm g$ and let $\bm X_{\bm g}^\varepsilon$ be its associated $\varepsilon$-greedy joint policy profile. For each $\varepsilon$, let $\tilde R_{\bm g}^{i,\varepsilon}(o,a)$ and $\tilde T_{\bm g}^{i,\varepsilon}(o,a,o')$ be the stationary effective reward and observation kernel induced by $\bm X_{\bm g}^\varepsilon$, and let $\bm Q_{\bm g}^\varepsilon$ be the unique solution of the following system
\begin{equation}
\label{eq:bellman-system-g-eps}
    Q_{\bm g}^{i,\varepsilon}(o,a)
    =
    \tilde R_{\bm g}^{i,\varepsilon}(o,a)
    +
    \gamma\sum_{o'\in\mathcal O^i}
    \tilde T_{\bm g}^{i,\varepsilon}(o,a,o')
    Q_{\bm g}^{i,\varepsilon}(o',g^i(o')).
\end{equation}
If
\[
    Q_{\bm g}^{i,0}(o,g^i(o))>Q_{\bm g}^{i,0}(o,a),
    \qquad
    i\in\mathcal N,\quad o\in\mathcal O^i,\quad a\ne g^i(o),
\]
then there exists $\varepsilon_0>0$ such that the same strict inequalities hold for every $0\leq\varepsilon<\varepsilon_0$. Consequently, $\bm Q_{\bm g}^\varepsilon$ is a strict solution of the coupled Bellman optimality equations with greedy profile $\bm g$ for every such $\varepsilon$. For $0<\varepsilon<\varepsilon_0$, it is a locally exponentially stable equilibrium of \eqref{eq:q-ode-stationary}.
\end{proposition}

\begin{proof}
Under fixed $\bm{g}$, $\bm X_{\bm g}^\varepsilon$ is continuous in $\varepsilon$, so the state transition probabilities between different states depend continuously on $\varepsilon$. For an ergodic chain on states, its stationary distribution $\tilde{p}_{\bm g}^\varepsilon$ is uniquely defined and continuous in $\varepsilon$ on a neighborhood of zero. The stationary observation probabilities $\tilde p_{\bm g}^{i,\varepsilon}(o)$ are therefore continuous. The effective rewards and observation kernels are finite sums of these conditional probabilities and the policy probabilities. Thus $\tilde{\bm R}_{\bm g}^{i,\varepsilon}$ and $\tilde{\bm T}_{\bm g}^{i,\varepsilon}$ are continuous in $\varepsilon$.

For each agent $i$, define the matrix $\tilde{\bm M}_{\bm g}^{i,\varepsilon}$ indexed by $(o,a)\in\mathcal O^i\times\mathcal A$ as
\[
    \bigl(\tilde M_{\bm g}^{i,\varepsilon}\bigr)_{(o,a),(o',a')}
    =
    \tilde T_{\bm g}^{i,\varepsilon}(o,a,o')
    \mathbb I\{a'=g^i(o')\}.
\]
Since this matrix is row-stochastic, $\rho(\gamma\tilde{\bm M}_{\bm g}^{i,\varepsilon})=\gamma<1$ and Eq.~\eqref{eq:bellman-system-g-eps} therefore has the unique solution
\[
    \bm Q_{\bm g}^{i,\varepsilon}
    =
    \bigl(\bm I-\gamma\tilde{\bm M}_{\bm g}^{i,\varepsilon}\bigr)^{-1}
    \tilde{\bm R}_{\bm g}^{i,\varepsilon}.
\]
Since matrix inversion is continuous on the set of invertible matrices, $\bm Q_{\bm g}^{i,\varepsilon}\to\bm Q_{\bm g}^{i,0}$ as $\varepsilon\to0$. Let
\[
    \Delta
    =
    \min_{i,o}\min_{a\ne g^i(o)}
    \left\{
        Q_{\bm g}^{i,0}(o,g^i(o))-Q_{\bm g}^{i,0}(o,a)
    \right\}
    >0.
\]
Due to continuity there exists an $\varepsilon_0>0$ such that
\[
    \left\|\bm Q_{\bm g}^{i,\varepsilon}-\bm Q_{\bm g}^{i,0}\right\|_\infty
    <\frac{\Delta}{2},
    \qquad
    i\in\mathcal N,\quad 0\leq\varepsilon<\varepsilon_0.
\]
For every $a\ne g^i(o)$,
\begin{equation*}
    Q_{\bm g}^{i,\varepsilon}(o,g^i(o))-Q_{\bm g}^{i,\varepsilon}(o,a)
    >
    Q_{\bm g}^{i,0}(o,g^i(o))-Q_{\bm g}^{i,0}(o,a)-\Delta
    \geq 0.
\end{equation*}
By Proposition~\ref{prop:self-consistency} $\bm Q_{\bm g}^\varepsilon$ is a strict solution of the coupled Bellman optimality equations with greedy profile $\bm g$. When $\varepsilon>0$, every action has positive policy probability and every observation retains positive stationary probability. The activation factors are positive, and by Theorem~\ref{thm:q-ode-stability} $\bm Q_{\bm g}^\varepsilon$ is asymptotically stable.
\end{proof}

Beyond strict Bellman solutions, two questions remain. Section~\ref{sec:demographic-stochasticity} studies which equilibrium the stochastic learning process selects. Boundary equilibria can also exist between greedy regions, where the right-hand side of \eqref{eq:q-ode-stationary} is discontinuous; Section~\ref{sec:normal-form-boundary-pseudo-equilibrium} treats them as Filippov solutions of a differential inclusion.

\subsection{Dynamics of Actor-Critic Learning}

By analogy with the analysis for Q-learning, we derive the actor-critic learning dynamics, leveraging the time-scale separation approach. For small learning rates $\alpha_X \ll \alpha_Q \ll 1$, the environment reaches the stationary distribution induced by the current joint policy $\bm X$, under which the critic then converges to its stationary value, and the actor evolves on the slowest time scale.

Specifically, on the slow time scale the state distribution converges to its stationary value $\tilde{p}(s)$ satisfying Eq.~\eqref{eq:stationary-ps}. On the intermediate time scale, the value estimate converges to the stationary value, which is the solution of the following system
\begin{equation}
\label{eq:ac-value-bellman}
    \bar Q^i(o, a)
    =
    \tilde R^i(o,a)
    +
    \gamma
    \sum_{o'\in\mathcal O^i}
    \tilde T^i(o,a,o')
    \sum_{a'\in\mathcal A}
    X^i(o',a')\bar Q^i(o', a').
\end{equation}
On the slowest time scale, with time measured in units of batch updates, the actor parameter $\theta^i(o,a)$ evolves according to
\begin{equation}
    \dot\theta^i(o,a)
    =
    \alpha_X
    \left[\bar Q^i(o,a)-\bar V^i(o)\right],
\end{equation}
where $\bar V^i(o)=\sum_{a\in\mathcal A}X^i(o,a)\bar Q^i(o,a)$. The dynamics of the policy $X^i(o,a)$ are then given by
\begin{equation}\label{eq:actor-critic-policy-ode}
	\begin{aligned}
        \frac{\mathrm d}{\mathrm dt}X^i(o,a)
        &=
        \frac{X^i(o,a)}{\tau}
        \left[
            \dot\theta^i(o,a)
            -
            \sum_{b\in\mathcal A}
            X^i(o,b)\dot\theta^i(o,b)
        \right] \\
        &=
        \frac{\alpha_X}{\tau}
        X^i(o,a)
        \left[\bar Q^i(o,a)-\bar V^i(o)\right].
	\end{aligned}
\end{equation}

The next theorem gives the stability criterion for the deterministic policy of actor-critic dynamics.

\begin{theorem}
\label{thm:ac-boundary-stability}
Consider the actor-critic policy dynamics \eqref{eq:actor-critic-policy-ode}. Let \(\bm g\) be a deterministic policy profile and let \(\bm X_{\bm g}\) be the associated boundary point of the policy simplex such that $X_{\bm g}^i(o,g^i(o))=1$. All stationary quantities in the statement are evaluated at \(\bm X_{\bm g}\). If the following inequality holds for all $(i, o, a)$ triplets with $a\ne g^i(o)$,
\begin{equation}
\label{eq:ac-boundary-gap}
    \bar Q_{\bm g}^i(o,g^i(o))>\bar Q_{\bm g}^i(o,a),
\end{equation}
where $\bar Q_{\bm g}$ denotes the stationary critic action values evaluated at $\bm{X}_{\bm g}$, then \(\bm X_{\bm g}\) is asymptotically stable.
\end{theorem}

\begin{proof}
We compute the Jacobian of the vector field of the reduced system obtained from \eqref{eq:actor-critic-policy-ode} by deleting the greedy-action equations at \(\bm X_{\bm g}\). Due to the normalization $\sum_{a}X^i(o,a) = 1$, the reduced system dynamics are equivalent to those of the full system on the policy simplex. In these non-greedy action coordinates,
\[
    \left.
    \frac{\partial }
         {\partial X^j(\bar o,b)}
    \frac{\mathrm d}{\mathrm dt}X^i(o,a)
    \right|_{\bm X_{\bm g}}
    =
    \frac{\alpha_X}{\tau}
    \left[\bar Q_{\bm g}^i(o,a)-\bar V_{\bm g}^i(o)\right]
    \mathbb{I}\{j=i,\bar o=o,b=a\}.
\]
The reduced Jacobian is therefore diagonal. Since \(\bar V_{\bm g}^i(o)=\bar Q_{\bm g}^i(o,g^i(o))\) under $\bm g$, the eigenvalues are
\[
    \lambda^i_{o,a}
    =
    \frac{\alpha_X}{\tau}
    \left[
        \bar Q_{\bm g}^i(o,a)
        -
        \bar Q_{\bm g}^i(o,g^i(o))
    \right],
    \qquad a\ne g^i(o).
\]
For ergodic chains the prefactor $\alpha_X/\tau$ is always positive. As a result, all eigenvalues are negative and $\bm X_{\bm g}$ is asymptotically stable when \eqref{eq:ac-boundary-gap} holds.
\end{proof}

For a pure policy profile, Eq.~\eqref{eq:ac-value-bellman} has the same form as Eq.~\eqref{eq:bellman-system-g}. Therefore, the deterministic profiles that are strictly best-response consistent are the common asymptotically stable equilibria of $\varepsilon$-greedy Q-learning under rare exploration by Corollary~\ref{cor:stability-criterion-epsilon-Q}, and actor-critic learning by Theorem~\ref{thm:ac-boundary-stability}. In the following sections we mainly consider $\varepsilon$-greedy Q-learning.

\section{Learning in One-Shot Normal-Form Games}
\label{sec:normal-form-games}

We first study the learning dynamics in one-shot games. Consider a two-player symmetric normal-form game with the following payoff matrix:
\begin{equation} \label{eq:payoff_one_shot}
    \begin{pNiceMatrix}[first-row, first-col]
    & \C & \D \\
    \C & R & S \\
    \D & T & P
    \end{pNiceMatrix},
\end{equation}
where we denote the two actions as $\C$ (Cooperate) and $\D$ (Defect), with the elements being the payoffs for the row player when the row player chooses the row action and the column player chooses the column action, respectively. 

In the stateless environment, each agent has a single observation and the Q-table reduces to two values $Q^i(\C)$ and $Q^i(\D)$, and the corresponding joint greedy profile is $(g^1, g^2)\in \{\C, \D\}^2$. We identify the stability of different joint strategy profiles using Corollary~\ref{cor:stability-criterion-epsilon-Q}. In the linear Bellman system \eqref{eq:bellman-system-g}, the next observation is exactly the current observation, so the future reward term cancels from the Q-gap \eqref{eq:greedy-gap}. The gap therefore equals the expected one-shot payoff advantage of $\C$ over $\D$. In particular, if the co-player's greedy action is $\C$,
\begin{equation}
    \Delta_{\C}
    =
    Q^i(\C) - Q^i(\D)
    =
    -(T - R) - \frac{\varepsilon}{2}\Big[(P - S) - (T - R)\Big],
    \label{eq:stateless-gap-c}
\end{equation}
and if the co-player's greedy action is $\D$,
\begin{equation}
    \Delta_{\D}
    =
    Q^i(\C) - Q^i(\D)
    =
    -(P - S) + \frac{\varepsilon}{2}\Big[(P - S) - (T - R)\Big].
    \label{eq:stateless-gap-d}
\end{equation}
Corollary~\ref{cor:stability-criterion-epsilon-Q} then classifies the four deterministic profiles. Mutual cooperation $(\C,\C)$ is best-response consistent if and only if $\Delta_{\C} > 0$; mutual defection $(\D,\D)$ is best-response consistent if and only if $\Delta_{\D} < 0$; and the anti-coordinated profiles $(\C,\D)$ and $(\D,\C)$ are best-response consistent if and only if $\Delta_{\D} > 0$ and $\Delta_{\C} < 0$. Every best-response-consistent profile is a locally exponentially stable equilibrium of the learning dynamics.

Note that both gaps depend on the payoffs only through the two differences $T - R$ and $P - S$, so the equilibrium structure can be represented in the $(T-R,\, P-S)$ plane. The two switching boundaries are obtained by setting these gaps to zero:
\begin{equation}
    \Delta_{\C} = 0
    \;\Longleftrightarrow\;
    (P-S) = -\frac{2-\varepsilon}{\varepsilon}\,(T-R),
    \quad
    \Delta_{\D} = 0
    \;\Longleftrightarrow\;
    (P-S) = -\frac{\varepsilon}{2-\varepsilon}\,(T-R).
    \label{eq:stateless-boundary-lines}
\end{equation}

In the donation game
\begin{equation} \label{eq:donation-game}
    \begin{pNiceMatrix}[first-row, first-col]
    & \C & \D \\
    \C & b-c & -c \\
    \D & b & 0
    \end{pNiceMatrix},
\end{equation}
where $b > c > 0$, the exploration terms in \eqref{eq:stateless-gap-c}--\eqref{eq:stateless-gap-d} vanish and $\Delta_{\C} = \Delta_{\D} = -c$ for all $\varepsilon \in [0, 1]$. Mutual defection is therefore the unique best-response-consistent profile of the one-shot donation game at every exploration rate. If cooperation is to appear at all in this game, it cannot do so as a strict equilibrium in the interior of a greedy cell; it can only arise on the boundary between greedy cells, where the learning dynamics are discontinuous.

\subsection{Boundary Pseudo-Equilibrium in the Donation Game}
\label{sec:normal-form-boundary-pseudo-equilibrium}

Theorem~\ref{thm:q-ode-stability} characterizes the equilibrium structure in the interior of the greedy regions of Q-space, where each agent's maximizer is unique. On the boundary between greedy regions, the right-hand side of the Q-dynamics is discontinuous and the solutions must be understood as Filippov solutions of a differential inclusion. Here we conduct this boundary analysis for the donation game and $\varepsilon$-greedy Q-learning. We show that the difference in update frequencies between actions can produce a pseudo-equilibrium on the boundary. We derive conditions for the existence of such an equilibrium and explicit expressions for the equilibrium and its cooperation rate. We also characterize the basins of attraction of equilibria under rare exploration.

Consider the one-shot donation game and symmetric initial Q-values for agent $1$ and agent $2$. As the symmetric Q-space subspace is invariant, we write
\[
    q_{\C}:=Q^1(\C)=Q^2(\C),
    \quad
    q_{\D}:=Q^1(\D)=Q^2(\D),
    \quad
    \bm{Q}=(q_{\C},q_{\D})^{\top}.
\]
Throughout this subsection we abbreviate
\begin{equation}
    \varepsilon^-=\frac{\varepsilon}{2},
    \qquad
    \varepsilon^+=1-\frac{\varepsilon}{2},
    \qquad
    \Delta_{\varepsilon}=\varepsilon^+-\varepsilon^-=1-\varepsilon,
    \label{eq:donation-parameters}
\end{equation}
where \(\varepsilon^-\) and \(\varepsilon^+\) are the \(\varepsilon\)-greedy choice probabilities of the non-greedy and the greedy action, and \(\Delta_{\varepsilon}\) is their difference. For batch size \(B=1\), the activation factor of each action is exactly its current choice probability. In the regions $\omega_{\C}=\{q_{\C}>q_{\D}\}$ and $\omega_{\D}=\{q_{\D}>q_{\C}\}$ in which cooperation and defection are greedy, respectively, the continuous-time symmetric Q-dynamics are
\begin{align}
    \frac{\mathrm{d}\bm{Q}}{\mathrm{d}t}
    &=
    \alpha
    \begin{bmatrix}
        \varepsilon^+\left(\varepsilon^+ b-c-(1-\gamma)q_{\C}\right)\\
        \varepsilon^-\left(\varepsilon^+ b+\gamma q_{\C}-q_{\D}\right)
    \end{bmatrix} := \bm{F}_{\C}(Q),
    \label{eq:donation-f-c}\\
    \frac{\mathrm{d}\bm{Q}}{\mathrm{d}t}
    &=
    \alpha
    \begin{bmatrix}
        \varepsilon^-\left(\varepsilon^- b-c+\gamma q_{\D}-q_{\C}\right)\\
        \varepsilon^+\left(\varepsilon^- b-(1-\gamma)q_{\D}\right)
    \end{bmatrix} := \bm{F}_{\D}(Q).
    \label{eq:donation-f-d}
\end{align}
The equilibrium of Eq.~\eqref{eq:donation-f-c} satisfies \(q_{\D}=q_{\C}+c\) which falls outside \(\omega_{\C}\). In other words, Eq.~\eqref{eq:donation-f-c} contains no equilibrium in $\omega_{\C}$. Setting the right-hand side of \eqref{eq:donation-f-d} to zero yields the strict-defection equilibrium
\begin{equation}
    Q_{\D}^{*}
    =
    \left(
        \frac{\varepsilon^- b}{1-\gamma}-c,
        \frac{\varepsilon^- b}{1-\gamma}
    \right)^{\top},
    \label{eq:donation-strict-defection-equilibrium}
\end{equation}
which lies in \(\omega_{\D}\). To analyze the switching dynamics, we define
\[
    h:=q_{\C}-q_{\D},
    \qquad
    \Gamma:=\{\bm Q:h=0\}.
\]
On \(\Gamma\), write
\[
    q_{\C}=q_{\D}=q,
    \qquad
    z=(1-\gamma)q.
\]
Since \(\nabla h=(1,-1)^{\top}\), the functions \(L_{\C}\) and \(L_{\D}\) are the one-sided Lie derivatives of \(h\) along \(\bm F_{\C}\) and \(\bm F_{\D}\), evaluated on \(\Gamma\):
\begin{align}
    L_{\C}(z)
    &:={\left.(\mathcal L_{\bm F_{\C}}h)\right|}_{\bm Q=(q,q)^{\top}}
    =(1,-1)\cdot\bm F_{\C}(q,q)
    =
    \alpha\Delta_{\varepsilon}(z_{\C}-z),
    \label{eq:donation-normal-c}\\
    L_{\D}(z)
    &:={\left.(\mathcal L_{\bm F_{\D}}h)\right|}_{\bm Q=(q,q)^{\top}}
    =(1,-1)\cdot\bm F_{\D}(q,q)
    =
    \alpha\Delta_{\varepsilon}(z-z_{\D}),
    \label{eq:donation-normal-d}
\end{align}
where
\begin{equation}
    z_{\C}
    =
    \frac{\varepsilon^+(\Delta_{\varepsilon} b-c)}{\Delta_{\varepsilon}},
    \qquad
    z_{\D}
    =
    \frac{\varepsilon^-(\Delta_{\varepsilon} b+c)}{\Delta_{\varepsilon}}.
    \label{eq:donation-switching-thresholds}
\end{equation}
Note that $z_{\C}+z_{\D}=b-c$, and the two thresholds are centered at $(b-c)/2$. The Filippov classification of \(\Gamma\) is read directly from the signs of \(L_{\C}\) and \(L_{\D}\):
\begin{equation}
\begin{array}{lll}
    L_{\C}(z)L_{\D}(z)>0
    &\Longleftrightarrow&
    \text{crossing segment},\\[1mm]
    L_{\C}(z)<0<L_{\D}(z)
    &\Longleftrightarrow&
    \text{attracting sliding segment},\\[1mm]
    L_{\C}(z)>0>L_{\D}(z)
    &\Longleftrightarrow&
    \text{escaping segment}.
\end{array}
\label{eq:donation-filippov-classification}
\end{equation}
If \(b\Delta_{\varepsilon}^2>c\), then \(z_{\D}<z_{\C}\), and the boundary decomposes as
\begin{equation}
\begin{array}{rclcl}
    z<z_{\D}
    &\Longrightarrow&
    L_{\C}>0>L_{\D}
    &\Longrightarrow&
    \text{escaping},\\[1mm]
    z_{\D}<z<z_{\C}
    &\Longrightarrow&
    L_{\C}>0,\ L_{\D}>0
    &\Longrightarrow&
    \text{crossing from }\omega_{\D}\text{ to }\omega_{\C},\\[1mm]
    z>z_{\C}
    &\Longrightarrow&
    L_{\C}<0<L_{\D}
    &\Longrightarrow&
    \text{attracting sliding}.
\end{array}
\label{eq:donation-boundary-decomposition-dc}
\end{equation}
If \(b\Delta_{\varepsilon}^2<c\), then \(z_{\C}<z_{\D}\), and
\begin{equation}
\begin{array}{rclcl}
    z<z_{\C}
    &\Longrightarrow&
    L_{\C}>0>L_{\D}
    &\Longrightarrow&
    \text{escaping},\\[1mm]
    z_{\C}<z<z_{\D}
    &\Longrightarrow&
    L_{\C}<0,\ L_{\D}<0
    &\Longrightarrow&
    \text{crossing from }\omega_{\C}\text{ to }\omega_{\D},\\[1mm]
    z>z_{\D}
    &\Longrightarrow&
    L_{\C}<0<L_{\D}
    &\Longrightarrow&
    \text{attracting sliding}.
\end{array}
\label{eq:donation-boundary-decomposition-cd}
\end{equation}
At equality \(b\Delta_{\varepsilon}^2=c\), the two thresholds $z_{\C}$ and $z_{\D}$ coincide at \((b-c)/2\) and the crossing interval vanishes.

On an attracting sliding segment, the Filippov vector field along \(\Gamma\) is the convex combination
\begin{equation}
    \bm{F}_{\Gamma}
    =
    w\bm{F}_{\C}+\bigl(1-w\bigr)\bm{F}_{\D},
    \label{eq:donation-gamma-sliding-vector-field}
\end{equation}
where \(w\in(0,1)\) is the fraction of time during which cooperation is the greedy action, chosen such that \(\bm{F}_{\Gamma}\) is tangent to \(\Gamma\). This requires \(h\) to remain constant along the boundary flow. We compute $w$ through the vanishing Lie derivative of $h$ along \(\bm F_{\Gamma}\)
\begin{equation}
\begin{aligned}
    0
    &=
    {\left.(\mathcal L_{\bm F_{\Gamma}}h)\right|}_{\bm Q=(q,q)^{\top}}\\
    &=
    \nabla h\cdot
    \left[w\bm F_{\C}(q,q)+(1-w)\bm F_{\D}(q,q)\right]\\
    &=
    w L_{\C}(z)+(1-w)L_{\D}(z)
\end{aligned}
    \label{eq:donation-tau-condition}
\end{equation}
to obtain
\begin{equation}
    w
    =
    \frac{L_{\D}(z)}{L_{\D}(z)-L_{\C}(z)}.
    \label{eq:donation-tau}
\end{equation}
On an attracting segment, \(L_{\C}<0<L_{\D}\) guarantees \(w\in(0,1)\) is well-defined. Substituting \eqref{eq:donation-tau} into \eqref{eq:donation-gamma-sliding-vector-field} shows that on a sliding segment the Filippov field is parallel to \(\Gamma\), with \(\dot q_{\C}=\dot q_{\D}=\dot q\) given by
\begin{equation}
    \dot q
    =
    -\alpha\,
    \frac{z^2-(b-c)z+b(b-c)\varepsilon^-\varepsilon^+}{2z-(b-c)},
    \qquad
    z=(1-\gamma)q.
    \label{eq:donation-sliding-flow}
\end{equation}
On the attracting segment \(z>z_{\C}>(b-c)/2\), the denominator is positive. Solving this numerator equation gives the two candidate roots
\begin{equation}
\begin{aligned}
    z_-
    &=
    \frac{
        b-c
        -
        \sqrt{(b-c)\left(b\Delta_{\varepsilon}^2-c\right)}
    }{2},\\
    z_+
    &=
    \frac{
        b-c
        +
        \sqrt{(b-c)\left(b\Delta_{\varepsilon}^2-c\right)}
    }{2}.
\end{aligned}
\label{eq:donation-boundary-roots}
\end{equation}
The condition for two distinct real roots is exactly $b\Delta_{\varepsilon}^2>c$, under which the boundary decomposes as in \eqref{eq:donation-boundary-decomposition-dc}. The roots and thresholds have the same sum, \(z_-+z_+=z_{\C}+z_{\D}=b-c\), so both pairs are centered at \((b-c)/2\). Since $(z_+-z_-)^2 - (z_{\C}-z_{\D})^2 > 0$ for $b\Delta_{\varepsilon}^2>c$, we have $z_-<z_{\D}<z_{\C}<z_+$.
In other words, \(z_-\) lies on the escaping part of the boundary, where it is a repelling pseudo-equilibrium, while \(z_+\) lies on the attracting sliding part. As a result, the unique attracting pseudo-equilibrium exists if and only if $b \Delta_{\varepsilon}^2>c$, or equivalently,
\begin{equation}
    \varepsilon < 1 - \sqrt{\frac{c}{b}}.
    \label{eq:donation-boundary-existence}
\end{equation}
At this equilibrium, the common Q-value profile is
\begin{equation}
    Q_{\Gamma}^{*}
    =
    (\frac{z_+}{1-\gamma},\frac{z_+}{1-\gamma})^{\top},
    \qquad
    z_+
    =
    \frac{b-c+\sqrt{(b-c)\left(b\Delta_{\varepsilon}^2-c\right)}}{2}.
    \label{eq:donation-boundary-equilibrium}
\end{equation}

To test the stability of \(z_+\), we factor the numerator as \((z-z_-)(z-z_+)\) under the condition of Eq.~\eqref{eq:donation-boundary-existence}, and the ordering \(z_-<z_{\D}<z_{\C}<z_+\) established above makes it negative for \(z_{\C}<z<z_+\) and positive for \(z>z_+\). The sliding motion therefore increases \(q\) below the pseudo-equilibrium and decreases \(q\) above it, so \(z_+\) attracts along the boundary. Since the surrounding segment also attracts from both greedy regions, \(z_+\) is an asymptotically stable equilibrium of the Filippov dynamics. If instead \(b\Delta_{\varepsilon}^2<c\), the numerator has no real roots and is positive for every \(z\). The sliding motion on the attracting segment \(z>z_{\D}\) then moves monotonically downward, and no pseudo-equilibrium exists.

The local-time weight of the cooperative side at the pseudo-equilibrium follows by substituting \eqref{eq:donation-boundary-equilibrium} into \eqref{eq:donation-tau}, which simplifies to the closed form
\begin{equation}
    w_{\ast}
    =
    \frac{1}{2}
    +
    \frac{1}{2\Delta_{\varepsilon}}
    \sqrt{\frac{b\Delta_{\varepsilon}^2-c}{b-c}}.
    \label{eq:donation-boundary-tau}
\end{equation}
Since \(0<b\Delta_{\varepsilon}^2-c<\Delta_{\varepsilon}^2(b-c)\), $w_{\ast}$ lies strictly between $1/2$ and $1$. For \(0<\varepsilon<1-\sqrt{c/b}\),
\begin{equation}
    w_{\ast}\in\left(\frac{1}{2},1\right),
    \qquad
    \lim_{\varepsilon\to 0}w_{\ast}=1,
    \quad
    \text{and}
    \quad
    \lim_{\varepsilon\to 1-\sqrt{c/b}}w_{\ast}=\frac{1}{2}.
    \label{eq:donation-boundary-tau-limits}
\end{equation}
At the pseudo-equilibrium the system thus spends more than half of its boundary local time on the cooperative side. Also in the limit \(\varepsilon\downarrow0\), one has \(z_+\to b-c\), and therefore \(Q_{\Gamma}^{*}\to (b-c)(1,1)^{\top}/(1-\gamma)\), in which the components are exactly the long-run value of mutual cooperation. As \(\varepsilon\) approaches the threshold in \eqref{eq:donation-boundary-existence}, one has \(z_+\to (b-c)/2\).

When $b \Delta_{\varepsilon}^2 < c$ the system contains only one equilibrium and it is globally stable. When $b \Delta_{\varepsilon}^2> c$ both the equilibrium in $\omega_{\D}$ and the pseudo-equilibrium on the boundary are stable. We next analyze the learning dynamics under $\varepsilon\to 0$ and further characterize the basins of attraction analytically. Under $\varepsilon\to 0$, the Q-value dynamics of the greedy and the non-greedy action are decoupled. The learning dynamics can be classified into five cases depending on the initial condition $\bm{q}$:
\begin{enumerate}[label=(\arabic*)]
    \item If $\bm{q}$ lies in $\omega_{\C}$, and $q_{\D} < \bm{Q}_{\Gamma}^*(\D)$, then $q_{\C}$ evolves on a faster time scale toward $Q_{\Gamma}^\ast$. $q_{\C}$ will first equilibrate at $Q_{\Gamma}^\ast(\C)$ and $q_{\D}$ evolves on the slower time scale until the system equilibrates at $\bm{Q}_{\Gamma}^*$.
    \item If $\bm{q}$ lies in $\omega_{\C}$, and $q_{\D} > \bm{Q}_{\Gamma}^*(\D)$, then $q_{\C}$ evolves on a faster time scale toward $Q_{\Gamma}^\ast$ until the system reaches the attracting sliding segment. The system will finally equilibrate at $\bm{Q}_{\Gamma}^*$.
    \item If $\bm{q}$ lies in $\omega_{\D}$, and $q_{\C} < Q_{\D}^\ast(\D)$, then $q_{\D}$ evolves on a faster time scale toward $Q_{\D}^\ast$. $q_{\D}$ will first equilibrate at $Q_{\D}^\ast(\D)$ and $q_{\C}$ evolves on the slower time scale until the system equilibrates at $\bm{Q}_{\D}^*$.
    \item If $\bm{q}$ lies in $\omega_{\D}$, and $Q_{\D}^\ast(\D) < q_{\C} < Q_{\Gamma}^\ast(\C)$, then $q_{\D}$ evolves on a faster time scale toward $Q_{\D}^\ast$ until the system reaches and passes through the crossing segment and goes into $\omega_{\C}$. The system then converges to $\bm{Q}_{\Gamma}^*$ according to (1).
    \item If $\bm{q}$ lies in $\omega_{\D}$, and $q_{\C} > Q_{\Gamma}^\ast(\C)$, then $q_{\D}$ evolves on a faster time scale toward $Q_{\D}^\ast$ until the system reaches the attracting sliding segment. The system will finally equilibrate at $\bm{Q}_{\Gamma}^*$.
\end{enumerate}

Intuition about the effect of update-rate asymmetry on condition \eqref{eq:donation-boundary-existence} can also be obtained from the above basin analysis. On the cooperative side of the boundary, $q_\C$ is updated at the high rate \(\varepsilon^+\), whereas $q_\D$ is re-estimated only at the rate \(\varepsilon^-\) and therefore adjusts slowly. Once the system crosses into the defection region, $q_\D$ is rapidly pulled down, while $q_\C$ is updated only through exploration and changes slowly. For small $\varepsilon$ satisfying condition~\eqref{eq:donation-boundary-existence}
these two opposing forces balance on the switching boundary and sustain the sliding pseudo-equilibrium. On the other hand, under frequent exploration, the two actions are updated at relatively similar rates, the asymmetry supporting this balance disappears, and defection remains the only equilibrium of the dynamics.

\section{Direct Reciprocity}
\label{sec:direct-reciprocity}

In the one-shot Prisoner's Dilemma, mutual defection constitutes the unique best-response-consistent profile at every exploration rate. When games are repeated, agents' actions can hinge on the history of previous interactions and thereby respond to the other agent's previous move. To study whether such repeated interactions promote cooperation, we study the MARL dynamics in the repeated donation game \eqref{eq:donation-game} between two memory-one agents. At each time step, each agent's observation $o$ is the joint action of the previous round. Each agent chooses either cooperation or defection according to its Q-table $Q^i(o, a)$. The current joint action determines the stage reward, after which the game is repeated.

\begin{table}[htbp]
    \centering
    \small
    \renewcommand{\arraystretch}{2.0}
    \begin{tabular}{p{1.8cm}p{2.9cm}p{9.0cm}}
        \toprule
        Strategy & $\varepsilon \to 0$ & $\varepsilon > 0$ \\
        \midrule
        ALLD & always stable & always stable \\
        GRIM & $\gamma > c/b$ &
        \begin{tabular}[t]{@{}l@{}}
            $\varepsilon < 1-\dfrac{c}{b}, ~~ \text{and} ~~ \dfrac{2 c}{b(1-\varepsilon)(2-\varepsilon)} < \gamma < \dfrac{2 c}{(1-\varepsilon)(b \varepsilon + 2 c)}$
        \end{tabular} \\
        WSLS & $\gamma > c/(b-c)$ & $\varepsilon < 1 - \dfrac{c}{b}, ~~ \text{and} ~~ \gamma > \dfrac{c}{(1-\varepsilon)\bigl(b(1-\varepsilon)-c\bigr)}$ \\
        ALLC & not an equilibrium & not an equilibrium \\
        TFT & not an equilibrium & not an equilibrium \\
        \bottomrule
    \end{tabular}
    \caption{Stability thresholds for different strategies in repeated donation games.}
    \label{tab:direct-reciprocity-thresholds}
\end{table}

\begin{figure}[H]
    \centering
    \includegraphics[width=0.99\textwidth]{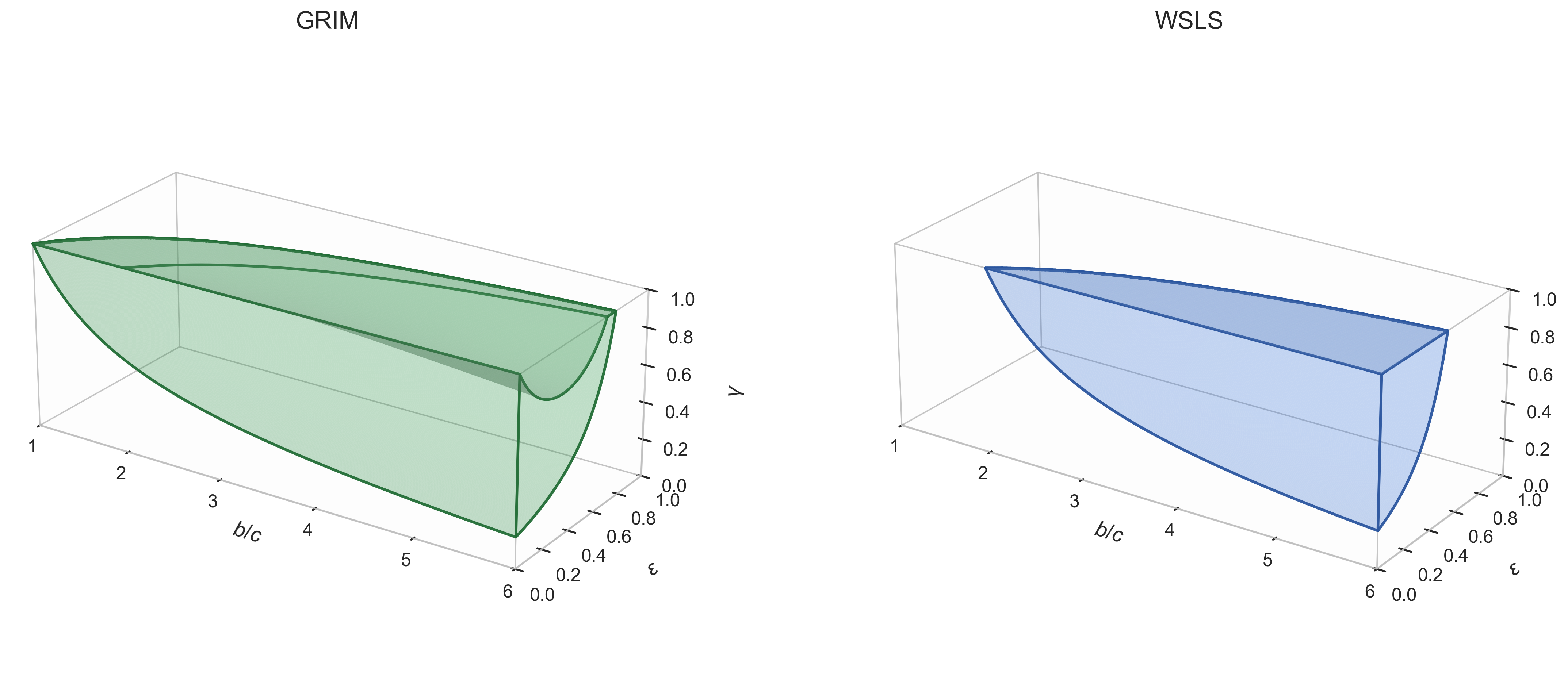}
    \caption{Stability regions for GRIM and WSLS in the repeated donation game as functions of the benefit-to-cost ratio $b/c$, the exploration rate $\varepsilon$, and the discount factor $\gamma$. The left panel shows the admissible parameter region for GRIM, and the right panel shows the corresponding region for WSLS.}
    \label{fig:mem1-gamma}
\end{figure}

\begin{figure}[H]
    \centering
    \includegraphics[width=0.99\textwidth]{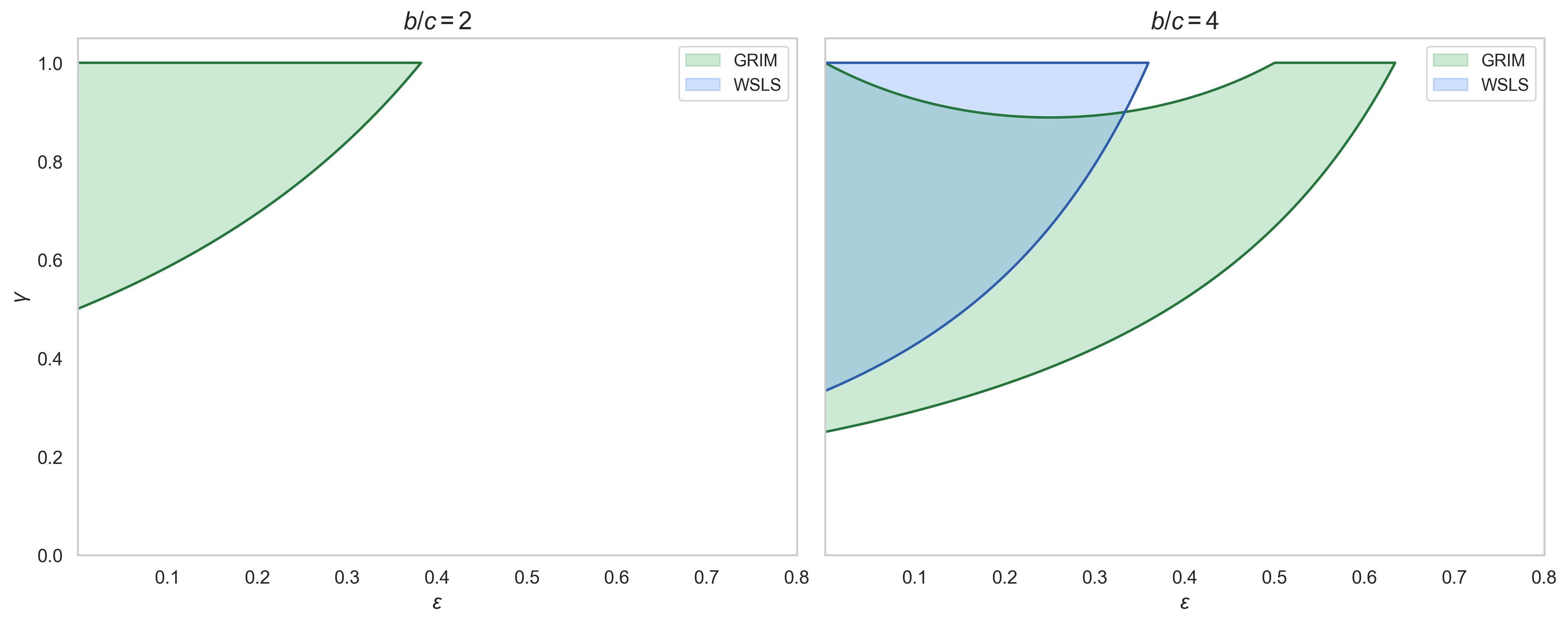}
    \caption{The stability regions of GRIM and WSLS in the $(\varepsilon,\gamma)$ plane at fixed benefit-to-cost ratios $b/c=2$ and $b/c=4$. At $b/c=2$, only GRIM admits a stable region. At $b/c=4$ GRIM becomes unstable for small exploration and a large discount factor, while WSLS is stable under a high discount factor and small exploration.}
    \label{fig:mem1-gamma-cross-section}
\end{figure}

Checking best-response consistency by Corollary~\ref{cor:stability-criterion-epsilon-Q} shows that under rare exploration the system has exactly mutual Always Defect (ALLD), mutual GRIM trigger (GRIM), and mutual Win-Stay Lose-Shift (WSLS) as possible best-response-consistent pure strategies. Table~\ref{tab:direct-reciprocity-thresholds} summarizes the resulting stability conditions for some classical strategies, and Figures~\ref{fig:mem1-gamma} and~\ref{fig:mem1-gamma-cross-section} visualize the parameter regions in which GRIM and WSLS are stable. For notational simplicity, we first define the Q-gap
\[
\Delta_o := Q(o, C) - Q(o, D),
\]
where we omit the agent index $i$ due to symmetry. For a candidate greedy strategy $g$, best-response consistency requires $\Delta_o > 0$ whenever $g(o)=C$ and $\Delta_o < 0$ whenever $g(o)=D$.

We denote $\bm{g}_{\text{name}} = (g(\CC), g(\CD), g(\DC), g(\DD))$ as the greedy strategy profile under the four possible observations. For both $\bm{g}_{\mathrm{ALLD}}=(\D,\D,\D,\D)$ and $\bm{g}_{\mathrm{ALLC}}=(\C,\C,\C,\C)$, agents fully cooperate or defect regardless of the past history. Their corresponding Bellman system Eq.~\eqref{eq:bellman-system-g} gives
\[
\Delta_{\CC}=\Delta_{\CD}=\Delta_{\DC}=\Delta_{\DD}=-c.
\]
ALLD is therefore always stable, whereas ALLC is always unstable.

For the Tit-For-Tat (TFT) strategy $\bm{g}_{\mathrm{TFT}}=(\C, \D, \C, \D)$, which always copies the opponent's previous move~\cite{axelrod1981evolution}, its four Q-value gaps always have the same expression and TFT is therefore always unstable:
\[
\Delta_{\CC}^{\mathrm{TFT}}
=
\Delta_{\CD}^{\mathrm{TFT}}
=
\Delta_{\DC}^{\mathrm{TFT}}
=
\Delta_{\DD}^{\mathrm{TFT}}
=
\frac{\gamma b(1-\varepsilon)-c}{1-\gamma^2(1-\varepsilon)}.
\]

For the GRIM strategy $\bm{g}_{\mathrm{GRIM}}=(\C,\D,\D,\D)$, agents cooperate only after mutual cooperation. Computing the Bellman gaps gives
\begin{align*}
\Delta_{\CC}^{\mathrm{GRIM}}
&=
\frac{\gamma b (1-\varepsilon)(2-\varepsilon)-2c}{2-\gamma(2-\varepsilon)},\\
\Delta_{\CD}^{\mathrm{GRIM}}
&=
\Delta_{\DC}^{\mathrm{GRIM}}
=
\Delta_{\DD}^{\mathrm{GRIM}}
=
\frac{\gamma (1-\varepsilon)(b\varepsilon+2c)-2c}{2-\gamma(2-\varepsilon)}.
\end{align*}
For $0 \le \gamma < 1$ and $0 < \varepsilon < 1$, the common denominator is always positive. The observation $\CC$ of mutual cooperation therefore gives the lower bound on $\gamma$, while the other three states give the upper bound
\[
\frac{2c}{b(1-\varepsilon)(2-\varepsilon)}
<
\gamma
<
\frac{2c}{(1-\varepsilon)(b\varepsilon+2c)}.
\]
The above condition additionally requires $\varepsilon < 1-c/b$ to ensure the left endpoint is smaller than the right endpoint. As $\varepsilon\to 0$, the Q-value gaps take the following form
\begin{equation}
    \begin{aligned}
        \lim_{\varepsilon \to 0}\Delta_{\CC}^{\mathrm{GRIM}}
        &=
        \frac{\gamma b-c}{1-\gamma},\\
        \lim_{\varepsilon \to 0}\Delta_{\CD}^{\mathrm{GRIM}}
        &=
        \lim_{\varepsilon \to 0}\Delta_{\DC}^{\mathrm{GRIM}}
        =
        \lim_{\varepsilon \to 0}\Delta_{\DD}^{\mathrm{GRIM}}
        = -c.
    \end{aligned}
\end{equation}
In this limit, the condition for GRIM to be stable reduces to $\gamma > c/b$.

Another candidate stable strategy is WSLS $\bm{g}_{\mathrm{WSLS}}=(\C,\D,\D,\C)$, where the agent cooperates after matching outcomes and defects after mismatches~\cite{nowak1993strategy}, with its Q-value gaps given by
\begin{align*}
\Delta_{\CC}^{\mathrm{WSLS}}
&=
\Delta_{\DD}^{\mathrm{WSLS}}
=
\gamma (1-\varepsilon)\bigl(b(1-\varepsilon)-c\bigr)-c,\\
\Delta_{\CD}^{\mathrm{WSLS}}
&=
\Delta_{\DC}^{\mathrm{WSLS}}
=
-\gamma (1-\varepsilon)\bigl(b(1-\varepsilon)-c\bigr)-c
<
0.
\end{align*}
The Q-value gaps of the mismatch states $\CD$ and $\DC$ are always negative. Stability is therefore decided by the two matching states $\CC$ and $\DD$, where the two gaps are positive only when the following two conditions hold
\[
\varepsilon < 1-\frac{c}{b},
\quad
\text{and}
\quad
\gamma
>
\frac{c}{(1-\varepsilon)\bigl(b(1-\varepsilon)-c\bigr)}.
\]
When $\varepsilon\to 0$, the two conditions reduce to $\gamma>c/(b-c)$.

\section{Indirect Reciprocity}
\label{sec:indirect-reciprocity}

Direct reciprocity explains cooperation in scenarios when agents interact with the same co-player repeatedly. However, in social encounters among agents who may never meet again, reciprocity depends less on these direct experiences and more on socially available information about others' standing or reputation in the community. To study the emergence of cooperation in such settings, we consider a population of agents with a reputation system in which each agent $i$ has a binary reputation $r^i_t \in \{\G,\B\}$. At each time step, a pair of agents $i$ and $j$ is sampled uniformly at random to play a donation game. Each player independently observes the opponent's current reputation with probability $q$. The observation space is therefore
\[
\mathcal O=\{\GG,\GB,\BG,\BB\},
\]
where the first and second coordinates represent its current reputation and its current assessment of the opponent, respectively. The opponent is treated as good if reputation is not observed. Both players then choose an action in $\{\C,\D\}$ according to the Q-table, interact, and receive rewards, after which the reputations of the two agents are updated by a prescribed assessment rule
\[
d:\{\G,\B\}\times\{\G,\B\}\times\{\C,\D\}\to\{\G,\B\},
\]
where $d(r,r^\prime,a)$ is the agent's new reputation, given the agent's current reputation $r$, its opponent's current reputation $r^\prime$, and the agent's action $a$.

\subsection{Stability of Discriminator against ALLD under Image Scoring}

We first restrict our attention to the assessment rule of \textbf{Image Scoring} $d_{\mathrm{IS}}$, under which the donor's new reputation depends only on the donor's action~\cite{nowak1998evolution}:
\[
d_{\mathrm{IS}}(x,y,\C)=\G,
\qquad
d_{\mathrm{IS}}(x,y,\D)=\B,
\qquad
\text{for all } x,y\in\{\G,\B\}.
\]
We first study whether agents can learn to cooperate with a good individual. We consider the action space $\mathcal A(\GG)=\mathcal A(\BG)=\{\C,\D\}, \mathcal A(\GB)=\mathcal A(\BB)=\{\D\}$ and check the stability of \textbf{Discriminator} (DISC):
\begin{equation}
    \begin{aligned}
    g_{\mathrm{DISC}}(\GG)&=g_{\mathrm{DISC}}(\BG)=\C \\
    g_{\mathrm{DISC}}(\GB)&=g_{\mathrm{DISC}}(\BB)=\D.
    \end{aligned}
\end{equation}
It therefore suffices to examine the best-response consistency (Corollary~\ref{cor:stability-criterion-epsilon-Q}) of discriminators at the $\GG$ and $\BG$ observations. Under next-encounter matching, the next observation uses the newly sampled opponent. At zero exploration on the all-good population branch, solving the Bellman equations under $g_{\mathrm{DISC}}$ gives the admissible Q-values
\begin{align*}
Q^i(\GG,\C) &= \frac{b-c}{1-\gamma},\\
Q^i(\GG,\D) &= Q^i(\GB,\D)
 = \frac{b-\gamma c}{1-\gamma}-\gamma bq,\\
Q^i(\BG,\C) &= \frac{b-c}{1-\gamma}-bq,\\
Q^i(\BG,\D) &= Q^i(\BB,\D)
 = \frac{b-\gamma c}{1-\gamma}-(1+\gamma)bq.
\end{align*}
Values at unvisited observations are understood as counterfactual one-step values on this branch. Define the gaps between cooperation and defection as
\begin{align*}
\Delta_{\GG}^{\mathrm{IS}}
&=
Q^i(\GG,\C)-Q^i(\GG,\D),
\\
\Delta_{\BG}^{\mathrm{IS}}
&=
Q^i(\BG,\C)-Q^i(\BG,\D).
\end{align*}
Direct calculation gives the strict best-response condition within the stated admissible action sets:
\[
\Delta_{\GG}^{\mathrm{IS}}
=
\Delta_{\BG}^{\mathrm{IS}}
=
b\gamma q-c>0,
\qquad\text{equivalently}\qquad
\gamma q>c/b.
\]

\subsection{Stability of the Leading Eight}

\begin{table}[h]
\centering
\large
\begin{tabular}{|c|c|c|c|c|}
\hline
Assessment Rule & GG & GB & BG & BB \\
\hline
C & G & $*$ & G & $*$ \\
\hline
D & B & G & B & $*$ \\
\hline
\hline
Policy  & C & D & C & $**$ \\
\hline
\end{tabular}
\caption{\textbf{The leading eight among third-order social norms.} Each configuration consists of (1) an assessment rule $d$, which specifies the donor's new reputation given the donor's reputation, the recipient's reputation, and the donor's action; and (2) the policy adopted by the population. The asterisk (*) in the assessment rule can be either $\G$ or $\B$, while the double asterisk (**) in the policy is determined by the assessment rule. Specifically, $**=\C$ if and only if $d(\BB,\C)=\G$ and $d(\BB,\D)=\B$, and $**=\D$ otherwise.}
\label{tab:leading-eight}
\end{table}

We next consider the full set of third-order social norms, which are social norms that depend on both the donor's and the recipient's reputations. Among these, there are eight combinations of social norms and behavioral strategies that are evolutionarily stable and can reach high levels of cooperation. These are known as the \textbf{leading eight} social norms~\cite{ohtsuki2006leading}, which are listed in Table~\ref{tab:leading-eight}.

\begin{figure}[H]
    \centering
    \begin{subfigure}[t]{0.24\textwidth}
        \centering
        \includegraphics[width=\textwidth]{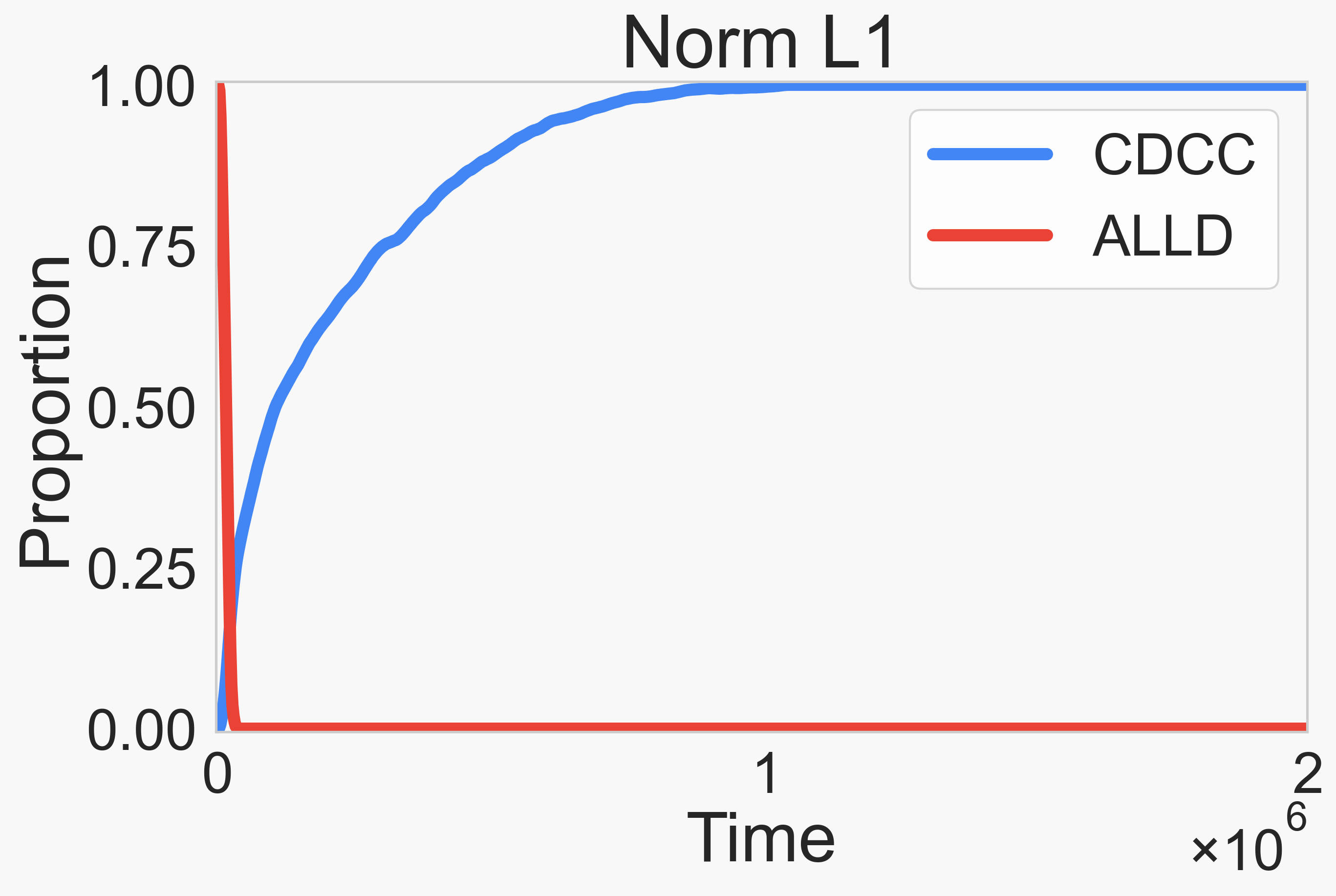}
    \end{subfigure}
    \hfill
    \begin{subfigure}[t]{0.24\textwidth}
        \centering
        \includegraphics[width=\textwidth]{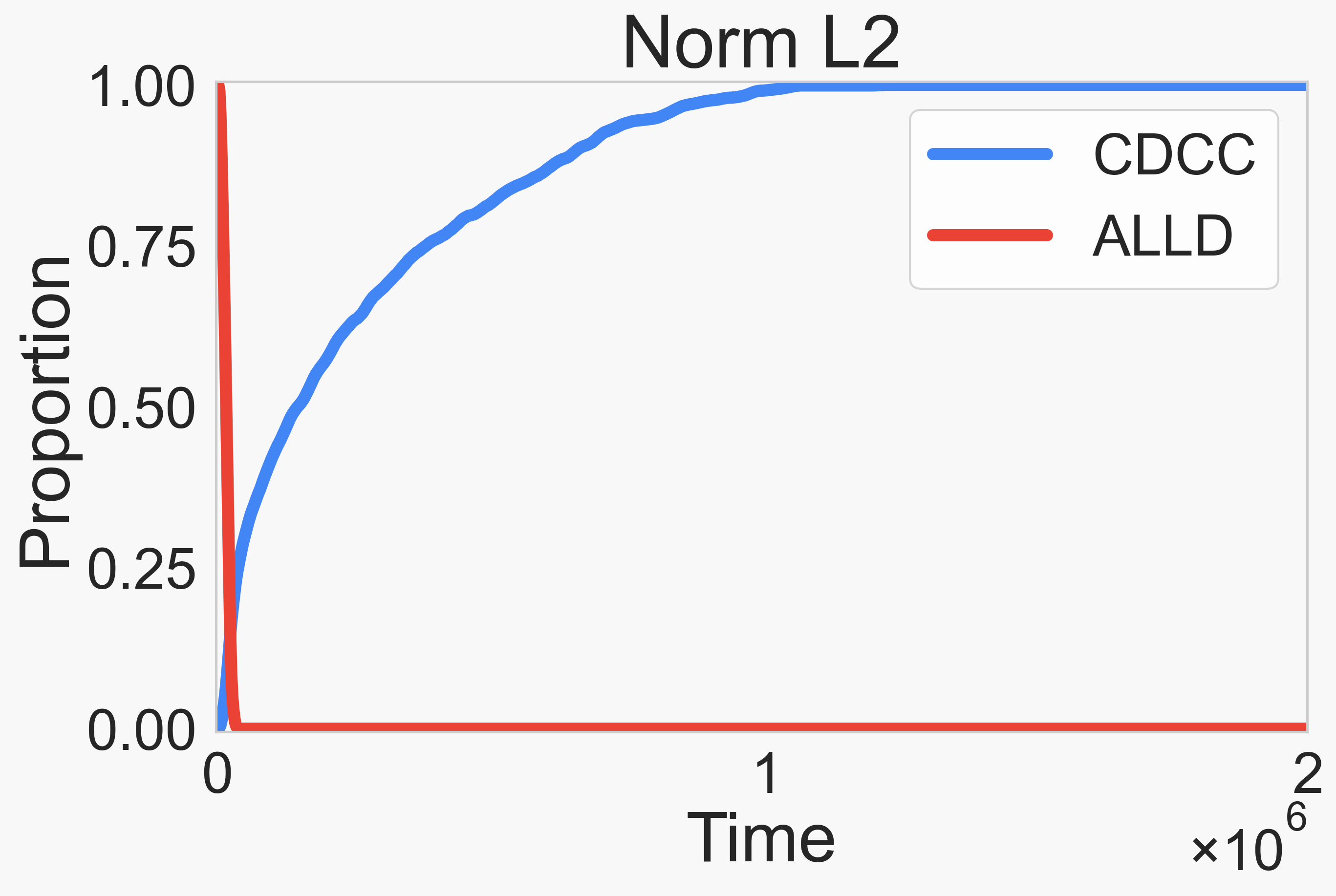}
    \end{subfigure}
    \hfill
    \begin{subfigure}[t]{0.24\textwidth}
        \centering
        \includegraphics[width=\textwidth]{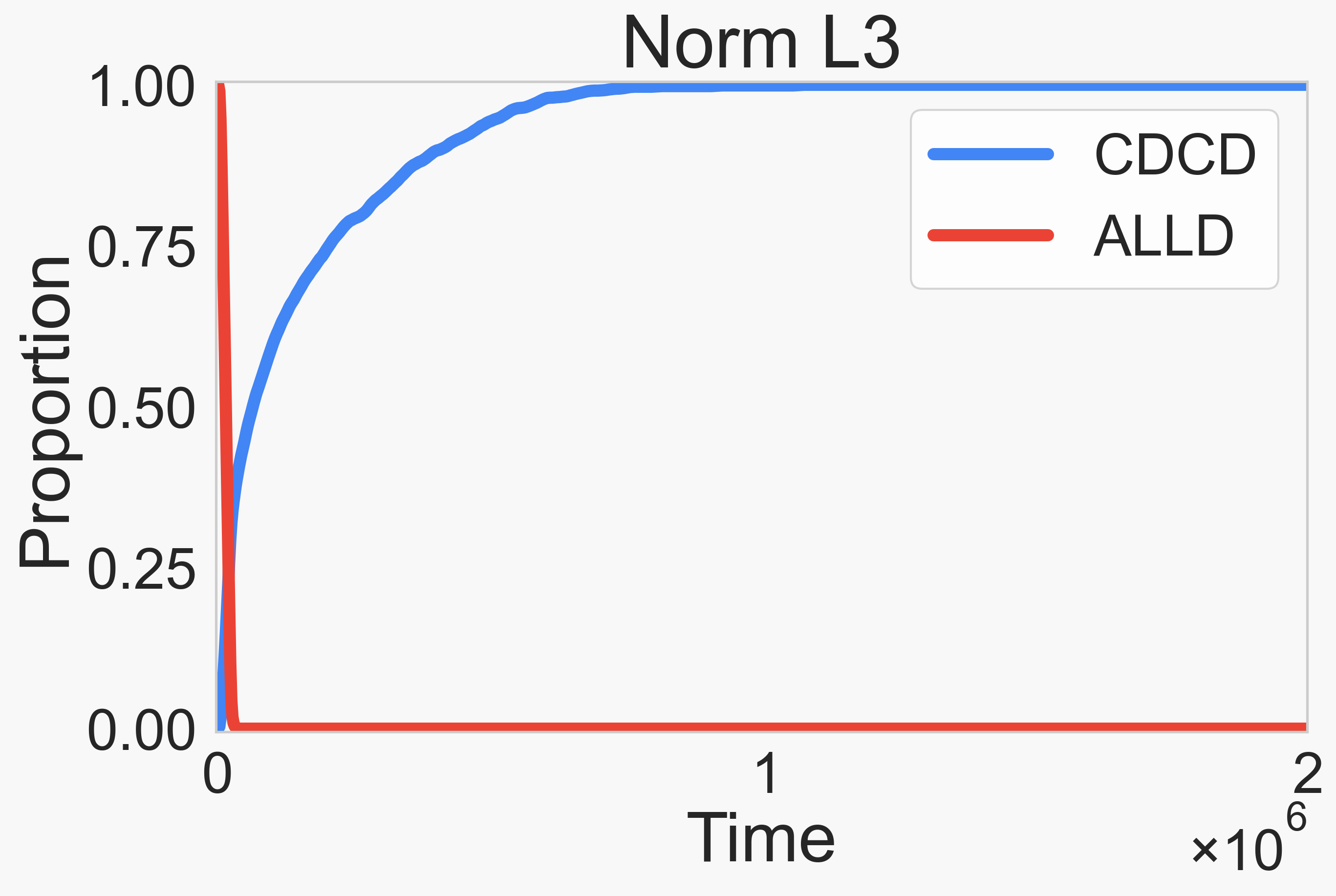}
    \end{subfigure}
    \hfill
    \begin{subfigure}[t]{0.24\textwidth}
        \centering
        \includegraphics[width=\textwidth]{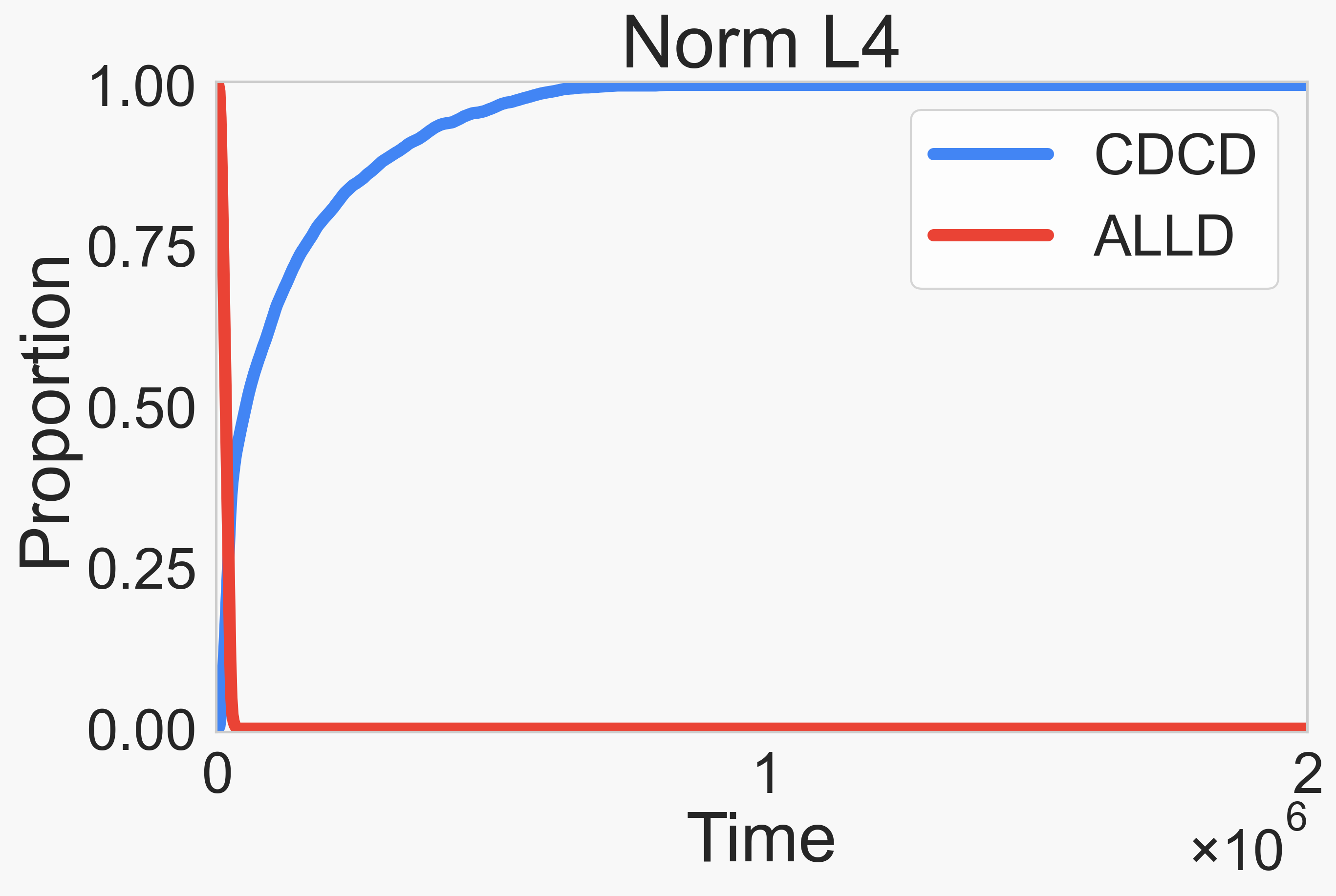}
    \end{subfigure}

    \medskip

    \begin{subfigure}[t]{0.24\textwidth}
        \centering
        \includegraphics[width=\textwidth]{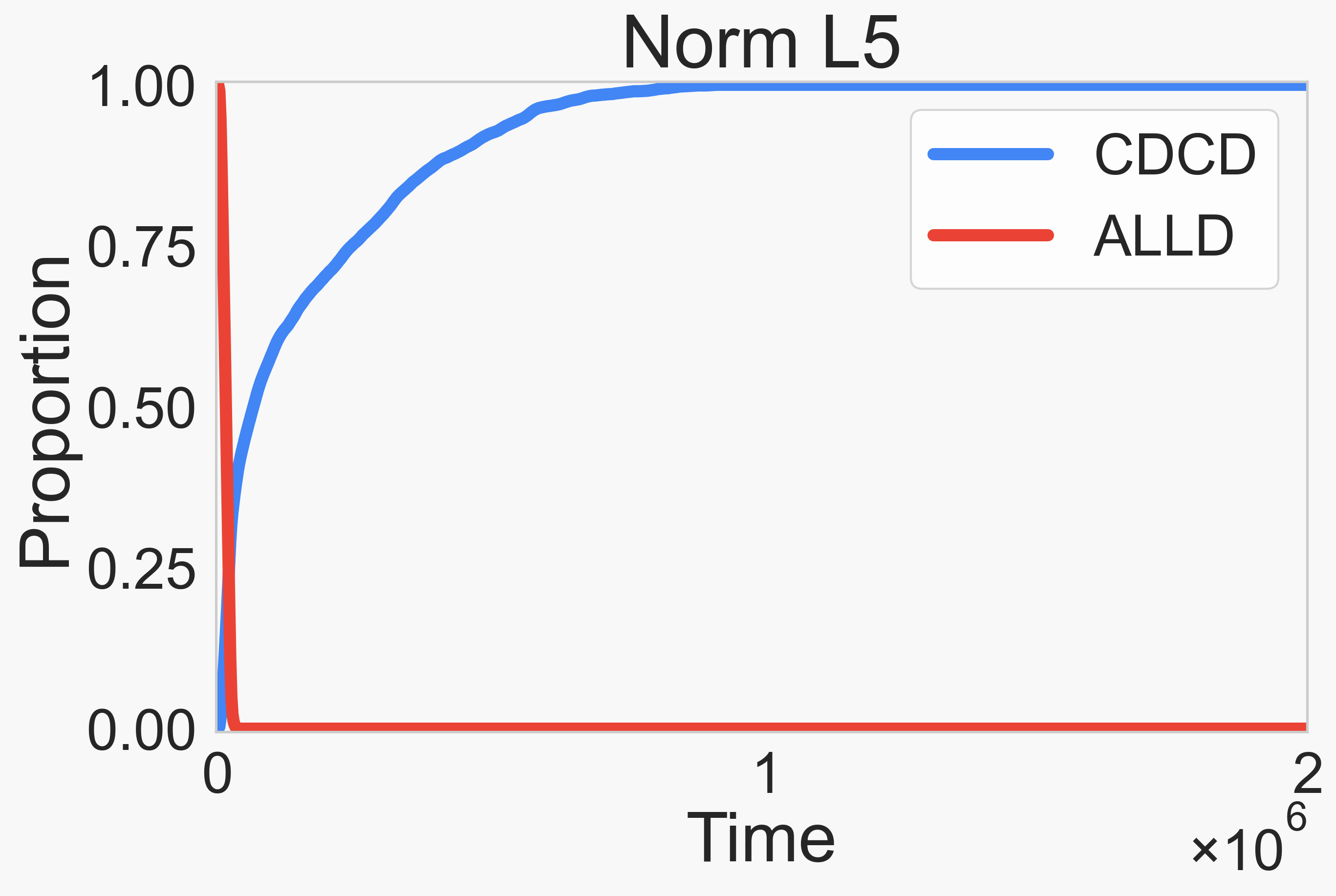}
    \end{subfigure}
    \hfill
    \begin{subfigure}[t]{0.24\textwidth}
        \centering
        \includegraphics[width=\textwidth]{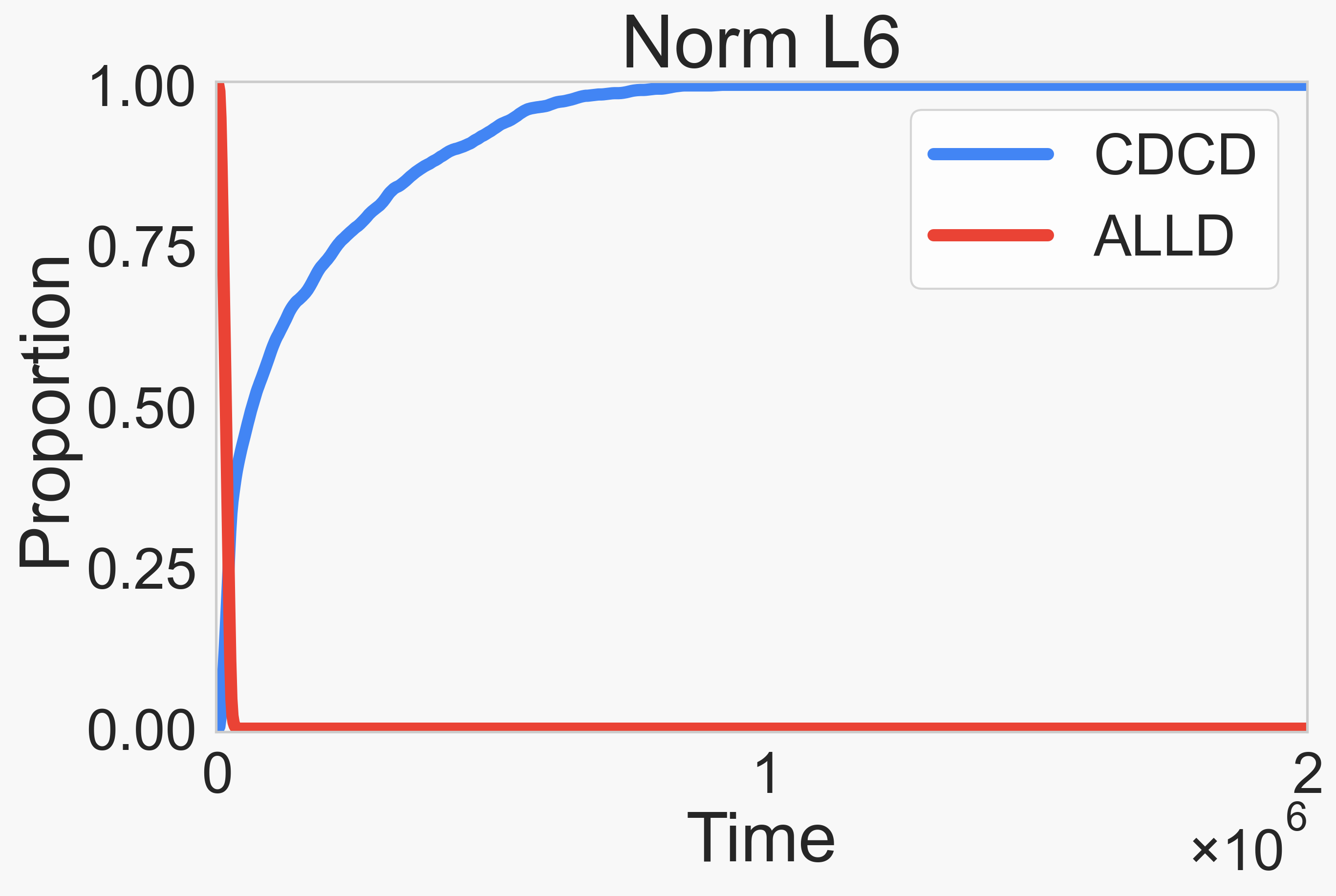}
    \end{subfigure}
    \hfill
    \begin{subfigure}[t]{0.24\textwidth}
        \centering
        \includegraphics[width=\textwidth]{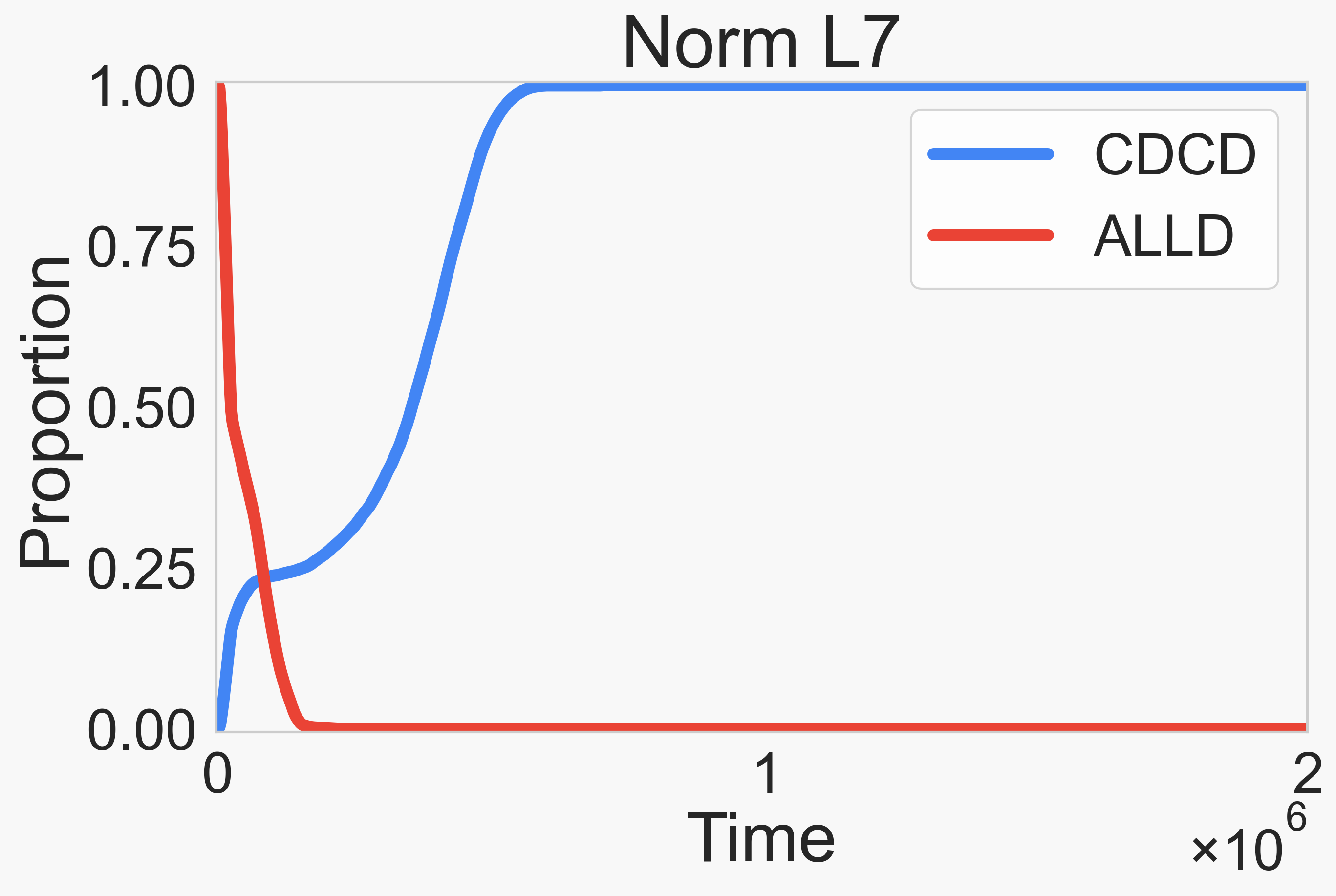}
    \end{subfigure}
    \hfill
    \begin{subfigure}[t]{0.24\textwidth}
        \centering
        \includegraphics[width=\textwidth]{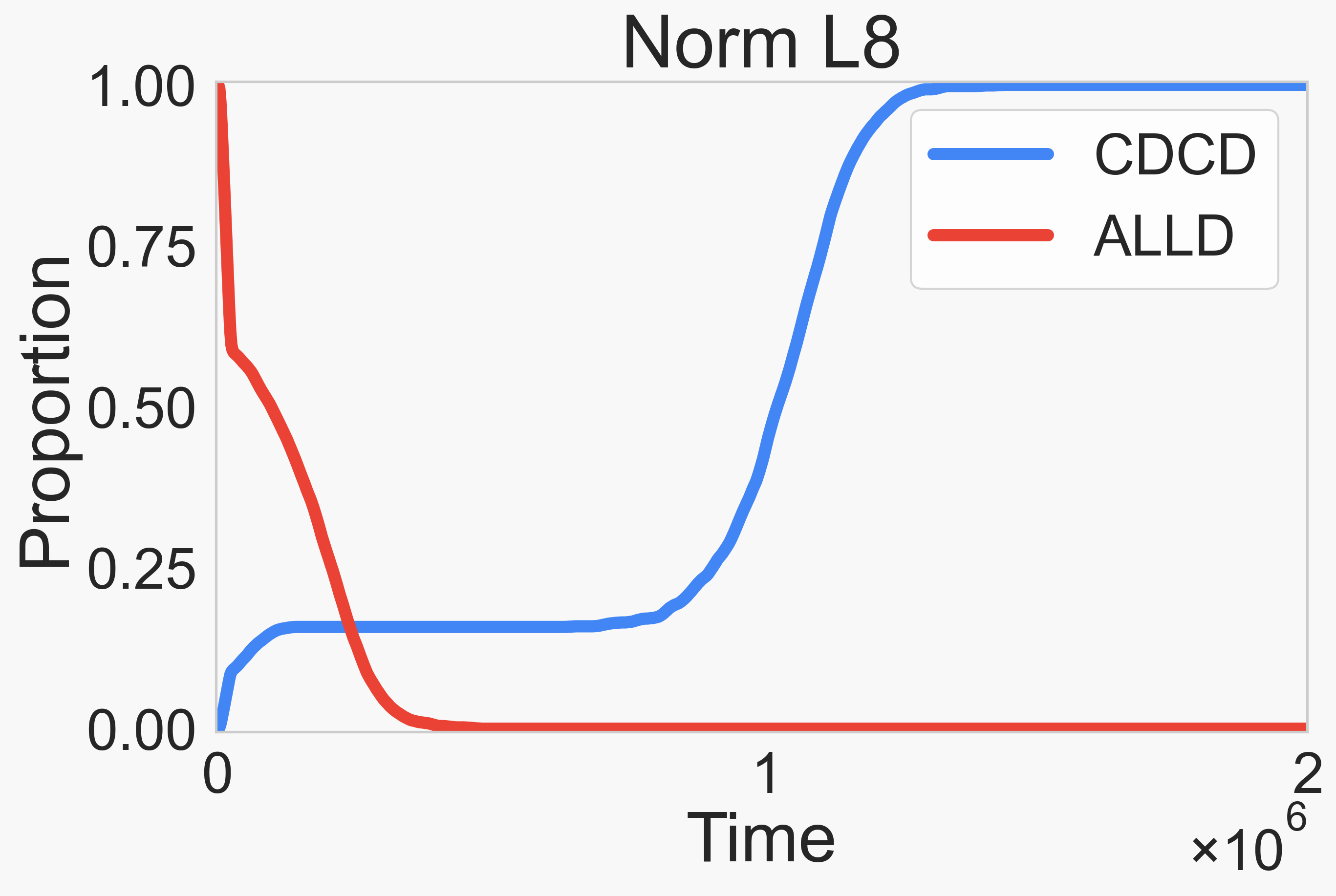}
    \end{subfigure}
    \caption{Q-learning simulation trajectories for the leading-eight. Each panel shows the proportions of one of the leading-eight strategies and the ALLD strategy over time.}
    \label{fig:ir-leading-eight-lines}
\end{figure}

By checking best-response consistency (Corollary~\ref{cor:stability-criterion-epsilon-Q}), we find that under $q = 1$ and $\varepsilon\to 0$, the leading eight are stable if and only if
\begin{equation}
    \gamma>\frac{c}{b}.
\end{equation}

\begin{proposition}[Continuity of Q-values at perfect observation]
\label{prop:ir-q-continuity}
Fix an assessment rule $d$, a candidate greedy policy $g$, an exploration rate $\varepsilon$ and a discount factor $0\leq\gamma<1$. The Q-values induced from the Bellman equations \eqref{eq:bellman-system-g} are continuous at $q=1$ from below:
\[
    \lim_{q\uparrow1}Q_g(o,a)=\left.Q_g(o,a)\right|_{q=1}.
\]
\end{proposition}

\begin{proof}
 Define the matrix
\[
    \bigl(\tilde{\bm M}_{\bm g}\bigr)_{(o,a),(o',a')}
    =\tilde T_{\bm g}(o,a,o')\,
    \mathbb I\{a'=g(o')\}.
\]
As the reward $\tilde R_{\bm g}(o,a)$ and observation kernel $\tilde T_{\bm g}(o,a,o')$ are continuous at $q=1$, $\tilde{\bm M}_{\bm g}$ is row-stochastic. Hence $\rho(\gamma\tilde{\bm M}_{\bm g})=\gamma<1$, so $\bm I-\gamma\tilde{\bm M}_{\bm g}$ is invertible and the unique Bellman solution is
\[
    \bm Q_{\bm g}
    =\bigl(\bm I-\gamma\tilde{\bm M}_{\bm g}\bigr)^{-1}
    \tilde{\bm R}_{\bm g}.
\]
The continuity of $\tilde T_{\bm g}$ established above and the fixed policy $g$ imply that $\tilde{\bm M}_{\bm g}$ is continuous at $q=1$ from below. Matrix inversion is continuous at the nonsingular matrix $\bm I-\gamma\tilde{\bm M}_{\bm g}(1)$. The continuity of $\tilde{\bm R}_{\bm g}$ then gives $\bm Q_{\bm g}\to\bm Q_{\bm g}(1)$ as $q\uparrow1$.
\end{proof}

\section{Environmental Stochasticity}
\label{sec:environmental-reciprocity}

The game environment in the previous sections is assumed to be stationary, in the sense that the reward structure is a donation game with fixed benefit $b$ and cost $c$ in every time step. However, environmental factors such as resource abundance can change in response to individual behavior. For instance, cooperation may enrich the environment, whereas defection may contribute to its degradation. Stochastic games provide a natural framework to model such environmental feedback~\cite{hilbe2018evolution}. In the following we focus on a two-state stochastic game model in which each state is a donation game with the same cost $c$ but different benefits $b_{\A}$ and $b_{\B}$ with $b_\A>b_\B$ in states $\A$ and $\B$, respectively. The system transitions to or remains in state $\A$ if both players cooperate, and state $\B$ otherwise.

The observation of each agent is the current state and we encode the symmetric greedy profile as $(g(\A),g(\B))$. Define the Q-gaps
\[
\Delta_\A:=Q^i(\A,\C)-Q^i(\A,\D),
\qquad
\Delta_\B:=Q^i(\B,\C)-Q^i(\B,\D).
\]
By Corollary~\ref{cor:stability-criterion-epsilon-Q}, stability requires \(\Delta_s>0\) when \(g(s)=\C\) and \(\Delta_s<0\) when \(g(s)=\D\).
The four symmetric greedy profiles are $(\C,\C)$, $(\C,\D)$, $(\D,\C)$, and $(\D,\D)$.

\subsection{Stability of $(\D,\D)$}

For \(g(\A)=g(\B)=\D\), the Bellman equations yield
\[
Q^i(\B,\D)=\frac{\varepsilon b_\B}{2(1-\gamma)},
\qquad
Q^i(\A,\D)=\frac{\varepsilon\bigl(b_\A(1-\gamma)+\gamma b_\B\bigr)}{2(1-\gamma)},
\]
and
\[
Q^i(\B,\C)=Q^i(\B,\D)+\Delta_\B,
\qquad
Q^i(\A,\C)=Q^i(\A,\D)+\Delta_\A,
\]
where
\[
\Delta_\A=\Delta_\B=-c+\frac{\gamma\varepsilon^2}{4}(b_\A-b_\B).
\]
Hence \((\D, \D)\) is stable if and only if
\[
\gamma\varepsilon^2(b_\A-b_\B)<4c.
\]
This condition is satisfied as $\varepsilon\to 0$. Therefore, \((\D, \D)\) is always stable under rare exploration.

\subsection{Stability of $(\C, \D)$}

For \(g(\A)=\C\) and \(g(\B)=\D\), we obtain
\[
\Delta_\A=
\frac{\gamma(2-\varepsilon)\bigl(b_{\A}(2-\varepsilon)-b_{\B}\varepsilon\bigr)-4c}
{2\bigl(2-\gamma(2-\varepsilon)\bigr)},
\]
\[
\Delta_\B=
\frac{\gamma\bigl(-b_{\A}\varepsilon^2+2b_{\A}\varepsilon-b_{\B}\varepsilon^2-4c\varepsilon+4c\bigr)-4c}
{2\bigl(2-\gamma(2-\varepsilon)\bigr)}.
\]
As the denominator is always positive, \((\C, \D)\) is stable if and only if
\[
\gamma(2-\varepsilon)\bigl(b_{\A}(2-\varepsilon)-b_{\B}\varepsilon\bigr)>4c,
\]
and
\[
\gamma\Bigl(b_{\A}\varepsilon(2-\varepsilon)-b_{\B}\varepsilon^2+4c(1-\varepsilon)\Bigr)<4c.
\]
As \(\varepsilon\to 0\), the second inequality reduces to \(\gamma<1\), and the stability depends on the first inequality, which reduces to
\[
\gamma>\frac{c}{b_{\A}}.
\]

\subsection{Stability of $(\D, \C)$}

For \(g(\A)=\D\) and \(g(\B)=\C\), we obtain
\[
\Delta_\A=
\frac{\gamma\bigl(b_{\A}\varepsilon^2-b_{\B}\varepsilon(2-\varepsilon)-4c(1-\varepsilon)\bigr)-4c}
{2\bigl(2+2\gamma-\gamma\varepsilon\bigr)},
\]
\[
\Delta_\B=
\frac{\gamma\bigl(b_{\A}\varepsilon(2-\varepsilon)-b_{\B}(2-\varepsilon)^2\bigr)-4c}
{2\bigl(2+2\gamma-\gamma\varepsilon\bigr)}.
\]
As the denominator is always positive, \((\D, \C)\) is stable if and only if
\[
\gamma\bigl(b_{\A}\varepsilon^2-b_{\B}\varepsilon(2-\varepsilon)-4c(1-\varepsilon)\bigr)<4c,
\]
and
\[
\gamma\bigl(b_{\A}\varepsilon(2-\varepsilon)-b_{\B}(2-\varepsilon)^2\bigr)>4c.
\]
As \(\varepsilon\to0\), we have \(\Delta_\A=-c\) and \(\Delta_\B=-(\gamma b_{\B}+c)/(1+\gamma)<0\), so \((\D, \C)\) cannot be stable.

\subsection{Stability of $(\C, \C)$}

For \(g(\A)=g(\B)=\C\), the Q-value gaps of the two states coincide:
\[
\Delta_\A=\Delta_\B=:\Delta
=\frac{\gamma(b_{\A}-b_{\B})}{4}(2-\varepsilon)^2-c.
\]
Therefore \((\C, \C)\) is stable if and only if
\[
\gamma(b_{\A}-b_{\B})(2-\varepsilon)^2>4c.
\]
This condition reduces to the following form under $\varepsilon\to 0$:
\[
\gamma>\frac{c}{b_{\A}-b_{\B}}.
\]

\begin{table}[ht]
    \centering
    \small
    \renewcommand{\arraystretch}{2.2}
    \begin{tabular}{>{\centering\arraybackslash}p{1.4cm} >{\centering\arraybackslash}p{4.0cm} >{\centering\arraybackslash}p{5.8cm}}
        \toprule
        State & \(Q(s,\C)\) & \(Q(s,\D)\) \\
        \midrule
        \(\A\) & \(\displaystyle \frac{b_{\A}-c}{1-\gamma}\) & \(\displaystyle b_{\A}+\frac{\gamma\left(b_{\B}-c+\gamma(b_{\A}-b_{\B})\right)}{1-\gamma}\) \\
        \(\B\) & \(\displaystyle \frac{b_{\B}-c+\gamma(b_{\A}-b_{\B})}{1-\gamma}\) & \(\displaystyle b_{\B}+\frac{\gamma\left(b_{\B}-c+\gamma(b_{\A}-b_{\B})\right)}{1-\gamma}\) \\
        \bottomrule
    \end{tabular}
    \caption{Q-values for the greedy policy \((\C,\C)\) in the two-state donation game.}
    \label{tab:env-cc-q-values-eps0}
\end{table}

\begin{table}[ht]
    \centering
    \small
    \renewcommand{\arraystretch}{1.8}
    \begin{tabular}{p{2.4cm}p{8.4cm}}
        \toprule
        Equilibrium point & Stability condition as \(\varepsilon\to0\) \\
        \midrule
        \((\D, \D)\) & always stable \\
        \((\C, \D)\) & \(\gamma>c/b_{\A}\) \\
        \((\D, \C)\) & always unstable \\
        \((\C, \C)\) & \(\gamma>c/(b_{\A}-b_{\B})\) \\
        \bottomrule
    \end{tabular}
    \caption{Thresholds for the stability of the four symmetric equilibria in the two-state donation game.}
    \label{tab:2pd-equilibria-eps0}
\end{table}

\begin{table}[ht]
    \centering
    \small
    \renewcommand{\arraystretch}{1.8}
    \begin{tabular}{p{2.4cm}p{8.4cm}}
        \toprule
        Equilibrium point & Condition to be stable for \(\varepsilon>0\) \\
        \midrule
        \((\D, \D)\) & \(\gamma\varepsilon^2(b_{\A}-b_{\B})<4c\) \\
        \((\C, \D)\) & \(\begin{aligned}[t]
            &\gamma(2-\varepsilon)\bigl(b_{\A}(2-\varepsilon)-b_{\B}\varepsilon\bigr)>4c,\ \text{and} \\
            &\gamma\bigl(-b_{\A}\varepsilon^2+2b_{\A}\varepsilon-b_{\B}\varepsilon^2-4c\varepsilon+4c\bigr)<4c
        \end{aligned}\) \\
        \((\D, \C)\) & \(\begin{aligned}[t]
            &\gamma\bigl(b_{\A}\varepsilon^2-b_{\B}\varepsilon(2-\varepsilon)-4c(1-\varepsilon)\bigr)<4c,\ \text{and} \\
            &\gamma\bigl(b_{\A}\varepsilon(2-\varepsilon)-b_{\B}(2-\varepsilon)^2\bigr)>4c
        \end{aligned}\) \\
        \((\C, \C)\) & \(\gamma(b_{\A}-b_{\B})(2-\varepsilon)^2>4c\) \\
        \bottomrule
    \end{tabular}
    \caption{Exact stability conditions for the four symmetric equilibria under positive exploration.}
    \label{tab:2pd-equilibria-eps}
\end{table}

\newpage

\section{Network Reciprocity}
\label{sec:network-reciprocity}
\label{sec:gr-model}

In previous sections, we considered dyadic interactions and well-mixed populations. In other words, each agent either interacted with a single partner or with every other agent in the population. However, real social interactions are often structured, in the sense that individuals meet and interact only with their neighbors, and benefits of cooperation are produced within local groups rather than globally shared. In this section we explore how networked group interactions affect the evolution of cooperation. We model the system as a hypergraph, in which each node represents an agent and each hyperedge collects the participants of a game~\cite{sheng2024strategy}. At each time step, each agent selects an action for each observation, interacts with the other participants in the hyperedge to which it belongs, and receives rewards. The state of the environment is then updated and the system transitions to the next time step. As before, agents perform batch updates after $\hat{B}$ time steps. We consider a two-state public goods game model with action-dependent state transitions. We show that the hypergraph order $k$ plays a critical role in the stability of cooperation, and derive analytical conditions for stability of cooperation.

We consider a setting in which each $k$-player group repeatedly plays a public goods game with state space $\{\A, \B\}$ and action space $\{\C, \D\}$. The synergy factors are $r_{\A}$ and $r_{\B}$ in states $\A$ and $\B$, respectively, with $k > r_{\A} > r_{\B} > 1$ and the cost $c$ is the same in both states. Writing $n_C$ and $n_D = k - n_C$ for the numbers of cooperators and defectors in the game, the immediate reward of player $i$ is
\begin{equation}
\label{eq:gr-payoff}
    R^i(s, \bm{a}) = \frac{r_s\, c\, n_C}{k} - c\,\mathbb{1}\{a_i = C\},
    \qquad s \in \{A, B\}.
\end{equation}
As contributing to the public pool only returns $r_s c / k$ of the cost to the contributor, the one-shot game is a social dilemma with defection as the best response.

The state transitions depend on the current state and the joint action. Defection degrades the environment in proportion to its prevalence, whereas recovery requires cooperation of the whole group:
\begin{equation}
\label{eq:gr-transitions}
    P(A \to B) = q_c \, \frac{n_D}{k},
    \qquad
    P(B \to A) = q_r \, \mathbb{1}\{n_C = k\},
\end{equation}
with degradation and recovery scales $q_c, q_r \in [0, 1]$. A single defector raises the collapse probability by $q_c/k$; a single defector in state $B$ removes the recovery prospect entirely.

As in Section~\ref{sec:environmental-reciprocity}, each player observes the current state of the game and we encode the symmetric profile as $(g(\A),g(\B))$ and define the Q-gaps
\[
\Delta_{\A} := Q(\A,\C)-Q(\A,\D),
\qquad
\Delta_{\B} := Q(\B,\C)-Q(\B,\D),
\]
where $Q(s,a)$ is the solution of the Bellman system \eqref{eq:bellman-system-g}. Similarly by Corollary~\ref{cor:stability-criterion-epsilon-Q}, best-response consistency requires $\Delta_s>0$ when $g(s)=\C$ and $\Delta_s<0$ when $g(s)=\D$.

Throughout this section $K := 1-\gamma+\gamma q_r$. We compute the Q-values and Q-gaps for the four symmetric greedy profiles $(\D, \D)$, $(\C, \D)$, $(\D, \C)$, and $(\C, \C)$ under $\varepsilon \to 0$. According to Proposition~\ref{prop:small-exploration-robustness}, the stability also holds for sufficiently small $\varepsilon > 0$.

\subsection{Stability of $(\D, \D)$}

Under full defection $(\D, \D)$, both continuation values vanish and the Q-value gaps are given by
\[
\Delta_{\A}=\frac{c\,(r_{\A}-k)}{k},
\qquad
\Delta_{\B}=\frac{c\,(r_{\B}-k)}{k}.
\]
As $k > r_{\A}$ always holds, $(\D, \D)$ is always stable.

\subsection{Stability of $(\C, \D)$}

Under $(\C, \D)$, agents cooperate in state $A$ and defect in state $B$. The corresponding Q-value gaps are
\[
\Delta_{\A}=\frac{c}{k}\left(r_{\A}-k+\frac{\gamma q_c\,(r_{\A}-1)}{1-\gamma}\right),
\qquad
\Delta_{\B}=\frac{c\,(r_{\B}-k)}{k},
\]
so the profile is stable if and only if
\[
k \;<\; r_{\A}+\frac{\gamma q_c\,(r_{\A}-1)}{1-\gamma}.
\]
This upper bound diverges as $\gamma \to 1$. Intuitively, compared with the prosperous state $A$ with a higher synergy factor and a higher cooperation rate, state $\B$ acts as an everlasting punishment state. As a result, environmental discipline supports cooperation in state $A$ for any group size.

\subsection{Stability of $(\D, \C)$}

For defection in the productive state and cooperation in the degraded state, the gap in state $B$ satisfies
\[
\Delta_{\B}
=\frac{c\left(r_{\B} \tilde K - k \hat K\right)}{k \tilde K},
\qquad
\tilde K := 1-\gamma+\gamma q_c+\gamma q_r,
\quad
\hat K := 1-\gamma+\gamma q_c+\gamma q_r r_{\B}.
\]
As $\hat K \ge \tilde K$ and $k > r_{\B}$, the profile cannot be self-consistent and therefore is never stable.

\subsection{Stability of $(\C, \C)$}
\label{sec:gr-full-cooperation}

Under full cooperation in both states, the four Q-values are
\begin{align*}
Q^i(A,C)
&=
v_A,
\\
Q^i(A,D)
&=
\frac{c\,r_{\A}\,(k-1)}{k}
{}+\gamma v_A
-\frac{\gamma q_c}{k}(v_A-v_B),
\\
Q^i(B,C)
&=
v_B,
\\
Q^i(B,D)
&=
\frac{c\,r_{\B}\,(k-1)}{k}
{}+\gamma v_B.
\end{align*}
Here we abbreviate the Q-values as
\[
v_A:=\frac{c\,(r_{\A}-1)}{1-\gamma},
\qquad
v_B:=\frac{c\left((1-\gamma)(r_{\B}-1)+\gamma q_r(r_{\A}-1)\right)}{(1-\gamma)K}.
\]
These values satisfy $v_A-v_B=c\,(r_{\A}-r_{\B})/K$. The corresponding Q-gaps are
\begin{equation}
\label{eq:gr-cc-gaps}
    \Delta_{\A}=\frac{c}{k}\left(r_{\A}-k+\frac{\gamma q_c\,(r_{\A}-r_{\B})}{K}\right),
    \qquad
    \Delta_{\B}=\frac{c}{k}\left(r_{\B}-k+\frac{\gamma q_r\,(r_{\A}-r_{\B})}{K}\,k\right).
\end{equation}
Rearranging $\Delta_{\B} > 0$ gives the condition
\[
k\,\Bigl(\gamma q_r\,(r_{\A}-r_{\B}-1) - (1-\gamma)\Bigr) \;>\; -\,r_{\B}\,K,
\]
which holds for every $k$ whenever $\gamma q_r\,(r_{\A}-r_{\B}-1) \ge 1-\gamma$; otherwise it caps $k$ at $r_{\B} K /\bigl((1-\gamma)-\gamma q_r(r_{\A}-r_{\B}-1)\bigr)$. This yields the group-size threshold of this section.

\begin{theorem}[Hyperedge-order threshold]
\label{thm:gr-threshold}
Consider the two-state public goods game with payoffs \eqref{eq:gr-payoff} and transitions \eqref{eq:gr-transitions} at $\varepsilon \to 0$, and suppose $\gamma q_r (r_{\A} - r_{\B} - 1) \ge 1-\gamma$. Full cooperation is best-response consistent if and only if
\begin{equation}
\label{eq:gr-threshold}
    k \;<\; k^* \;=\; r_{\A} + \frac{\gamma q_c\,(r_{\A}-r_{\B})}{1-\gamma+\gamma q_r}.
\end{equation}
\end{theorem}

Since the second term of \eqref{eq:gr-threshold} is positive, we have $k^* > r_{\A}$ and the system contains a nonempty hyperparameter range $r_{\A} < k < k^*$ of hyperedge orders that support full cooperation. For $q_c = 0$ the threshold reduces to the trivial bound $r_{\A}$. In the $\gamma\to 1$, $q_c\to 1$ and $q_r\to 1$ limit,
\begin{equation}
\label{eq:gr-threshold-gamma1}
    k^* \;\rightarrow\; 2r_{\A}-r_{\B}.
\end{equation}

\begin{table}[ht]
    \centering
    \small
    \renewcommand{\arraystretch}{1.8}
    \begin{tabular}{>{\centering\arraybackslash}p{3.4cm}p{8.6cm}}
        \toprule
        Equilibrium point & Condition to be stable at \(\varepsilon=0\) \\
        \midrule
        \((\D, \D)\) & always stable \\
        \((\C, \D)\) & \(k < r_{\A}+\gamma q_c (r_{\A}-1)/(1-\gamma)\) \\
        \((\D, \C)\) & never stable \\
        \((\C, \C)\) & \(k < r_{\A}+\gamma q_c (r_{\A}-r_{\B})/K\) and, if \(\gamma q_r (r_{\A}-r_{\B}-1) < 1-\gamma\), additionally \(k < r_{\B} K/\bigl((1-\gamma)-\gamma q_r(r_{\A}-r_{\B}-1)\bigr)\) \\
        \bottomrule
    \end{tabular}
    \caption{Zero-noise stability conditions for the four symmetric profiles of the two-state public goods game with proportional degradation and unanimous recovery, \eqref{eq:gr-transitions}, under the standing dilemma assumption $k > r_{\A} > r_{\B}$; $K = 1-\gamma+\gamma q_r$.}
    \label{tab:gr-equilibria-eps0}
\end{table}

\begin{table}[H]
    \centering
    \small
    \renewcommand{\arraystretch}{2.2}
    \begin{tabular}{>{\centering\arraybackslash}p{1.4cm} >{\centering\arraybackslash}p{3.2cm} >{\centering\arraybackslash}p{8.6cm}}
        \toprule
        State & \(Q(s,\C)\) & \(Q(s,\D)\) \\
        \midrule
        \(\A\) & \(\displaystyle \frac{c\,(r_{\A}-1)}{1-\gamma}\) & \(\displaystyle \frac{c\,r_{\A}\,(k-1)}{k}+\frac{\gamma}{k}\left((k-1)Q^i(\A,\C)+Q^i(\B,\C)\right)\) \\
        \(\B\) & \(\displaystyle \frac{c\left(r_{\B}-1+\gamma(r_{\A}-r_{\B})\right)}{1-\gamma}\) & \(\displaystyle \frac{c\,r_{\B}\,(k-1)}{k}+\gamma Q^i(\B,\C)\) \\
        \bottomrule
    \end{tabular}
    \caption{Q-values for the greedy policy $(\C,\C)$ in the two-state public goods game under $q_c=q_r=1$, and $\varepsilon \to 0$.}
    \label{tab:gr-cc-q-values-q1-eps0}
\end{table}

\begin{figure}[H]
    \centering
    \includegraphics[width=0.5\textwidth]{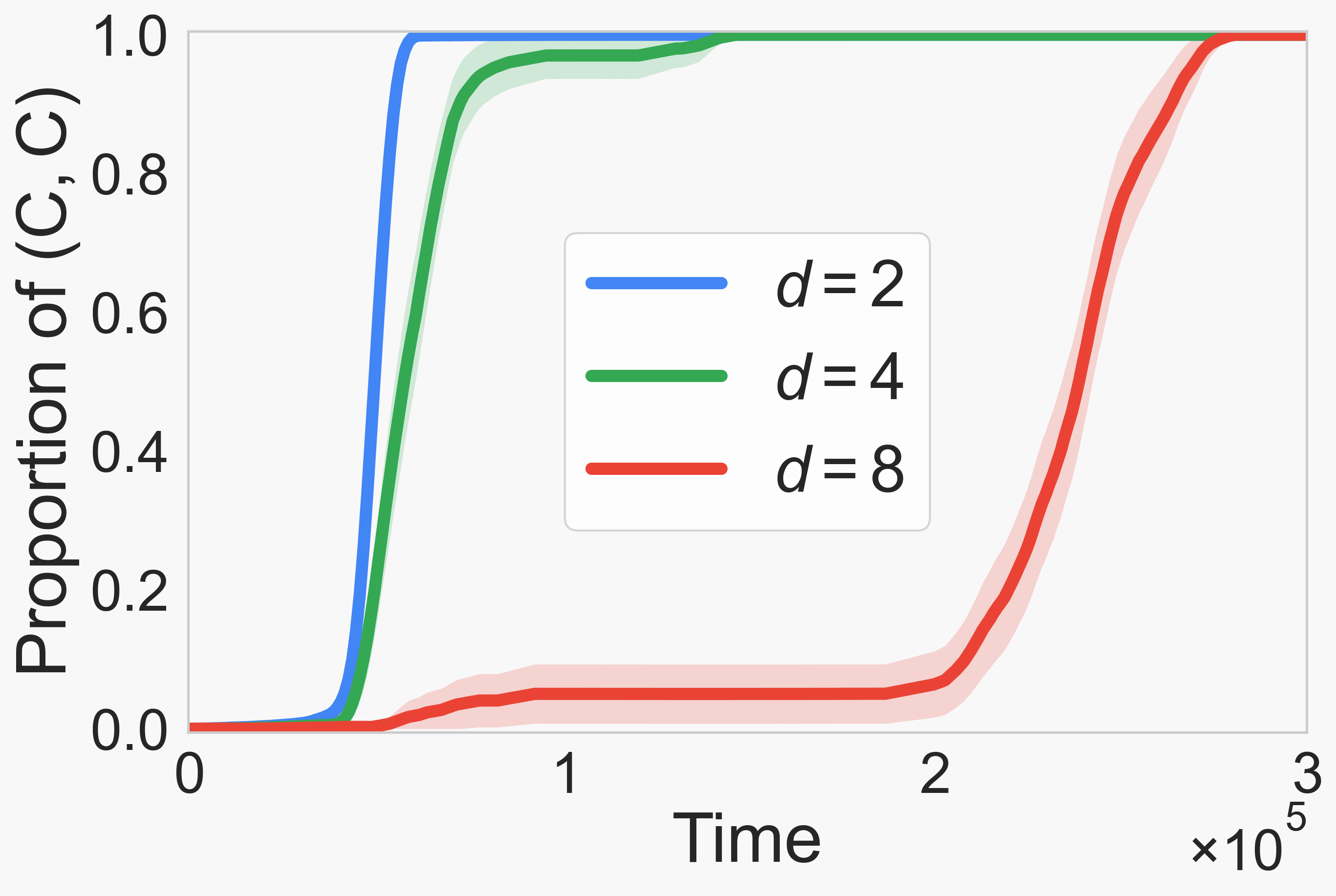}
    \caption{Agents play the two-state public goods game on random regular hypergraphs with fixed order $k=3$ and degrees $d\in\{2,4,8\}$. Curves show the mean proportion of agents whose greedy policy is $(\C,\C)$, cooperating in both environmental states, over 100 independent runs. Shaded bands indicate 95\% confidence intervals; the plotted means and standard deviations are smoothed over ten recorded observations. Parameters: $N=105$, $(r_{\A},r_{\B})=(2.8,1.2)$, $c=1$, $q_c=q_r=1$, $\alpha=0.1$, $\gamma=0.999$.}
    \label{fig:gr-degree-sweep}
\end{figure}

\newpage

\section{Demographic Stochasticity and Equilibrium Selection}
\label{sec:demographic-stochasticity}

The deterministic Q-learning dynamics of Section~\ref{sec:general-theory} explain which
Bellman-optimality solutions are locally stable. For such a framework, once a trajectory
enters the basin of attraction of a stable
equilibrium, it remains there forever. However, fluctuations around each deterministic equilibrium exist due to
finite learning rates, finite batch sizes, and exploration, which can push the
system across the boundary of a basin of attraction and induce transitions
between different stable equilibria~\cite{barfuss2023intrinsic}. We quantify the stochastic fluctuations and
derive the analytical results for one-shot coordination games introduced in
Section~\ref{sec:normal-form-games}.

\subsection{Finite-batch Q-learning dynamics as stochastic differential equations}

From the perspective of stochastic systems, the joint Q-learning dynamics are modeled as a set of stochastic differential equations (SDEs)
\begin{equation}
    \mathrm d\bm Q_t
    =
    \bm\mu\,\mathrm dt
    +\sqrt{\bm\Sigma}\,\mathrm d\bm W_t.
    \label{eq:q-sde}
\end{equation}
Here, $\bm\mu$ is the drift term representing the expected change in Q-values per unit time. Its component $\mu^i(o,a)$ has the same form as in Equation~\eqref{eq:q_ode}. The matrix \(\bm\Sigma\) is the conditional covariance of one batch update.

To derive \(\bm\Sigma\), we assume a separation between the interaction and
Q-update time scales. Between successive Q-updates, the interaction process
reaches its stationary distribution under the current joint Q-profile, and
each agent accumulates a sufficiently large replay buffer from which it
independently draws a random batch of \(B\) transitions. Conditional on the
current joint Q-profile, these agent-specific batches are treated as mutually
independent. The latent states underlying the sampled transitions therefore
follow the stationary distribution \(\tilde p(s)\). To simplify notation, we
omit the tilde from all quantities in the derivation below.

For $(i, o, a) = (j, \bar{o}, \bar{a})$, suppose that $n_{ioa}$ is the number of samples in the batch that visit entry $(i, o, a)$. Define $d(i, s, \bm{a}, o, o')$ as
\begin{equation}
    d(i, s, \bm{a}, o, o')
    =
    R^i(s, \bm{a})
    + \gamma \max_{a' \in \mathcal{A}(o')} Q^i(o', a')
    - Q^i(o, a).
\end{equation}
The probability that one sample visits entry $(i,o,a)$ is
\begin{equation}
    \pi^i_{oa}
    =
    X^i(o,a)
    \sum_{s\in\mathcal S}p(s)p^i(o\mid s).
    \label{eq:ds-entry-probability}
\end{equation}
Given that a sample visits $(i, o, a)$, the conditional variance $v^i(o, a)$ of its TD error is computed as
\begin{equation}
\begin{aligned}
    v^i_{oa}
    ={}&
    \frac{1}{\pi^i_{oa}}\sum_{s\in\mathcal S} p(s)p^i(o\mid s)X^i(o,a)
    \\
    &\times
    \sum_{\bm o^{-i}\in\mathcal O^{-i}}
    \sum_{\bm a^{-i}\in\mathcal A^{N-1}}
    \left[
        \prod_{\ell\ne i}
        p^\ell(o^\ell\mid s)X^\ell(o^\ell,a^\ell)
    \right]
    \\
    &\times
    \sum_{s^\prime\in\mathcal S}
    T\!\left(s,(a,\bm a^{-i}),s^\prime\right)
    \sum_{o_i^\prime\in\mathcal O^i}
    p^i(o_i^\prime\mid s^\prime)
    \\
    &\times
    \left[
        d\!\left(i, s,(a,\bm a^{-i}),o, o_i^\prime\right)
        -\bar\delta(i, o, a)
    \right]^2.
    \label{eq:ds-count-expanded-variance}
\end{aligned}
\end{equation}
The number of samples \(n^i_{oa}\) that visit \((i,o,a)\) therefore obeys
\(\operatorname{Binomial}(B,\pi^i_{oa})\). If \(n^i_{oa}=0\), the
update and both of its conditional moments are zero.
The law of total variance then gives the diagonal covariance as
\begin{equation}
\begin{aligned}
    \operatorname{Var}\!\left(\Delta Q^i(o,a)\right)
    ={}&
    \mathbb E\!\left[
        \operatorname{Var}\!\left(
            \Delta Q^i(o,a)\mid n^i_{oa}
        \right)
    \right]
    +
    \operatorname{Var}\!\left(
        \mathbb E\!\left[
            \Delta Q^i(o,a)\mid n^i_{oa}
        \right]
    \right)
    \\
    ={}&
    \mathbb E\!\left[
        \alpha^2v^i_{oa}
        \frac{\mathbb I\{n^i_{oa}>0\}}{n^i_{oa}}
    \right]
    +
    \operatorname{Var}\!\left(
        \alpha\bar\delta(i,o,a)
        \mathbb I\{n^i_{oa}>0\}
    \right)
    \\
    ={}&
    \alpha^2 v^i_{oa}
    \sum_{n=1}^{B}
    \frac{1}{n}
    \binom{B}{n}
    (\pi^i_{oa})^n
    (1-\pi^i_{oa})^{B-n}
    \\
    &+
    \alpha^2\bar\delta(i,o,a)^2
    \left[1-(1-\pi^i_{oa})^B\right]
    (1-\pi^i_{oa})^B.
    \label{eq:ds-total-variance-diagonal}
\end{aligned}
\end{equation}
For fixed \(\pi^i_{oa}>0\) and large \(B\), a Taylor expansion of
Equation~\eqref{eq:ds-total-variance-diagonal} gives
\begin{equation}
\begin{aligned}
    \operatorname{Var}\!\left(\Delta Q^i(o,a)\right)
    ={}&
    \alpha^2v^i_{oa}
    \left[
        \frac{1}{B\pi^i_{oa}}
        +
        \frac{1-\pi^i_{oa}}
             {B^2(\pi^i_{oa})^2}
        +
        O(B^{-3})
    \right]
    \\
    &+
    \alpha^2\bar\delta(i,o,a)^2
    \left[
        (1-\pi^i_{oa})^B
        -
        (1-\pi^i_{oa})^{2B}
    \right]
    \\
    ={}&
    \frac{\alpha^2v^i_{oa}}{B\pi^i_{oa}}
    +
    O(B^{-2})
    +
    O\!\left((1-\pi^i_{oa})^B\right)
    =
    O(B^{-1}).
    \label{eq:ds-total-variance-large-b}
\end{aligned}
\end{equation}
When \(v^i_{oa}>0\), the variance is \(\Theta(B^{-1})\), so the
corresponding diffusion amplitude is \(\Theta(B^{-1/2})\).

For $i = j$ and $(o, a) \neq (\bar{o}, \bar{a})$, let \(n^i_{\bar o\bar a}\) be the number of samples that visit
\((i,\bar o,\bar a)\), whose one-sample visitation probability is
\begin{equation}
    \pi^i_{\bar o\bar a}
    =
    X^i(\bar o,\bar a)
    \sum_{s\in\mathcal S}p(s)p^i(\bar o\mid s).
\end{equation}
A single record contains only one observation-action pair for agent \(i\), so
it cannot visit \((i,o,a)\) and \((i,\bar o,\bar a)\) simultaneously.
Conditional on \(n^i_{oa}\) and \(n^i_{\bar o\bar a}\), the two updates average
TD errors from disjoint independent records. Their conditional covariance is
therefore zero. The law of total covariance gives
\begin{equation}
\begin{aligned}
    &\operatorname{Cov}\!\left(
        \Delta Q^i(o,a),
        \Delta Q^i(\bar o,\bar a)
    \right)
    \\
    ={}&
    \mathbb E\!\left[
        \operatorname{Cov}\!\left(
            \Delta Q^i(o,a),
            \Delta Q^i(\bar o,\bar a)
            \mid n^i_{oa},n^i_{\bar o\bar a}
        \right)
    \right]
    \\
    &+
    \operatorname{Cov}\!\left(
        \mathbb E\!\left[
            \Delta Q^i(o,a)
            \mid n^i_{oa},n^i_{\bar o\bar a}
        \right],
        \mathbb E\!\left[
            \Delta Q^i(\bar o,\bar a)
            \mid n^i_{oa},n^i_{\bar o\bar a}
        \right]
    \right)
    \\
    ={}&
    0
    +
    \operatorname{Cov}\!\left(
        \alpha\bar\delta(i,o,a)
        \mathbb I\{n^i_{oa}>0\},
        \alpha\bar\delta(i,\bar o,\bar a)
        \mathbb I\{n^i_{\bar o\bar a}>0\}
    \right)
    \\
    ={}&
    \alpha^2
    \bar\delta(i,o,a)
    \bar\delta(i,\bar o,\bar a)
    \operatorname{Cov}\!\left(
        \mathbb I\{n^i_{oa}>0\},
        \mathbb I\{n^i_{\bar o\bar a}>0\}
    \right)
    \\
    ={}&
    \alpha^2
    \bar\delta(i,o,a)
    \bar\delta(i,\bar o,\bar a)
    \left[
        (1-\pi^i_{oa}-\pi^i_{\bar o\bar a})^B
        -
        (1-\pi^i_{oa})^B
        (1-\pi^i_{\bar o\bar a})^B
    \right]
    \\
    ={}&
    o(B^{-1}).
    \label{eq:ds-same-agent-off-diagonal}
\end{aligned}
\end{equation}

For $i \neq j$, as each agent independently draws samples from its buffer, the covariance $\mathrm{Cov}\!\left[\Delta Q^i(o,a), \Delta Q^j(\bar o,\bar a)\mid\bm Q\right]$ for $i\neq j$ is zero.

Combining the three cases, define the diagonal matrix
\begin{equation}
    \bm\Omega\!\left[
        (i,o,a),(j,\bar o,\bar a)
    \right]
    =
    \begin{cases}
        \displaystyle
        \dfrac{v^i_{oa}}{\pi^i_{oa}},
        & (i,o,a)=(j,\bar o,\bar a),\\[3mm]
        0,
        & (i,o,a)\ne(j,\bar o,\bar a).
    \end{cases}
    \label{eq:ds-leading-covariance-matrix}
\end{equation}
We have
$\displaystyle \bm\Sigma
=
\frac{\alpha^2}{B}\bm\Omega
+o(B^{-1})$
and
$\displaystyle \sqrt{\bm\Sigma}
=
\frac{\alpha}{\sqrt B}\sqrt{\bm\Omega}
+o(B^{-1/2})$. In other words, the diffusion term is of order $1/\sqrt{B}$.

\subsection{Stochastically stable equilibria in one-shot games}

Consider a representative Q-learner playing against an independent copy of
its current policy in the symmetric one-shot game
\eqref{eq:payoff_one_shot}. Define
\begin{equation}
    z_n=Q_n(\C)-Q_n(\D),
    \qquad
    \zeta=1-\frac{\varepsilon}{2},
    \qquad
    \eta=\frac{\varepsilon}{2}.
    \label{eq:ds-self-play-coordinates}
\end{equation}
The cooperative region is \(z_n>0\), where both copies choose \(\C\) with
probability \(\zeta\). The defection region is \(z_n<0\), where they choose
\(\C\) with probability \(\eta\). The corresponding payoff gaps from
Equations~\eqref{eq:stateless-gap-c} and \eqref{eq:stateless-gap-d} are
\begin{equation}
    \Delta_{\C}
    =
    \zeta(R-T)+\eta(S-P),
    \qquad
    \Delta_{\D}
    =
    \eta(R-T)+\zeta(S-P).
    \label{eq:ds-self-play-gaps}
\end{equation}
Both self-play equilibria are locally stable when
\begin{equation}
    \Delta_{\C}>0,
    \qquad
    \Delta_{\D}<0.
    \label{eq:ds-coordination-conditions}
\end{equation}

Fix \(\varepsilon>0\) and consider \(B\eta\gg1\). Each Q-entry is then absent
from a batch with a probability that is exponentially small in \(B\). Up to
such events, both entries are updated, and their common continuation term
\(\gamma\max_a Q_n(a)\) cancels from the gap update. Hence
\begin{equation}
    \mathbb E[\Delta z_n\mid z_n]
    =
    \begin{cases}
        \alpha(\Delta_{\C}-z_n), & z_n>0,\\
        \alpha(\Delta_{\D}-z_n), & z_n<0,
    \end{cases}
    +O\!\left(\alpha(1-\eta)^B\right).
    \label{eq:ds-self-play-gap-drift}
\end{equation}

The covariance calculation in the preceding subsection determines the gap
noise without a separate batch-count derivation. Conditional on either action,
the one-sample payoff variances are
\begin{equation}
    v_{\C}
    =
    \zeta\eta(R-S)^2,
    \qquad
    v_{\D}
    =
    \zeta\eta(T-P)^2.
    \label{eq:ds-self-play-action-variances}
\end{equation}
In the cooperative region, the visitation probabilities of \(\C\) and \(\D\)
are \(\zeta\) and \(\eta\). Projecting
Equation~\eqref{eq:ds-leading-covariance-matrix} onto the gap direction
\((1,-1)\) gives
\begin{equation}
\begin{aligned}
    \operatorname{Var}(\Delta z_n\mid z_n>0)
    &={}
    \frac{\alpha^2}{B}
    \left(
        \frac{v_{\C}}{\zeta}
        +
        \frac{v_{\D}}{\eta}
    \right)
    +o(B^{-1})
    \\
    &={}
    \frac{\alpha^2}{B}
    \left[
        \eta(R-S)^2
        +
        \zeta(T-P)^2
    \right]
    +o(B^{-1}).
    \label{eq:ds-self-play-coop-noise}
\end{aligned}
\end{equation}
In the defection region, the visitation probabilities are \(\eta\) and
\(\zeta\), so
\begin{equation}
\begin{aligned}
    \operatorname{Var}(\Delta z_n\mid z_n<0)
    &={}
    \frac{\alpha^2}{B}
    \left(
        \frac{v_{\C}}{\eta}
        +
        \frac{v_{\D}}{\zeta}
    \right)
    +o(B^{-1})
    \\
    &={}
    \frac{\alpha^2}{B}
    \left[
        \zeta(R-S)^2
        +
        \eta(T-P)^2
    \right]
    +o(B^{-1}).
    \label{eq:ds-self-play-def-noise}
\end{aligned}
\end{equation}
The same-agent off-diagonal covariance contributes only \(o(B^{-1})\), as
shown in Equation~\eqref{eq:ds-same-agent-off-diagonal}. Define
\begin{equation}
    \sigma_{\C}^2
    =
    \eta(R-S)^2+\zeta(T-P)^2,
    \qquad
    \sigma_{\D}^2
    =
    \zeta(R-S)^2+\eta(T-P)^2.
    \label{eq:ds-self-play-noise-coefficients}
\end{equation}

On the stochastic-approximation time \(\tau=\alpha n\), the leading gap
diffusion is
\begin{equation}
    \mathrm dz_\tau
    =
    \begin{cases}
        (\Delta_{\C}-z_\tau)\,\mathrm d\tau
        +\sqrt{\dfrac{\alpha}{B}}\,\sigma_{\C}\,\mathrm dW_\tau,
        & z_\tau>0,\\[3mm]
        (\Delta_{\D}-z_\tau)\,\mathrm d\tau
        +\sqrt{\dfrac{\alpha}{B}}\,\sigma_{\D}\,\mathrm dW_\tau,
        & z_\tau<0.
    \end{cases}
    \label{eq:ds-self-play-piecewise-ou}
\end{equation}
Condition~\eqref{eq:ds-coordination-conditions} places each mean inside its
corresponding greedy region.

The distances from these means to the switching boundary \(z=0\) are
\begin{equation}
    G_{\C}=\Delta_{\C},
    \qquad
    G_{\D}=-\Delta_{\D}.
    \label{eq:ds-self-play-barriers}
\end{equation}
For \(g\in\{\C,\D\}\), measure the gap from the switching boundary toward
the local mean. In this coordinate, both branches take the form
\begin{equation}
    \mathrm dy_\tau
    =
    (G_g-y_\tau)\,\mathrm d\tau
    +
    \sqrt{\frac{\alpha}{B}}\,\sigma_g\,\mathrm dW_\tau.
    \label{eq:ds-self-play-reflected-ou}
\end{equation}
Let \(m_g(y)\) be the expected time to hit \(y=0\). Its backward equation is
\begin{equation}
    (G_g-y)m_g'(y)
    +
    \frac{\alpha\sigma_g^2}{2B}m_g''(y)
    =-1,
    \qquad
    m_g(0)=0,
    \qquad
    \lim_{y\to\infty}m_g'(y)=0.
    \label{eq:ds-self-play-backward-equation}
\end{equation}
For the initial condition \(y=G_g\), the expected hitting time is
\begin{equation}
    \mathbb E_g[\tau_0]
    =
    \sqrt{\pi}
    \int_0^{G_g\sqrt{B}/(\sqrt{\alpha}\sigma_g)}
    e^{u^2}\left[1+\operatorname{erf}(u)\right]\,\mathrm du.
    \label{eq:ds-self-play-exact-passage-time}
\end{equation}
For non-degenerate \(\sigma_g>0\), as \(B/\alpha\to\infty\), its logarithmic
weak-noise asymptotic is
\begin{equation}
    \log\mathbb E_g[\tau_0]
    \sim
    \frac{B G_g^2}{\alpha\sigma_g^2}.
    \label{eq:ds-self-play-passage-exponent}
\end{equation}
The corresponding switching rates satisfy
\begin{equation}
    \lambda_{\C\to\D}
    \asymp
    \exp\!\left(
        -\frac{B G_{\C}^2}{\alpha\sigma_{\C}^2}
    \right),
    \qquad
    \lambda_{\D\to\C}
    \asymp
    \exp\!\left(
        -\frac{B G_{\D}^2}{\alpha\sigma_{\D}^2}
    \right).
    \label{eq:ds-self-play-transition-rates}
\end{equation}
Approximating the rare switching process by a two-state Markov chain yields
\begin{equation}
    \frac{\rho_{\C}}{\rho_{\D}}
    \asymp
    \exp\!\left[
        \frac{B}{\alpha}
        \left(
            \frac{G_{\C}^2}{\sigma_{\C}^2}
            -
            \frac{G_{\D}^2}{\sigma_{\D}^2}
        \right)
    \right].
    \label{eq:ds-self-play-stationary-odds}
\end{equation}
Within this self-play reduction, mutual cooperation is selected in the
weak-noise limit when
\begin{equation}
    \frac{G_{\C}^2}{\sigma_{\C}^2}
    >
    \frac{G_{\D}^2}{\sigma_{\D}^2}.
    \label{eq:ds-self-play-selection-criterion}
\end{equation}
The reverse inequality selects mutual defection, while equality gives the two
branches the same leading exponential weight. The relevant robustness measure
is the squared gap normalized by its local noise coefficient.

As \(\varepsilon\to0\), one has
\(\zeta\to1\), \(\eta\to0\), and
\begin{equation}
    G_{\C}\to R-T,
    \qquad
    G_{\D}\to P-S,
    \qquad
    \sigma_{\C}^2\to(T-P)^2,
    \qquad
    \sigma_{\D}^2\to(R-S)^2.
\end{equation}
Equation~\eqref{eq:ds-self-play-selection-criterion} therefore has the formal
limit, provided \(T\ne P\) and \(R\ne S\),
\begin{equation}
    \frac{(R-T)^2}{(T-P)^2}
    >
    \frac{(P-S)^2}{(R-S)^2}.
    \label{eq:ds-self-play-zero-exploration-selection}
\end{equation}
Under the coordination conditions \(R>T\) and \(P>S\), this is equivalent to
\begin{equation}
    (R-T)\lvert R-S\rvert
    >
    (P-S)\lvert T-P\rvert.
    \label{eq:ds-self-play-zero-exploration-coordination}
\end{equation}

\begin{figure}[H]
    \centering
    \includegraphics[width=0.5\textwidth]{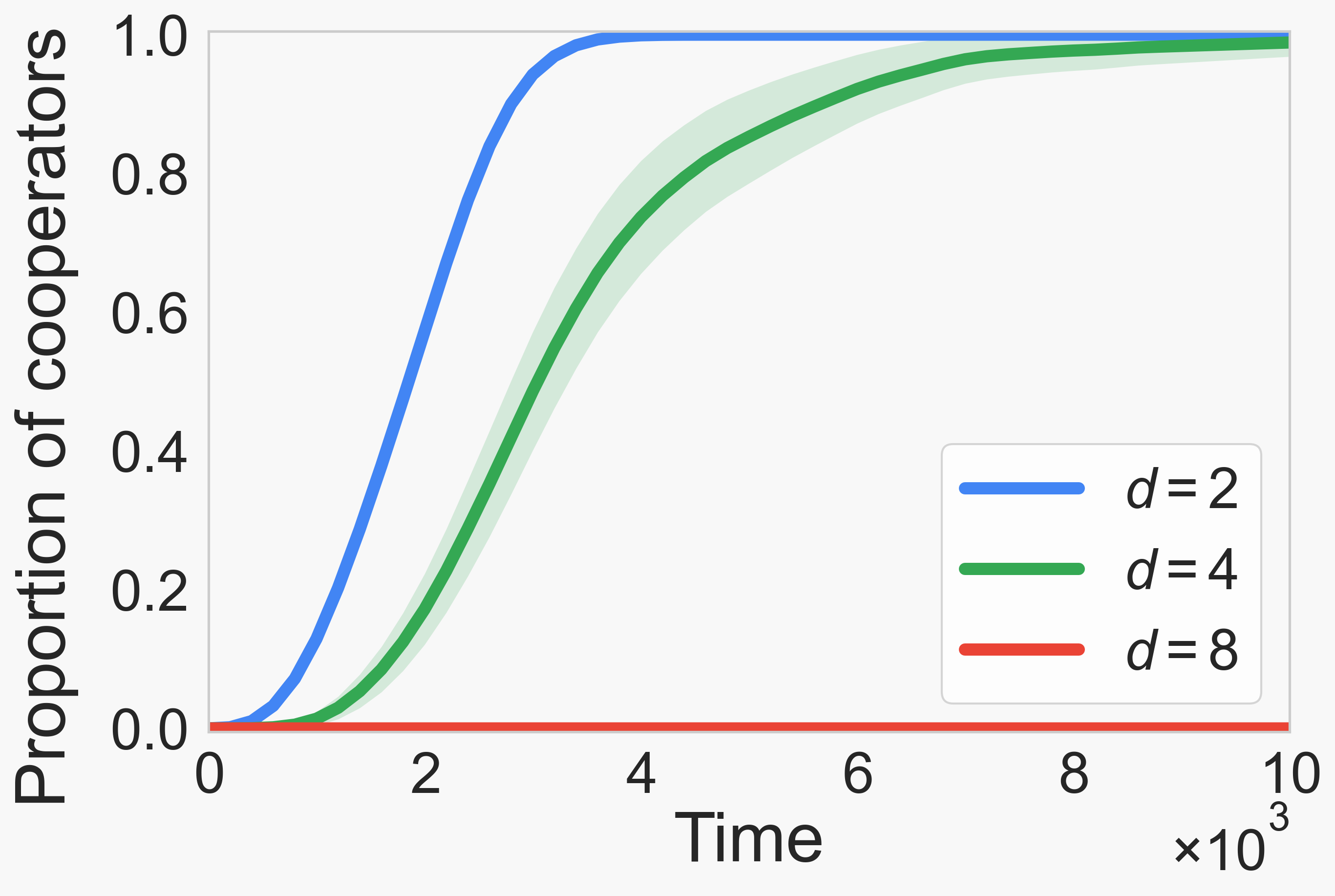}
    \caption{A population of Q-learning agents play the Stag Hunt game on random regular graphs. Curves indicate the mean proportion of agents whose greedy action is $\C$ over 100 independent runs and shaded bands indicate 95\% confidence intervals. Parameters: population size $N=100$ agents, network degrees $d\in\{2,4,8\}$, payoffs $(R,S,T,P)=(8,0,2,1)$, learning rate $\alpha=0.1$, discount factor $\gamma=0$, exploration rate $\varepsilon=0.1$, and agents' initial greedy policies are all defection.}
    \label{fig:ds-regular-graph-stag-hunt}
\end{figure}

\section{Meta-Policy Learning over Reactive Strategies}
\label{sec:reactive-meta-policy}

In previous sections, each agent learns a policy over primitive actions. Beyond this, the learning process can happen on different hierarchical levels. For example, in social learning, a learner may select a behavioral rule by observing the behavior of others.

Drawing a parallel, we test the validity of direct reciprocity under such a meta-policy learning process. In particular, we consider a homogeneous well-mixed population in which individuals use a meta-policy over a finite set of reactive strategies. The meta-action $a_m$ is parameterized by the reactive-strategy pair $\bm p_m=(p_{m1},p_{m2})$, where $p_{m1}$ and $p_{m2}$ are the probabilities of cooperating after the opponent previously chose cooperation and defection, respectively~\cite{nowak1992tit}.

The payoff $A(a_m,a_n)$ of meta-action $\bm p_m=(p_{m1},p_{m2})$ interacting with
$\bm p_n=(p_{n1},p_{n2})$ is computed as follows. Denote their stationary
cooperation probabilities by $p_{mn}$ and $p_{nm}$, respectively.
Reactivity gives
\begin{equation}
    p_{mn}
    =p_{m1}p_{nm}+p_{m2}(1-p_{nm}),
    \qquad
    p_{nm}
    =p_{n1}p_{mn}+p_{n2}(1-p_{mn}).
    \label{eq:meta-stationary-cooperation-system}
\end{equation}
Solving these equations gives
\begin{align}
    p_{mn}
    &=
    \frac{p_{m2}+(p_{m1}-p_{m2})p_{n2}}
    {D(\bm p_m,\bm p_n)},
    \\
    p_{nm}
    &=
    \frac{p_{n2}+(p_{n1}-p_{n2})p_{m2}}
    {D(\bm p_m,\bm p_n)}.
    \label{eq:meta-stationary-cooperation-rates}
\end{align}
Here
$D(\bm p_m,\bm p_n)=1-(p_{m1}-p_{m2})(p_{n1}-p_{n2})$. The long-run
average payoff of meta-action $a_m$, parameterized by $\bm p_m$, against
meta-action $a_n$, parameterized by $\bm p_n$, is
\begin{equation}
    A(a_m,a_n)
    =
    b\,p_{nm}
    -c\,p_{mn}.
    \label{eq:meta-reactive-payoff}
\end{equation}
Let $\bm X=\bigl(X(a_1),\ldots,X(a_M)\bigr)$ be the  meta-policy, where $X(a_m)\geq0$ and
$\sum_m X(a_m)=1$. The expected payoff of meta-action $a_m$ against the
current population and the population-average payoff are
\begin{equation}
    r(a_m)
    =
    \sum_{n=1}^M A(a_m,a_n)X(a_n),
    \qquad
    \overline r
    =
    \sum_{m=1}^M X(a_m)r(a_m)
    =
    \bm X^{\mathsf T}\bm A\bm X.
    \label{eq:meta-population-payoffs}
\end{equation}
The corresponding learning dynamics of the meta-policy are
\begin{equation}
    \frac{\mathrm d X(a_m)}{\mathrm dt}
    =
    \frac{\alpha_X}{\tau}
    X(a_m)
    \left[r(a_m)-\overline r\right],
    \qquad
    m=1,\ldots,M.
    \label{eq:meta-self-play-replicator}
\end{equation}


\clearpage

\bibliographystyle{unsrt}
\bibliography{refs.bib}